\documentclass[
	a4paper, % Paper size, use either a4paper or letterpaper
	10pt, % Default font size, can also use 11pt or 12pt, although this is not recommended
]{LTJournalArticle}

\usepackage[space]{grffile} 
\usepackage{textcomp, subcaption, longtable, multirow, multicol, booktabs, amsfonts, amssymb, hyperref, cleveref, etoolbox, here, xr, arydshln}  
\usepackage[detect-all=true,group-minimum-digits=4]{siunitx} 
\DeclareSIUnit\calorie{cal} 
\newcommand{\xtb}{\textit{x}\texttt{TB}\,}
\newcommand{\gfn}{GFNi-\xtb}
\newcommand{\crest}{\texttt{CREST}\,}
\newcommand{\gfna}{GFN0-\xtb} 
\newcommand{\gfnb}{GFN1-\xtb} 
\newcommand{\gfnc}{GFN2-\xtb} 
\newcommand{\gfnf}{GFN-FF\,}
\newcommand{\bm}[1]{\mathbf{#1}}
\newcommand{\centered}[1]{\begin{tabular}{l} #1 \end{tabular}}
\newsavebox\ltmcbox

\newcommand{\newsecnumstyle}{
	\renewcommand{\thesection}{S} 
	\renewcommand{\thesubsection}{\thesection.\arabic{subsection}}
	\renewcommand{\thetable}{\thesection. \arabic{table}~}
	\renewcommand{\thefigure}{\thesection. \arabic{figure}~}
	\setcounter{page}{1}
	\renewcommand*{\thepage}{S\arabic{page}}
}
\title{Comparative analysis of GFN methods in geometry optimization of small organic semiconductor molecules: A DFT benchmarking study} % Article title, use manual lines breaks (\\) to beautify the layout

\author{%
	Steve Cabrel Teguia Kouam\textsuperscript{1}, Jean-Pierre Tchapet Njafa\textsuperscript{2}, Raoult Dabou Teukam\textsuperscript{3},\\ Patrick Mvoto Kongo\textsuperscript{2}, Jean-Pierre Nguenang\textsuperscript{1} and Serge Guy Nana Engo\textsuperscript{2}\thanks{Corresponding author: \href{mailto:steve.teguia@facsciences-uy1.cm}{steve.teguia@facsciences-uy1.cm}} \\
	}

\date{\footnotesize\textsuperscript{\textbf{1}}Department of Physics, Faculty of Science, University of Douala, Po. Box 24157, Douala, Cameroon\\ \textsuperscript{\textbf{2}}Department of Physics, Faculty of Science, University of Yaounde 1, Po. Box 812, Yaounde, Cameroon\\ \textsuperscript{\textbf{3}}Department of Engineering, University of Quebec in Abitibi-Témiscamingue, Quebec, Canada}

\DeclareUnicodeCharacter{2009}{\textendash}

\begin{document}
\onecolumn
\maketitle % Output the title section

\begin{abstract}
	\noindent This study benchmarks the GFN family of semi-empirical methods (\gfnb, \gfnc, \gfna, and \gfnf) against density functional theory (DFT) for the evaluation of optimized molecular geometries and electronic properties of small organic semiconductor molecules. This work offers a systematic assessment of these computationally efficient quantum chemical methods and their accuracy-cost profiles when applied to a challenging class of systems, characterized, for instance, by extended $\pi$-conjugation, conformational flexibility, and sensitivity of properties to subtle structural changes. Two datasets are evaluated: a QM9-derived subset of small organic molecules and the Harvard Clean Energy Project (CEP) database of extended $\pi$-systems relevant to organic photovoltaics. Structural agreement is quantified using heavy-atom RMSD, equilibrium rotational constants, bond lengths, and angles, while electronic property prediction is assessed via HOMO–LUMO energy gaps. Computational efficiency is assessed via CPU time and scaling behavior. \gfnb and \gfnc demonstrate the highest structural fidelity, while \gfnf offers an optimal balance between accuracy and speed, particularly for larger systems. The results indicate that GFN-based methods are suitable for high-throughput molecular screening of small organic semiconductors, with the choice of method depending on accuracy-cost trade-offs. The findings support the deployment of GFN approaches in computational pipelines for the discovery of organic electronics and materials, providing information on their strengths and limitations relative to established DFT methods. \\
	
\end{abstract}

\section{Introduction}\label{sec:intro}

Accurate yet computationally tractable molecular geometries are paramount for understanding and predicting the properties of chemical systems, particularly in applications where structure fundamentally dictates function. The rational design and discovery of novel organic semiconductor molecules stand as a contemporary and challenging application domain, a domain fraught with complexities, including managing the conformational flexibility of larger systems, accurately capturing subtle electronic effects within extended $\pi$-conjugated networks, and judiciously balancing the description of covalent versus non-covalent interactions, a critical aspect. Their physical, chemical and electronic properties, essential for device performance in areas ranging from energy harvesting to optoelectronics, are intimately linked to their precise molecular geometry. For decades, quantum chemistry (QC) methods have provided the theoretical bedrock for simulating material properties. However, while highly accurate, traditional \textit{ab initio} methods such as Hartree-Fock (HF) and density functional theory (DFT) often present a significant computational bottleneck, their inherent resource intensity and time demands pose a significant hurdle for high-throughput screening and large-scale complex systems in modern materials research. This bottleneck has spurred the development of multilevel computational strategies \cite{bannwarth_extended_2021}, where computationally lighter methods are used for initial screening, followed by more accurate calculations for refinement. Semiempirical (SE) methods, rooted in simplified HF theory (e.g., NDDO approximations with neglect of differential-diatomic overlap \cite{pople_approximate_1965-1}) or tight-binding (TB) approximations to DFT \cite{porezag_construction_1995}, have historically served as essential components within these multi-scale modeling frameworks for striking a balance between computational cost and accuracy in quantum chemical calculations. The rigorous assessment of these approximate quantum chemical approaches remains critical for guiding their reliable application and informing future methodological advancements.

While offering substantial speed advantages, earlier generations of SE methods frequently exhibited limitations in overall reliability across diverse chemical spaces. For instance, older methods such as AM1 or PM6 often struggled with non-covalent interactions, accurate representation of thermochemical and reaction energies, or electron delocalization, particularly for extended $\pi$-conjugated systems relevant to organic semiconductors. Extensive benchmarking efforts have compared the performance of DFT and SE methods for various molecular properties, including geometry, optical gaps, and conformational energies, in diverse systems such as organic photovoltaics, protein-ligand complexes, and nanostructures \cite{zheng_performance_2005, schenker_assessment_2011, yilmazer_comparison_2013, tortorella_benchmarking_2016}. Self-consistent charge density functional tight binding methods (SCC-DFTB) \cite{elstner_self-consistent-charge_1998, gaus_dftb3_2011}, in particular, have proven valuable for approximating \textit{ab initio} calculations with significantly reduced computational cost. However, persistent challenges in accurately modeling non-covalent interactions, the demanding parameterization for diverse chemical systems, and achieving consistent accuracy across varying molecular structures and reaction profiles highlighted the persistent need for methodological advancements \cite{christensen_semiempirical_2016} within approximate quantum chemical frameworks.

It was in this context that Grimme \textit{et al.} introduced the \emph{Geometry, vibrational Frequency, Noncovalent interactions, eXtended Tight-Binding} (GFN-\xtb) family of methods \cite{grimme_robust_2017, bannwarth_gfn2-xtbaccurate_2019, pracht_robust_2019, spicher_robust_2020}. These methods represent a modern evolution of tight-binding approaches, specifically designed to address many of the limitations of older SE models through advanced parameterization, improved dispersion corrections, and a more rigorous treatment of self-consistent charge interactions. They are engineered to strike a compelling balance between computational efficiency and accuracy across a broad spectrum of target properties. The GFN framework encompasses several levels of theory, including \gfnb, \gfnc, \gfna, and \gfnf. These methods are rapidly gaining traction for efficient computational investigations in a wide array of chemical systems, from large transition-metal complexes \cite{bursch_structure_2019} to complex biomolecular assemblies \cite{schmitz_quantum_2020}. Previous work has shown the utility of GFN methods in the realm of organic electronic systems, including studies on molecular muscles, covalent organic frameworks, and electronic coupling integrals \cite{kohn_quickstart_2022, kohn_quantum_2023, kohn_efficient_2023, kohn_semi-automated_2024}. In addition, their reliability in predicting molecular geometry and electronic structure for the design of effective organic solar cells (OSCs), organic light-emitting diodes (OLEDs), and organic field-effect transistors (OFETs) has also been demonstrated \cite{menzel_silico_2022, kohn_quickstart_2022, kohn_efficient_2023, kohn_quantum_2023, kohn_semi-automated_2024}. Their integration into machine learning-driven materials discovery pipelines, enabling tasks such as geometry optimization, conformational analysis, and understanding complex interactions, further highlights their growing importance \cite{chen_reorganization_2022, chen_physics-inspired_2023, li_machine_2023, nigam_tartarus_2023, anstine_aimnet2_2024}. While these applications underscore GFN's compelling potential, a systematic, side-by-side evaluation across diverse organic semiconductor molecules, specifically focusing on geometry optimization against high-level DFT, remains a critical necessity.

Despite their considerable success, self-consistent GFN methods still grapple with inherent self-interaction errors (SIE) resulting from the absence of exact Fock exchange in the underlying DFT approximation. This is particularly problematic in systems with significant charge delocalization or polarity, leading to potential failures such as overdelocalization, inaccurate energy barriers, and distorted bond lengths. Such issues can also impede the convergence of self-consistent field (SCF) calculations. Consequently, non-iterative methods such as \gfna and force-field approaches such as \gfnf are often employed as practical alternatives. Given the increasing demand for efficient computational screening tools in materials science, especially for organic electronics, where performance metrics (e.g., power conversion efficiency, singlet-triplet gaps, and oscillator strengths) are intricately linked to both electronic structure and, importantly, molecular geometry, a critical assessment of GFN methods' reliability specifically for molecular geometry optimization and electronic structure prediction is therefore essential. For the quantum chemistry community, a thorough benchmarking exercise of these widely used approximate methods provides valuable insights into their capabilities and limitations when applied to challenging chemical systems, and sheds light on the efficacy of different theoretical approximations.

In this paper, we present a systematic comparative analysis of the performance of the GFN family of methods against DFT for the optimization of the geometry and prediction of the electronic properties of small organic molecules, with a particular focus on systems relevant to organic photovoltaics (OPVs). These molecules collectively serve as a demanding yet highly relevant benchmark set, as both their structural and electronic properties are critical, and computational efficiency is highly desirable, making them ideal test cases for approximate quantum chemical methods. To this end, we have curated and evaluated two different reference datasets: a subset of small organic molecules derived from the QM9 database \cite{ramakrishnan_quantum_2014}, specifically filtered to mimic semiconductor behavior based on HOMO-LUMO gap criteria, i.e., selecting systems with smaller band gaps often associated with $\pi$-conjugated character, analogous to features found in larger organic semiconductors, and a selection of extended $\pi$ systems from the well-established Harvard Clean Energy Project (CEP) database \cite{hachmann_harvard_2011}. The QM9 dataset offers access to readily available high-accuracy DFT benchmark geometries and properties \cite{ramakrishnan_quantum_2014}, derived from computations at the B3LYP/6-31G(2df,p) level in gas phase, while the CEP dataset provides larger systems relevant to real-world OPV applications. Structural agreement is quantified using multiple metrics, including the heavy atom RMSD, the radius of gyration, equilibrium rotational constants, bond lengths and angles, and the energy gap between the highest occupied molecular orbital (HOMO) and the lowest unoccupied molecular orbital (LUMO), a key electronic descriptor. Our primary objective is to rigorously quantify the extent to which GFN methods can achieve DFT-level accuracy in optimized structures and relevant electronic properties for organic semiconductor molecules, while simultaneously evaluating the computational speed advantages they offer. This work aims to provide clear guidance on the accuracy-cost trade-offs associated with GFN methods, informing their appropriate deployment in computational pipelines for organic materials discovery. More broadly, this study contributes to the ongoing effort within quantum chemistry to assess and understand the performance envelope of computationally efficient methods for structural and electronic property prediction across diverse chemical spaces.

The remainder of this paper is structured as follows: \Cref{sec:methodology} details the computational methodology used, beginning with a description of the data sources and describing the study workflow. Then, it presents strategies for curation and molecular selection techniques of the data set. This is followed by details on the quantum chemistry (QC) calculations, focusing on both semiempirical and DFT methods, and concludes with a description of the benchmarking metrics utilized. \Cref{sec:results} presents and discusses the results, analyzing the performance of the different levels of GFN in the metrics evaluated, including representativeness of data sampling, computational efficiency and reliability. Finally, \Cref{sec:conclusions} offers concluding remarks, summarizing key findings and discussing their implications for future research in computational materials science and the development of quantum chemical methods. 

%------------------------------------------------
\section{Computational Methods}\label{sec:methodology}

To provide a rigorous foundation for benchmarking the GFN methods as efficient approximate quantum chemical tools, we carefully established the datasets, molecular sampling strategies, quantum chemistry protocols and metrics to evaluate performance. This section elucidates the technical framework that underpins our comparative analysis. The general workflow of the study, from data collection to benchmarking, is depicted in \Cref{fig:workflow}. Subsequent subsections delve into the specific theoretical levels employed for these benchmark calculations.

\subsection{Dataset description}\label{subsec:dataset}

The charge transport properties critical to the functionality of organic semiconductors are intimately linked to their electronic structure, particularly the HOMO-LUMO energy gap, which typically falls below \qty{3}{\electronvolt} for these materials \cite{costa_optical_2016}. Using this criterion, we curated a subset of \num{216} small $\pi$-systems filtered from the extensive QM9 database to serve as benchmarks for geometry optimization. The QM9 database \cite{ramakrishnan_quantum_2014}, itself readily available on \emph{Figshare}, is a valuable resource containing \num{130e3} stable small organic molecules composed primarily of carbon, hydrogen, nitrogen, oxygen and fluorine (CHNOF), along with established DFT reference data, specifically calculated at the B3LYP/6-31G(2df,p) level in gas phase. This subset was specifically filtered to mimic semiconductor behavior based on a HOMO-LUMO gap criterion below \qty{3}{\electronvolt}, a common threshold for such materials. This ensures that the selected QM9 molecules, though small, exhibit relevant electronic characteristics, e.g., the presence of conjugated rings and frontier orbital energies conductive to semiconductor-like behavior for organic semiconductors, making them a suitable benchmark for evaluating methods on localized $\pi$-systems.

For evaluation of GFN methods on larger and more extended systems directly relevant to organic photovoltaics (OPVs), we leveraged a subset of molecules from the Harvard Clean Energy Project (CEP) database \cite{hachmann_harvard_2011}. The CEP database is a comprehensive collection specifically focused on organic semiconductors for photovoltaic applications. The CEP subset used in this study comprises \num{29978} extended $\pi$-systems, conveniently encoded in the Simplified Molecular Input Line Entry System (SMILES) format \cite{weininger_smiles_1988} and includes associated power conversion efficiency (PCE) data. It is accessible via a public Git repository at \url{http://github.com/HIPS/neural-fingerprint}.

\subsection{Molecular sampling strategy}\label{subsec:sampling}

\begin{figure}[H]
	\centering
	\includegraphics[width=.9\linewidth]{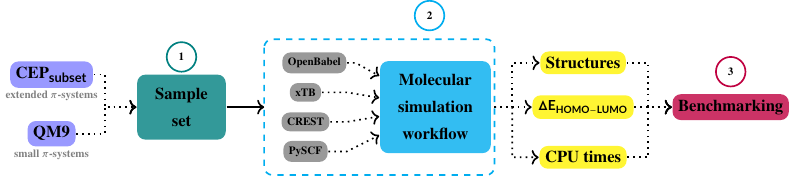}
	\caption{Overall workflow illustration. \textbf{1} The entire molecular chemical space is explored using molecular clustering and stratified sampling, resulting in a smaller representative subset. \textbf{2} Semiempirical quantum chemistry calculations are performed to optimize the atomic Cartesian coordinates using GFN methods (\gfnb, \gfnc, \gfna, and \gfnf). \textbf{3} The optimized geometries against DFT-based structures, as well as HOMO-LUMO bandgap energies and CPU times, are used for benchmarking to evaluate the quality and computational efficiency of each GFN theoretical level.}
	\label{fig:workflow}%
\end{figure}

An effective exploration of the chemical space is essential to ensure that the selected sample set accurately represents the diversity of the parent databases for robust method benchmarking. This not only optimizes computational resource utilization, but also enhances the generalizability and interpretability of subsequent molecular simulation results. As illustrated in \Cref{fig:workflow} (Step 1), the sampling process was the pivotal initial step. Although large-scale testing would ideally involve entire databases, computational constraints necessitated the selection of smaller, representative subsets.

To achieve this representativeness, we employed distinct unsupervised machine learning and stratified sampling techniques for the QM9 and CEP datasets, respectively. For QM9 small $\pi$-systems, we utilized a clustering algorithm, specifically the \emph{k}-means method \cite{macqueen_methods_1967} implemented in the scikit-learn package (version 1.2.2) \cite{pedregosa2011scikit}. This approach was guided by a molecular featurization strategy detailed in \ref{sec:SI_Featurization} of the supplementary information (SI). Our molecular characterization incorporated a blend of local atomic and bond features, e.g., atom types, hybridization states, partial charges, and bond orders, along with global molecular descriptors that encompass various levels of complexity \cite{nicolotti_molecular_2018}. These features were integrated into unified molecular feature vectors using principal component analysis (PCA) \cite{ringner_what_2008, greenacre_principal_2022}. The \emph{k}-means algorithm was chosen for its computational efficiency and scalability. To determine the optimal number of clusters, we used the Silhouette index \cite{rousseeuw_silhouettes_1987}, a widely used metric based on pairwise differences in intra-cluster distances and inter-cluster distances \cite{liu_understanding_2010}. \ref{fig:SI_S1_Silhouette} shows the Silhouette scores obtained for various numbers of clusters, which guide our selection of optimal partitioning. \ref{fig:SI_S2_Density}(a) illustrates the distribution of molecules across the resulting QM9 clusters.

In contrast to the clustering approach for QM9, the CEP subset, characterized by a higher degree of homogeneity in the initial features (despite its size), faced challenges for distinct cluster separation as observed by Hadipour \textit{et al.} \cite{hadipour_deep_2022}. Consequently, a stratified sampling strategy was adopted, partitioning the database into non-overlapping strata based on a specific stratification variable \cite{singh_stratified_1996}. We chose the number of molecular atoms as this variable, creating strata analogous to the QM9 clusters. This choice aimed to simplify and streamline the sampling, minimize potential biases from complex feature combinations, and enhance interpretability. \ref{fig:SI_S2_Density}(b) shows the distribution of molecules in the CEP strata based on the atom count.

The sample size assigned to each cluster or stratum was determined proportionally to its inherent variance, using a variation of the standard Neyman allocation method \cite{singh_stratified_1996}. This approach prioritizes sampling more heavily from clusters or strata that exhibit greater internal variability based on the selected features, which contributes more significantly to the overall uncertainty of population estimates. This strategy bypasses the need for a preliminary pilot study. The proportion $P_k$ of the total sample size allocated to the $k^{th}$ cluster or stratum is given by:
\begin{equation}
	P_k = \frac{n_k \sigma^2_k}{\sum_{i=1}^{K} n_i \sigma^2_i},
\end{equation}
where $n_k$ and $\sigma_k$ represent the number of molecules and the variance within the $k^{th}$ cluster or stratum, respectively. The total sample size $n$ was limited by computational resources, and the individual cluster/stratum sizes $N_k$, were calculated as $N_k = n \times P_k$.

\begin{figure}[htbp]
	\begin{center}
		\includegraphics[width=0.6\linewidth]{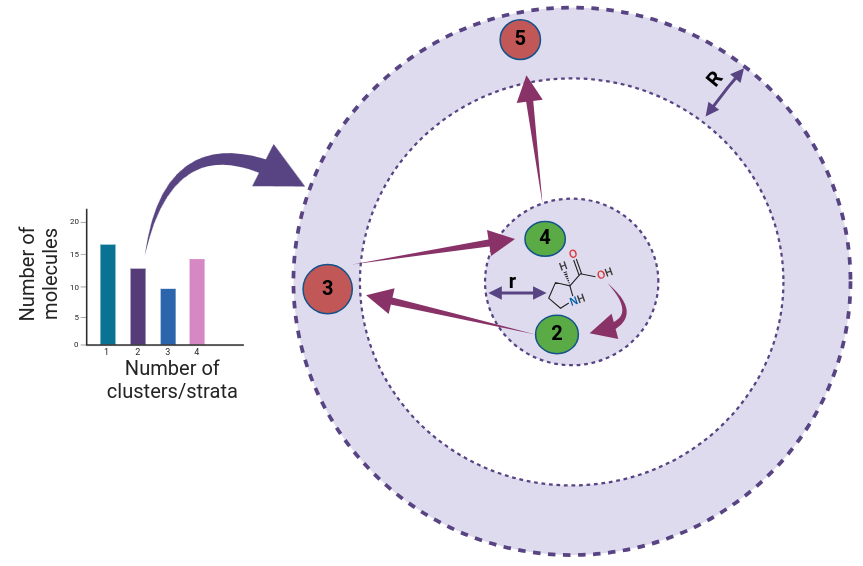}
		\caption{Random sampling illustration. \textbf{1} The first candidate selected is either the molecular centroid for a QM9 cluster or a random molecule for a CEP stratum. \textbf{2} If more than one molecule is required for the sample set to be representative, additional molecules are added by alternating between molecules more and less similar to the initial candidate. $r \leq 0.2$ and $R \geq 0.65$ represent the Tanimoto similarity ranges for the larger and smaller similarity regions, respectively.}
		\label{fig:RandSel}
	\end{center}
\end{figure}

Finally, molecules from each cluster or stratum were selected using a random sampling without replacement. For subgroups with one single molecule, the centroid cluster (QM9 cluster) or a random molecule (CEP stratum) was chosen. For larger subgroups, additional molecules were selected by alternating between candidates exhibiting high similarity and those exhibiting lower similarity to the initial selection, as illustrated in \Cref{fig:RandSel}. This alternating approach aimed to balance representativeness with the capture of molecular diversity within each subgroup. The Tanimoto similarity score \cite{rogers_computer_1960}, recognized as a robust metric for the comparison of molecular similarity based on fingerprints \cite{bajusz_why_2015}, and an extension of the Jaccard coefficient \cite{bero_similarity_2017}, was used. We used extended-connectivity fingerprints (ECFP) \cite{rogers_extended-connectivity_2010} (radius=1, 2048 bits) to calculate Tanimoto scores, focusing on localized atomic environments. Molecular similarity maps \cite{riniker_similarity_2013} (using ECFP, radius=2, 2048 bits) were constructed to visually validate clustering and stratification, confirming that molecules within the same group generally exhibit higher similarity. \ref{fig:SI_S3_Heatmap} presents the Tanimoto similarity matrix for selected QM9 clusters, and \ref{tab:SI_S3_QM9_SimMap} and \ref{tab:SI_S4_CEP_SimMap} show example similarity maps for QM9 and CEP samples, respectively, demonstrating the structural similarities and differences captured by our sampling. This comprehensive sampling process resulted in benchmark sets comprising \num{64} small $\pi$-systems from QM9 and \num{76} extended $\pi$-systems from CEP.

\subsection{Quantum chemistry protocols}\label{subsec:qc_protocols}

Following the careful selection of benchmark molecules to serve as representative test cases for our method evaluation, quantum chemistry calculations were performed to generate reference data and evaluate GFN methods. All calculations were executed on a high performance computing cluster, using Intel Xeon Gold 6142@2.6 GHz processors (8 cores per task with 20 GB RAM).

\subsubsection{Semiempirical quantum chemistry calculations}

\begin{figure}[H]
	\centering
	\includegraphics[width=.9\linewidth]{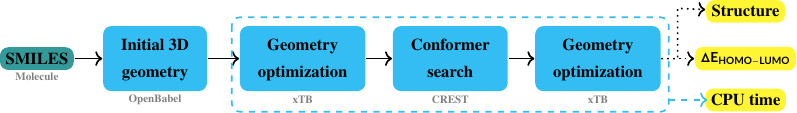}
	\caption{Semiempirical Quantum Chemistry Workflow. \textbf{1} The initial molecular geometry is derived from the molecular SMILES string using the \texttt{OpenBabel} program. \textbf{2} The initial atomic Cartesian coordinates are preoptimized using a GFN level of theory (1, 2, 0, or FF). \textbf{3} The conformer-rotamer ensemble of the preoptimized structure is obtained through conformational sampling using the \crest program. \textbf{4} The best molecular conformation is finally optimized at the same GFN preoptimization level.}
	\label{fig:SemiempiricalQC}
\end{figure}

For semi-empirical calculations (depicted in \Cref{fig:workflow}, Step 2, and detailed in \Cref{fig:SemiempiricalQC}), we utilized the GFN family of methods (\gfnb, \gfnc, \gfna, and \gfnf) as implemented in the package \xtb (version 6.7.1). The workflow, inspired by benchmark platforms like TARTARUS \cite{nigam_tartarus_2023}, began with generating initial 3D Cartesian coordinates from molecular SMILES strings using the \texttt{OpenBabel} software (version 3.1.0) \cite{oboyle_open_2011, yoshikawa_fast_2019}. The initial minimization of energy was performed using the force field MMFF94s \cite{halgren_mmff_1999} for the QM9 subset. Given the limited parameter coverage of MMFF94s for extended $\pi$-systems, the initial minimization of the CEP subset used the more broadly applicable Universal Force Field (UFF) \cite{rappe_uff_1992}. Both force field minimizations used a steepest descent algorithm with tight energy convergence criteria. These force field optimized structures then served as input for pre-optimization using the respective level of theory of the GFN within \xtb. This was followed by robust conformational sampling using the \texttt{CREST} program (version 3.0.2) \cite{pracht_crest_2024}. \ref{sec:SI_QC_Theories} provides further details on the theoretical levels \xtb (\ref{sec:SI_xTB}) and the \texttt{CREST} conformer search algorithm (\ref{sec:SI_CREST}).

Finally, the lowest-energy conformation identified by \texttt{CREST} was subjected to a high-precision geometry optimization at the same GFN level, incorporating the analytical linearized Poisson-Boltzmann (ALPB) solvation model \cite{ehlert_robust_2021} with toluene. It should be noted that these GFN calculations with implicit solvation are compared against gas-phase DFT reference data (see \Cref{sec:dft}). This difference in modeled environments might introduce systematic deviations in the absolute structural and electronic parameters. However, the primary objective of this study is to benchmark the relative performance and structural fidelity of GFN methods against commonly used DFT references for these classes of molecules, and the chosen ALPB model represents a typical approach to perform realistic simulations and approximate condensed phase effects with GFN methods. For the non-electronic method \gfnf, additional SCC calculations were performed using the integrated GFN2-\xtb functionality within \xtb on optimized geometries to obtain electronic structure information. This post-processing step ensures that electronic properties like HOMO-LUMO gaps can be consistently obtained for GFN-FF-optimized structures, enabling a holistic comparison across all GFN methods for both structural and electronic metrics. The optimized geometries were saved as structural data files (SDF) for downstream analysis using \texttt{OpenBabel} \cite{oboyle_open_2011}. CPU execution times and HOMO/LUMO energies were recorded for each step.

\subsubsection{DFT calculations} \label{sec:dft}

\begin{figure}[htbp]
	\begin{center}
		\includegraphics[width=0.7\linewidth]{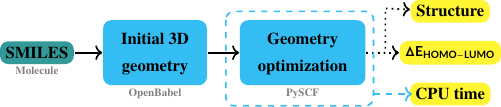}
		\caption{Workflow for DFT-based geometry optimization and electronic structure calculations. \textbf{1} The initial molecular geometry is derived from the molecular SMILES string using the \texttt{OpenBabel} program. \textbf{2} Geometry optimization is performed using the PySCF software package. \textbf{3} Optimized structures are analyzed for electronic properties and CPU time.}
		\label{fig:dft_workflow}
	\end{center}
\end{figure}

DFT reference data was obtained from two sources. For the QM9 subset, reference geometries and properties were taken from Ramakrishnan \textit{et al.} \cite{ramakrishnan_quantum_2014}, which utilized the B3LYP/6-31G(2df,p) level of theory in gas phase. B3LYP was chosen for its proven reliability in describing bond lengths, angles, and electronic properties of small organic molecules, particularly due to its inclusion of exact Hartree-Fock exchange. While other hybrid functionals like $\omega$B97X-D might offer slightly higher accuracy for specific properties, B3LYP provides a robust and widely accepted benchmark at a reasonable computational cost for a dataset of this size. For the CEP subset, DFT calculations were performed for the optimization of the geometry and analysis of the electronic structure using the restricted Kohn-Sham method within the \texttt{PySCF} software package (version 2.8.0) \cite{sun_pyscf_2018} in gas phase. Geometry optimizations utilized the generalized gradient approximation (GGA) functional BP86 \cite{porezag_construction_1995, Perdew1996} in conjunction with the double-$\zeta$ \texttt{def2-SVP} basis set \cite{weigend_balanced_2005, schuchardt_basis_2007}, mirroring the approach reported for this dataset in the literature \cite{hachmann_harvard_2011, lopez_harvard_2016} to ensure comparability (\Cref{fig:workflow}, step 2, and detailed in \Cref{fig:dft_workflow}). BP86 was selected for the larger CEP $\pi$-systems due to its efficiency and known performance for delocalized systems, offering a good balance between accuracy and computational cost for a dataset of this scale, and maintaining consistency with previous work on this database. While hybrid GGA or meta-GGA functionals might offer improved accuracy for some specific electronic properties, particularly those sensitive to exact exchange, their significantly higher computational cost would have rendered the systematic benchmarking of a large dataset like CEP impractical for this study. Therefore, BP86's established track record for similar extended $\pi$-systems makes it a pragmatic and well-justified choice for establishing a relevant and computationally feasible reference for this benchmarking study.

Geometry optimization calculations explicitly included collective molecular translations and rotations as degrees of freedom \cite{wang_geometry_2016}, improving the robustness for molecular structure relaxation. Interatomic bonds were initialized on the basis of distances scaled by covalent radii. The energy minimization proceeded iteratively, with a maximum of \num{300} cycles and subject to stringent convergence criteria for the total energy change ( $\Delta E < \qty{1.00e-06}{Hartree}$), forces (root-mean-square (RMS) gradient $< \qty{3.00e-04}{Hartree/Bohr}$ and maximum gradient threshold $< \qty{4.50e-04}{Hartree/Bohr}$) and spatial displacement (RMS displacement $< \qty{1.20e-03}{\angstrom}$ and maximum displacement $< \qty{1.80e-03}{\angstrom}$). The SCF optimization process was iterated until all five convergence criteria were met. From the optimized state function, the HOMO/LUMO energies and relaxation elapsed times were extracted. Optimized DFT geometries were initially output as XYZ files and subsequently converted to SDF format using \texttt{OpenBabel} \cite{oboyle_open_2011}. CPU execution times and HOMO/LUMO energies were recorded for each step.

\subsection{Benchmarking Metrics}\label{subsec:metrics}

To quantify the precision and computational efficiency of the GFN methods relative to the DFT references within a quantum chemical context, we employ a comprehensive set of benchmarking metrics (\Cref{fig:workflow}, step 3). Structural accuracy was assessed by comparing optimized GFN geometries with their DFT counterparts using the following metrics:
\begin{itemize}
	\item center of mass (CMA) deviations;
	\item heavy-atom root-mean-square deviation (hRMSD) after optimal superposition;
	\item radius of gyration (Rg), providing a measure of molecular compactness;
	\item equilibrium rotational constants ($A_e, B_e, C_e$), which are sensitive probes of the overall molecular shape and mass distribution;
	\item bond lengths and bond angles, assessing local geometric fidelity.
\end{itemize}
The equilibrium rotational constants $B_e^i$ are calculated using the relationship with the principal moments of inertia:
\begin{equation}\label{eq:rotational_constant}
	B_e^i = \frac{\hbar}{4\pi c I_{ii}}, \quad i \in \{a, b, c\},
\end{equation}
where the subscript $i$ indicates the inertial axis $a$, $b$ or $c$ ($B_e^a = A_e, B_e^b = B_e, B_e^c = C_e$, and $A_e > B_e > C_e$), $c$ is the light speed, and $I_{ii}$ stands for the $i^{th}$ diagonal element of the inertia tensor $I$, which is defined as:
\begin{equation}\label{eq:inertia_tensor}
	I = \sum_K M_K (R_K^\intercal R_K \mathbb{I} - R_K R_K^\intercal),
\end{equation}
in which the sum relates to all atoms in the molecule, $R_K$ and $M_K$ denoting atomic coordinates relative to the center of mass and atomic mass, respectively, and $\mathbb{I}$ is the identity matrix. A high level of accuracy in computational predictions (ranging from \qty{0.01}{\percent} to \qty{0.1}{\percent}, depending on the dedicated task) is required to support the experimental data $B_0$ obtained from rotational spectroscopy. However, Puzzarini and Stanton have reported that the prediction of the equilibrium contribution ($B_e^i$) is more significant than that of the vibrational contribution ($\Delta B_{vib}$), for which errors are estimated to be no greater than around \qty{0.05}{\percent} of the total value of ground state constants $B_0$. \cite{puzzarini_connections_2023}.

The precision of the electronic structure, critical for the properties of semiconductors and a key observable in quantum chemical calculations, was evaluated by comparing HOMO-LUMO energy gaps. Finally, computational cost was measured on the basis of average CPU time per molecule and analyzed in terms of algorithmic scaling behavior with respect to system size, providing insight into the practical applicability of each method for high-throughput workflows in computational quantum chemistry.

\section{Results and discussion}\label{sec:results}

The preceding sections outlined the meticulous preparation of benchmark datasets and the quantum chemistry protocols used. In this section, we now delve into the outcomes of our computational analysis, presenting a thorough examination of molecular sampling robustness, the accuracy of GFN methods in geometry optimization and HOMO-LUMO gap prediction compared to DFT references, and their associated computational costs. Ultimately, our aim is to furnish a comprehensive picture of the performance and trade-offs offered by different levels of GFN for organic semiconductor molecules as representative challenging chemical systems for approximate quantum mechanical methods.

\subsection{Molecular sampling analysis}\label{subsec:sampling_results}

A robust and representative sample set is fundamental to drawing reliable conclusions from computational benchmarking of theoretical methods. For the QM9 dataset, the chemical space was explored using \emph{k}-means clustering based on a rich set of molecular descriptors (detailed in \ref{sec:SI_Featurization}). We analyze the silhouette index \cite{rousseeuw_silhouettes_1987} for varying numbers of clusters (from \numrange{5}{200}), as shown in \ref{fig:SI_S1_Silhouette}. This analysis indicated that a cluster size of \num{25} provided relatively stable Silhouette scores (\num{0.278}), thereby suggesting a meaningful partitioning of the chemical space based on our chosen features. The distribution of molecules in these 25 clusters is illustrated in \ref{fig:SI_S2_Density}(a), revealing a relatively uniform distribution with $\qty{56}{\percent}$ of clusters containing eight or more molecules.

In contrast, the CEP dataset, characterized by larger $\pi$-systems, exhibited greater homogeneity in the initial features (despite its size), faced challenges for distinct cluster separation, as observed by Hadipour \textit{et al.} \cite{hadipour_deep_2022}. For CEP, we employed stratified sampling based on the number of atoms per molecule. \ref{fig:SI_S2_Density}(b) shows the distribution of molecules across these strata, which exhibits an approximately normal bell-shaped pattern centered around \num{42} atoms, with some right skewness and a high average density of \num{908} molecules per stratum.

To validate the effectiveness and diversity of our sampling strategies in capturing the relevant chemical space for method evaluation, we examined molecular similarities using Tanimoto scores based on ECFP. For QM9, pairwise similarities within and between five representative randomly selected clusters (C0, C13, C17, C18, C24) were evaluated. As depicted in \ref{fig:SI_S3_Heatmap}, the Tanimoto similarity matrix shows higher scores within clusters (indicated by red rectangles on the diagonal), confirming that the clustering successfully grouped structurally similar molecules. Conversely, lower scores between clusters demonstrated significant inter-cluster dissimilarity, supporting the notion that the sampling captures distinct regions of chemical space. Example similarity maps for selected molecules from QM9 clusters are provided in \ref{tab:SI_S3_QM9_SimMap}, further illustrating these structural similarities within groups.

\begin{figure}[htbp]
	\begin{center}
		\subfloat[\centering QM9]{\includegraphics[width=.49\textwidth]{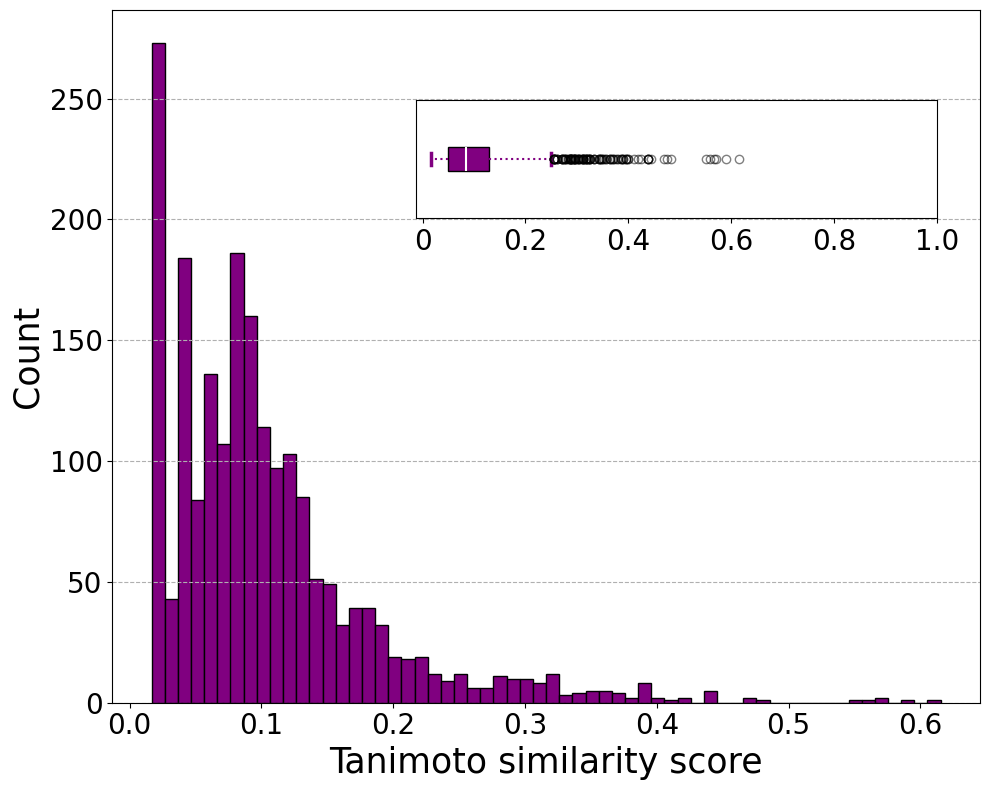}}
		\subfloat[\centering CEP]{\includegraphics[width=.49\textwidth]{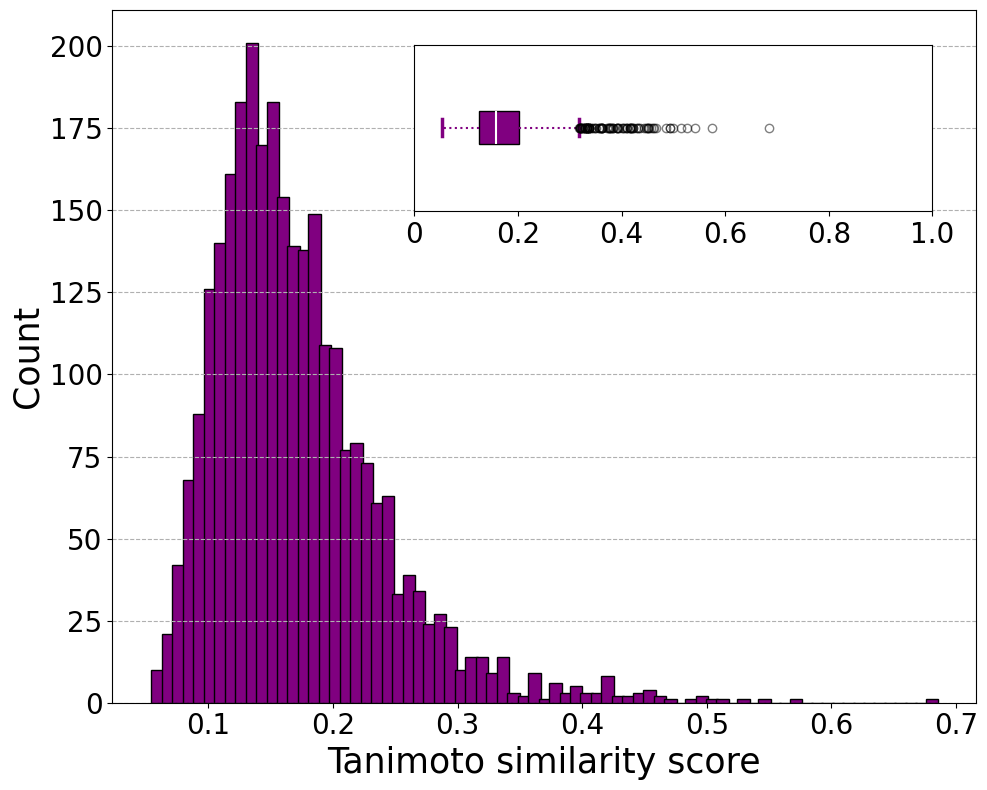}}
		\caption{The distribution of Tanimoto similarity scores in the similarity matrix calculated for the sample sets of small and extended $\pi$-systems used in this study.}
		\label{fig:Mol_diversity}
	\end{center}
\end{figure}

Similarly, molecular similarity maps were constructed for representative molecules from selected CEP strata (\ref{tab:SI_S4_CEP_SimMap}). These maps highlight common atomic environments (green) and dissimilar features (red), underscoring structural similarities within strata based on atom count, particularly among the extended $\pi$-systems. For instance, specific heterocyclic substructures were frequently observed within molecules belonging to the same atom-count stratum. Overall, the distribution of Tanimoto similarity scores for the final benchmark sets (\Cref{fig:Mol_diversity}) shows average scores of \num{0.104} (QM9) and \num{0.170} (CEP), indicating that both sets capture a substantial range of molecular diversity present in the original large databases, with the smaller QM9 set exhibiting slightly higher structural variation relative to its size.

\begin{figure}[htbp]
	\begin{center}
		\subfloat[\centering QM9 clusters]{\includegraphics[width=.9\textwidth]{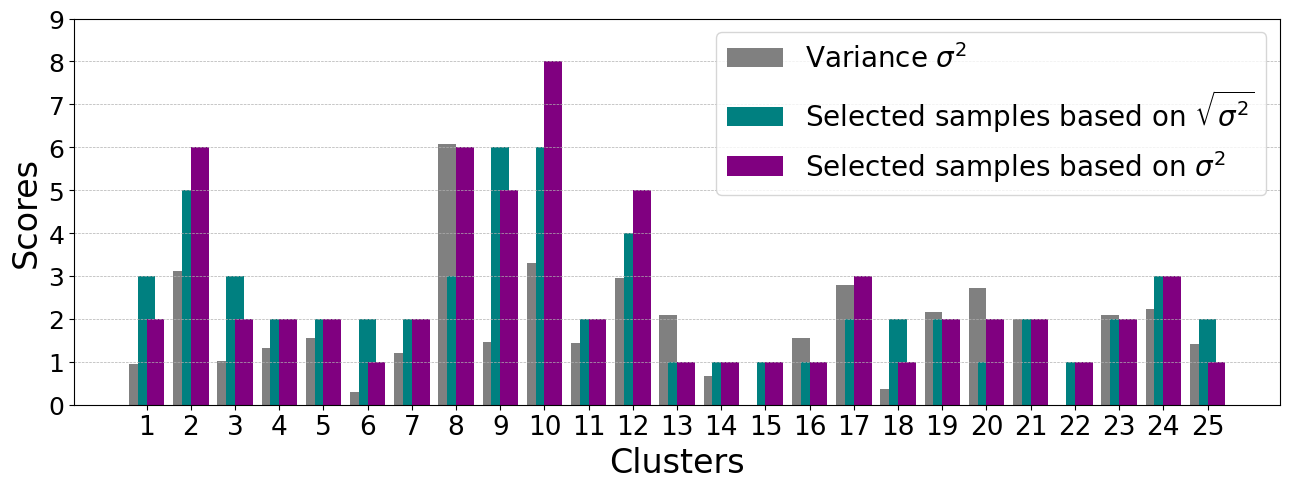}} \\
		\subfloat[\centering CEP strata]{\includegraphics[width=.9\textwidth]{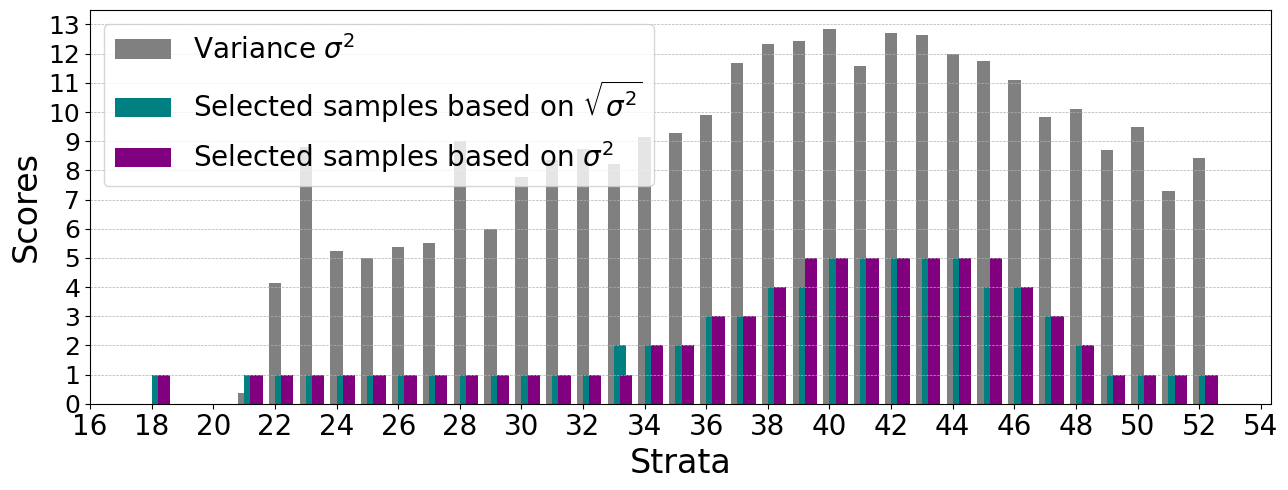}}
		\caption{The amount of variance (indicated by the gray bars) and the number of molecules selected for each cluster (a) and stratum (b), based on the standard Neyman allocation (teal bars) and the variance-based Neyman allocation (purple bars).}
		\label{fig:Mol_Selection}
	\end{center}
\end{figure}

Finally, we evaluated the allocation of sample sizes using both the standard- and variance-based Neyman methods (\Cref{fig:Mol_Selection}). This analysis guided our selection to ensure that the clusters or strata that contributed the most significantly to the overall variance were adequately represented. For QM9 clusters (\Cref{fig:Mol_Selection}(a)), the variance-based approach indeed allocated larger sample sizes to clusters with higher variability (e.g. C8 $\sigma^2 \approx 6.0$, $N_8 = 6$) compared to those with lower variance (e.g., C9 $\sigma^2 \approx 1.5$, $N_9 = 5$), effectively prioritizing sampling from more diverse regions, while standard-based Neyman approach allocated three samples to C8 and six samples to C9. In contrast, for the CEP strata (\Cref{fig:Mol_Selection}(b)), the large population size within each stratum appeared to dominate the allocation, leading to near convergence between the variance-based and standard Neyman methods, even when the variances in the strata differed. This sensitivity of the Neyman allocation to population density can, however, dilute the impact of variance signals in large strata. Despite these nuances, the systematic sampling process resulted in the final benchmark sets of \num{64} QM9 and \num{76} CEP molecules, designed to balance representativeness, diversity, and computational feasibility.

\subsection{Quantum chemistry calculation results}\label{subsec:QC_Results}

With the benchmark sets established to test the selected quantum chemical methods, we performed geometry optimizations using the four GFN methods and compared the resulting structures and calculated properties with the DFT reference data. This comprehensive comparison aims to elucidate the accuracy and reliability of each GFN level for small organic semiconductor molecules particularly given their demanding nature as challenging systems for approximate quantum mechanical methods.

\begin{figure}[htbp]
	\leavevmode
	\subfloat[QM9$_\text{ID}$ 6683]{\includegraphics[width=.15\textwidth]{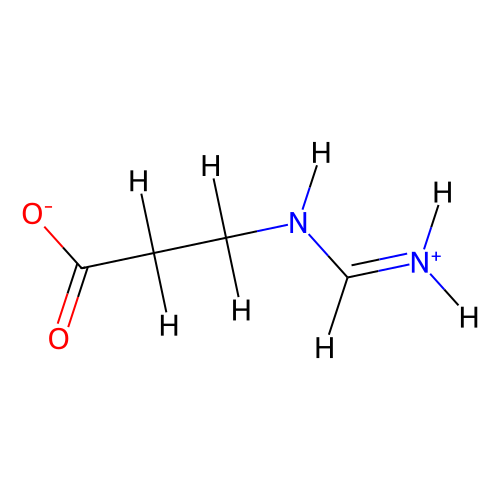}}\hfill
	\subfloat[QM9$_\text{ID}$ 26912]{\includegraphics[width=.15\textwidth]{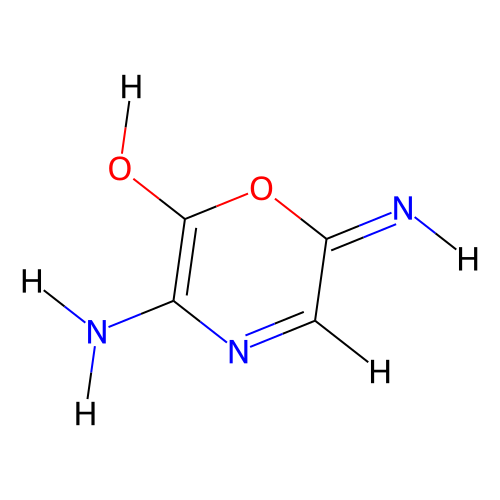}}\hfill
	\subfloat[QM9$_\text{ID}$ 66599]{\includegraphics[width=.15\textwidth]{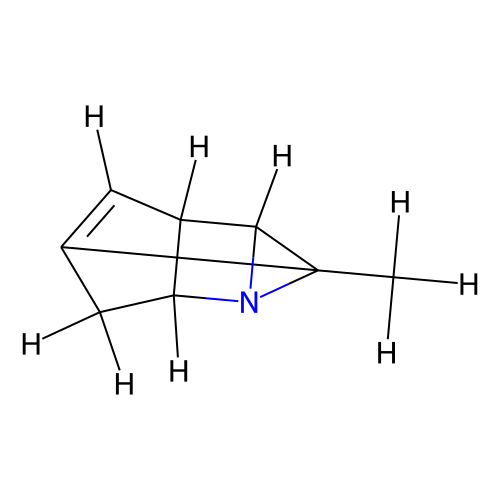}}\hfill
	\subfloat[QM9$_\text{ID}$ 74202]{\includegraphics[width=.15\textwidth]{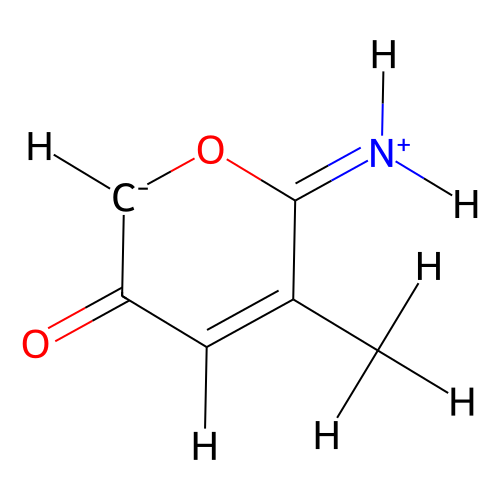}}\hfill
	\subfloat[QM9$_\text{ID}$ 130511]{\includegraphics[width=.15\textwidth]{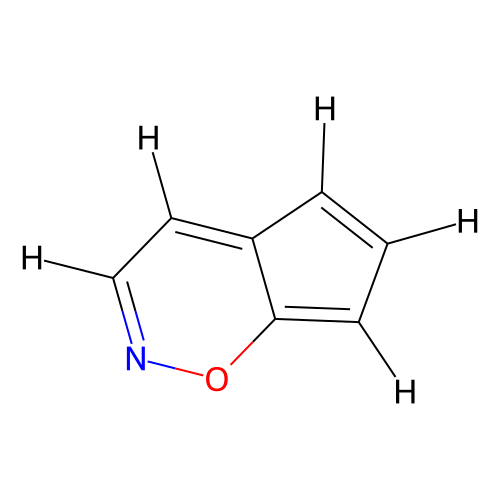}}\hfill
	\subfloat[QM9$_\text{ID}$ 130518]{\includegraphics[width=.15\textwidth]{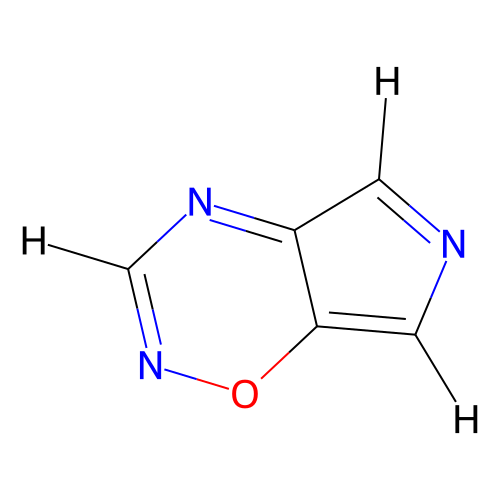}}\\
	\subfloat[CEP$_\text{ID}$ 161]{\includegraphics[width=.2\textwidth]{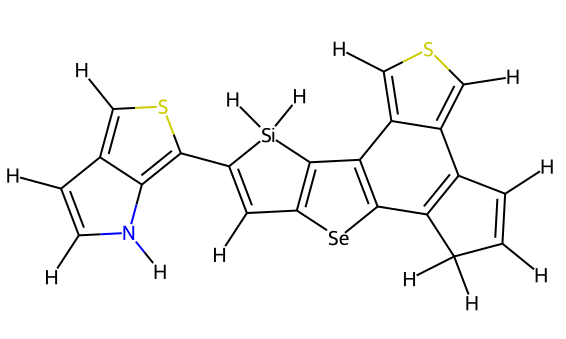}}
	\subfloat[CEP$_\text{ID}$ 18406]{\includegraphics[width=.2\textwidth]{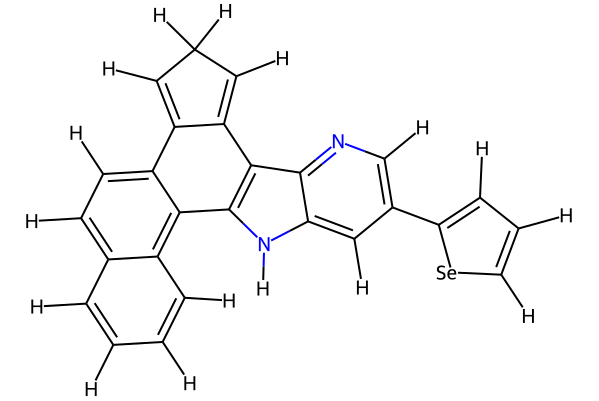}}
	\subfloat[CEP$_\text{ID}$ 23308]{\includegraphics[width=.2\textwidth]{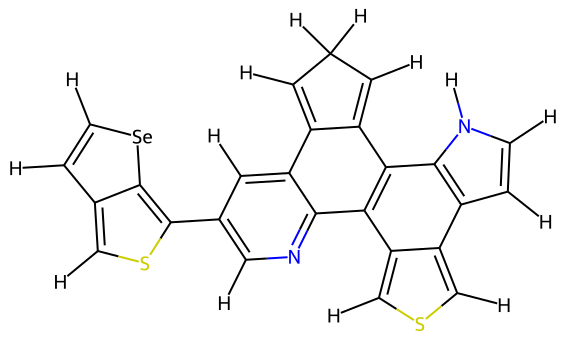}}
	\caption{Molecules from the QM9 and CEP set samples excluded from the statistical process.}
	\label{fig:Outliers}
\end{figure}

A necessary preliminary step involved identifying and excluding molecules that failed the geometry optimization process (e.g., CREST or \xtb convergence errors) or generated critical computational errors. These molecules (specifically QM9 ID 6683, CEP IDs 161, 18406, 23308 visualized in \Cref{fig:Outliers} and listed in \ref{tab:SI_S5_Optimization_Exclusions}) were completely excluded from the subsequent statistical analysis.

Furthermore, a few molecules generated warnings or exhibited persistent issues during structure processing (e.g., QM9 IDs 26912, 66599, 74202, 130511, 130518 with \texttt{OpenBabel} visualized in \Cref{fig:Outliers} and listed in \ref{tab:SI_S6_VF2_Exclusions}) or resulted in failures of the VF2 algorithm when comparing GFN and DFT structures. \ref{tab:SI_S6_VF2_Exclusions} lists the molecules that were excluded \emph{specifically} from the bond length and angle analyzes, to ensure a valid comparison, due to the failures of the VF2 mapping algorithm to establish atom-to-atom correspondence, indicating significant topological differences. These failures often point to challenging structural characteristics, such as high rigidity, significant steric hindrance, complex potential energy surfaces, or inherent limitations in the handling of specific bonding types by computational tools. Such cases, while excluded from statistical analysis for consistency, provide valuable feedback on the practical limitations of these computational tools and the underlying GFN parametrizations for certain chemical structures, potentially indicating biases in their applicability for molecules with very unusual bonding patterns or high conformational complexity not well captured by the current models. This suggests potential biases in the applicability of the methods for molecules with very unusual bonding or high conformational complexity.

\subsubsection{Optimized molecular structures}

We first assessed the quality of the optimized geometries by examining several structural metrics, including measures of global shape and local geometry. Analysis of CMA deviations from DFT (as shown in \ref{fig:SI_S4_CMA}) revealed that all GFN methods exhibit equivalent geometric translational accuracy in 3D space across both benchmark sets. This global agreement is encouraging and likely stems from common features within the GFN parameterization.

\begin{figure}[htbp]
	\centering
	\leavevmode
	\subfloat[\centering QM9]{\includegraphics[width=.49\textwidth]{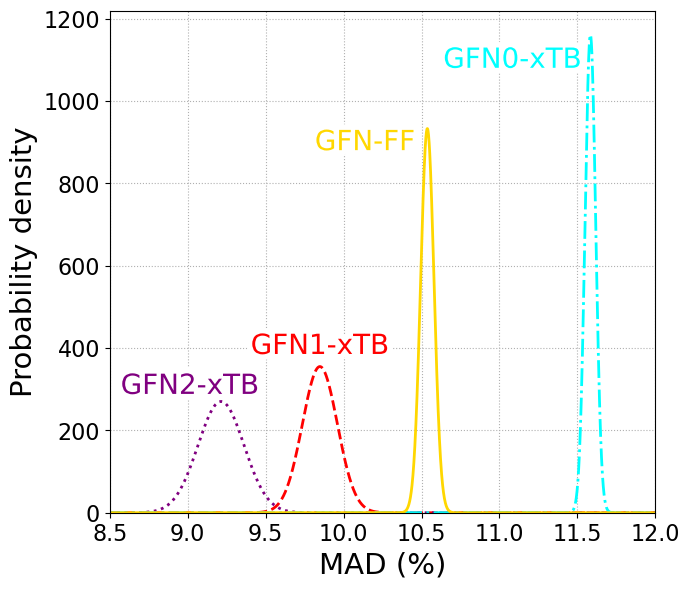}}
	\subfloat[\centering CEP]{\includegraphics[width=.49\textwidth]{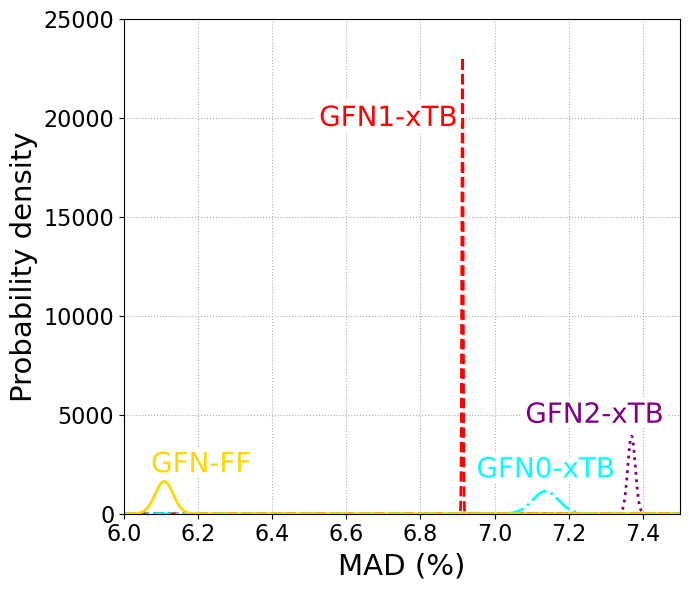}}
	\caption{The normal distribution plots for the absolute errors in radii of gyration calculated using the GFN methods. Graph (a) refers to the small $\pi$-systems of the QM9 sample set, while graph (b) refers to the extended $\pi$-systems of the CEP sample set.}
	\label{fig:RG_Spect}
\end{figure}

More detailed insights into molecular shape were obtained from the radius-of-gyration (Rg) deviations. The CEP set, comprising larger $\pi$-systems, exhibits a wider range of absolute Rg values (up to \qty{6.0}{\angstrom}, as seen in \ref{fig:SI_S5_Rg}) compared to the smaller, more rigid QM9 molecules, underscoring their greater spatial extent and conformational flexibility. Upon examining the absolute errors in Rg (\Cref{fig:RG_Spect}), we observed that for the QM9 set (\Cref{fig:RG_Spect}(a)), \gfnc and \gfnb yielded the lowest mean absolute deviations (MADs) of \qty{0.0921}{\angstrom} and \qty{0.0984}{\angstrom}, respectively. Although \gfnf (MAD =\qty{0.1054}{\angstrom}) and \gfna (MAD=\qty{0.1158}{\angstrom}) showed slightly higher errors, they also had lower standard deviations, indicating less variability in their predictions despite lower overall accuracy. For the extended CEP $\pi$-systems (\Cref{fig:RG_Spect}(b)), \gfnf achieved the higher accuracy with the lowest MAD (\qty{0.0611}{\angstrom}), followed closely by \gfnb (MAD=\qty{0.0691}{\angstrom}), which also showed notably higher precision (lower standard deviation, \qty{0.00002}{\angstrom}). These results, summarized in \Cref{tab:ResultsSummary}, collectively suggest that GFN methods, particularly \gfnf and \gfnb, competently reproduce the overall compactness of extended systems, though their performance on smaller and more rigid systems exhibits greater variability. The differences observed highlight how the underlying parametrization and treatment of interactions in each GFN variant affect global structural predictions.

\begin{figure}[htbp]
	\centering
	\leavevmode
	\subfloat[\centering QM9]{\includegraphics[width=.49\textwidth]{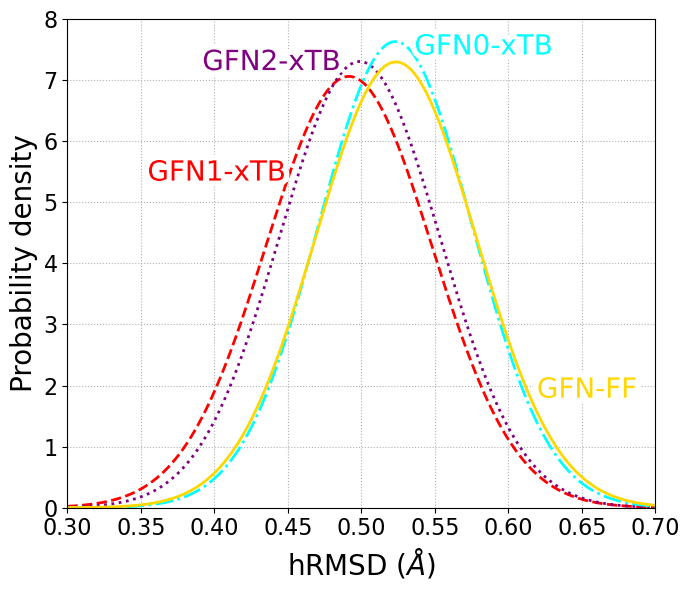}}
	\subfloat[\centering CEP]{\includegraphics[width=.49\textwidth]{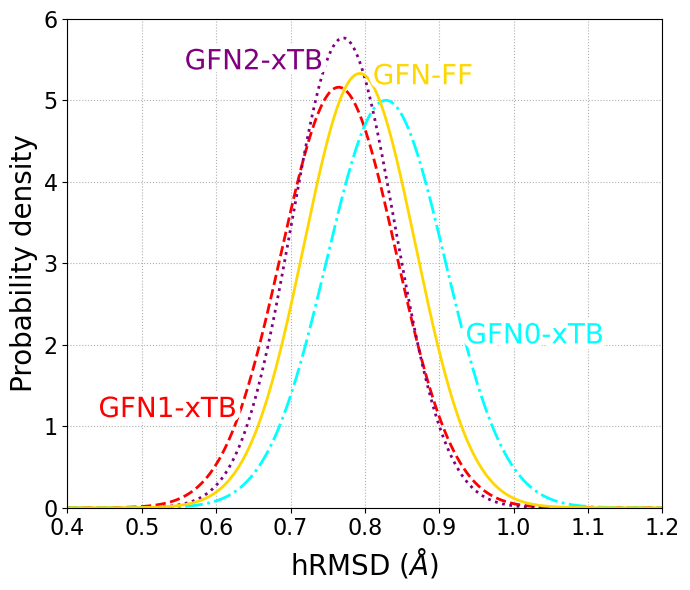}}
	\caption{The normal distribution plots for the heavy-atoms root-mean-square deviations obtained using the GFN methods. Graph (a) refers to the small $\pi$-systems of the QM9 sample set, while graph (b) refers to the extended $\pi$-systems of the CEP sample set.}
	\label{fig:hRMSD_Spect}
\end{figure}

Heavy-atom root-mean-square deviation (hRMSD) provides a robust measure of overall 3D structural agreement after optimal superposition. The hRMSD distributions (\Cref{fig:hRMSD_Spect}) show that GFN methods generally yield better spatial agreement with DFT for QM9 (hRMSD typically below \qty{1.0}{\angstrom}, peaks around \qty{1.75}{\angstrom}) compared to the more flexible CEP systems (hRMSD extending beyond \qty{2.0}{\angstrom}). As detailed in \Cref{fig:hRMSD_Spect}, \gfnb consistently exhibits the smallest deviations for both sets, with modes centered around \qty{0.491}{\angstrom} (QM9) and \qty{0.764}{\angstrom} (CEP). It is closely followed by the \gfnc (\qty{0.498}{\angstrom} and \qty{0.770}{\angstrom} modes). This superior performance of the SCC GFN methods (\gfnb, \gfnc) in reproducing heavy-atom positions can be attributed to their self-consistent treatment of charge interactions, which more effectively captures the subtle electronic effects influencing molecular structure compared to non-iterative methods (\gfna, \gfnf). However, while \gfnc and \gfnf showed greater consistency (lower standard deviations of \qty{6.91}{\percent} and \qty{7.48}{\percent} for CEP hRMSD, \Cref{tab:ResultsSummary}), no single GFN method demonstrated a substantial advantage over all others in both datasets, suggesting that the choice of the method could depend on the specific characteristics of the system. The hRMSD values per molecule are shown in \ref{fig:SI_S8_hRMSD}. While this metric (hRMSD), which emphasizes overall molecular shape, provides useful insights, it offers an incomplete picture of local geometric accuracy, which is further explored through analyses of bond lengths and angles.

\begin{figure}[htbp]
	\centering
	\leavevmode
	\subfloat[\centering QM9]{\includegraphics[width=.49\textwidth]{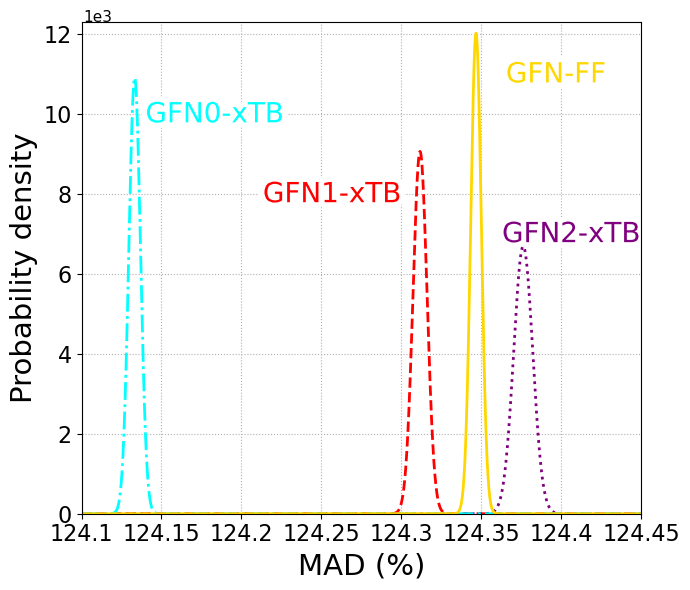}}
	\subfloat[\centering CEP]{\includegraphics[width=.49\textwidth]{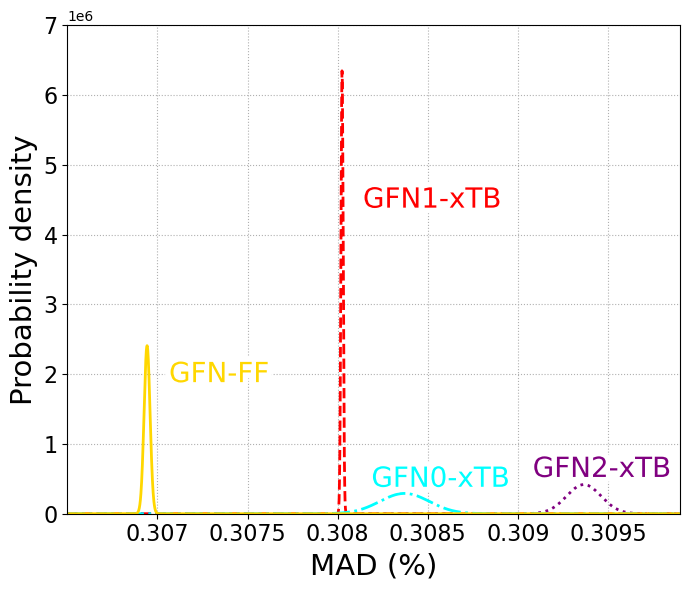}}
	\caption{The normal distribution plots for the absolute deviations in equilibrium rotational constants $B_e$ computed using the GFN methods. Graph (a) refers to the small $\pi$-systems of the QM9 sample set, while graph (b) refers to the extended $\pi$-systems of the CEP sample set.}
	\label{fig:Rot_B}
\end{figure}

While hRMSD provides a global measure, subtle but important local geometry deviations (bond lengths, angles, and torsions) can significantly impact molecular properties. Equilibrium rotational constants ($A_e, B_e, C_e$), being functions of the moments of inertia (Eq. \ref{eq:rotational_constant} and \ref{eq:inertia_tensor}), are particularly sensitive to these fine structural details and conformation. High-accuracy computational predictions for rotational constants ($B_e^i$) are essential for comparison with experimental spectroscopic data \cite{puzzarini_connections_2023}. As shown in \ref{fig:SI_S6_QM9_Rot} (QM9) and \ref{fig:SI_S7_CEP_Rot} (CEP), GFN methods generally reproduce the trends in rotational constants, but systematic errors are apparent. For small QM9 systems, all GFN methods systematically underestimate the rotational constants, hinting at potential overestimations of molecular size or biased molecular shapes. This likely stems from the absence of exact HF exchange in GFN, which is critical to accurately describing the localized electron density in smaller, more rigid systems. For the CEP set, the GFN methods show good agreement for $B_e$ compared to BP86/def2-SVP (MAD $\sim \qty{0.003}{\per\centi\meter}$, \Cref{tab:ResultsSummary}), while $A_e$ and $C_e$ are slightly overestimated, suggesting a tendency to underpredict planarity in extended systems. \Cref{fig:Rot_B} shows the error distributions for $B_e$. For QM9, GFN methods produce significantly larger errors ($\sim \qty{1.2}{\per\centi\meter}$) compared to CEP ($\sim \qty{0.003}{\per\centi\meter}$), reinforcing their comparatively lower accuracy for the specific structural nuances of smaller, more rigid systems. Among GFNs, \gfnb shows the lowest MAD (\qty{1.2431}{\per\centi\meter} for QM9, \qty{0.0031}{\per\centi\meter} for CEP), closely followed by \gfna for QM9 and \gfnf for CEP (\Cref{tab:ResultsSummary}). These findings underscore the challenge of accurately capturing fine structural details with tight-binding methods, especially for systems where electron localization effects play a significant role, and highlight areas for potential method improvement. 

To probe local geometry more directly, we analyzed bond lengths and angles after establishing atom-to-atom correspondence using the VF2 algorithm \cite{cordella_subgraph_2004}. As noted previously, molecules where VF2 failed to find an isomorphism, indicating significant topological differences, were excluded from this analysis (listed in \ref{tab:SI_S6_VF2_Exclusions}). In particular, \gfnf resulted in the fewest of such discrepancies (\num{0} for QM9, \num{1} for CEP), suggesting a better preservation of overall connectivity, while \gfnb and \gfnc showed more outliers (\num{9}/\num{2} and \num{8}/\num{3}, respectively), possibly due to issues such as spurious bond formation / breaking or significant distortions. Analyzing such failures provides critical information about the limitations of the underlying methods in handling complex bonding environments. For instance, the higher number of VF2 failures for \gfnb and \gfnc in the QM9 set, compared to \gfnf, suggests that while these SCC methods aim for higher accuracy, they might occasionally predict geometries with bonding patterns (e.g., unexpected ring formations or breakages) that deviate significantly from the reference topology. These discrepancies can arise from challenges in navigating complex potential energy surfaces or from inherent parameterization limitations of GFN methods for specific strained or unusual structural motifs, particularly in smaller and more constrained systems.

\begin{figure}[htbp]
	\begin{center}
		\includegraphics[width=.49\textwidth]{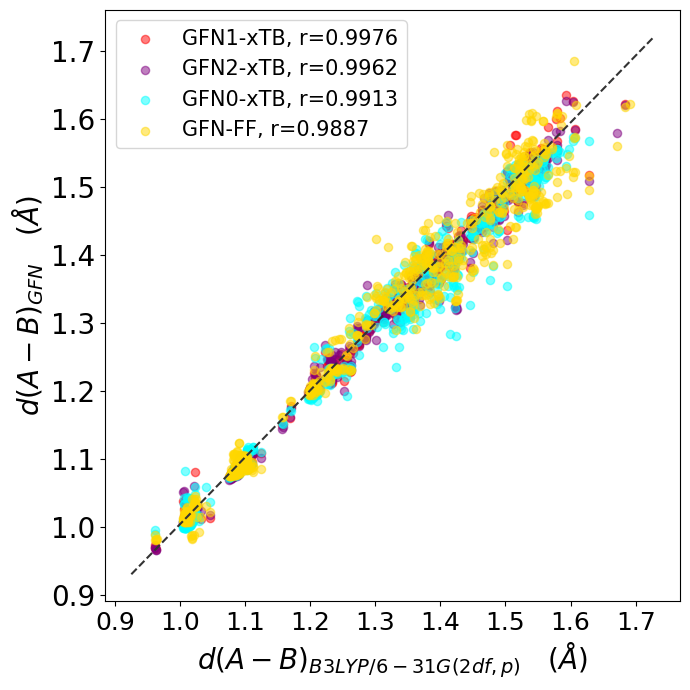}
		\subfloat[QM9 samples]{\includegraphics[width=.49\textwidth]{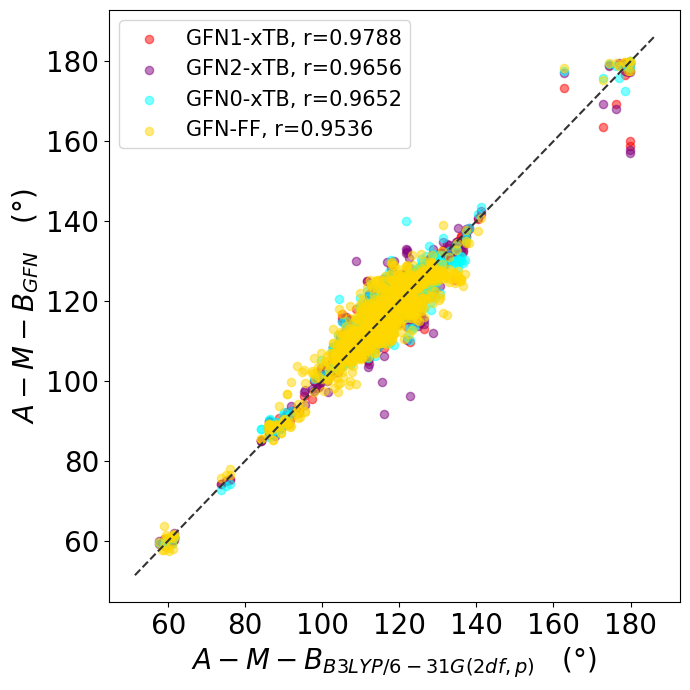}}\\
		\includegraphics[width=.49\textwidth]{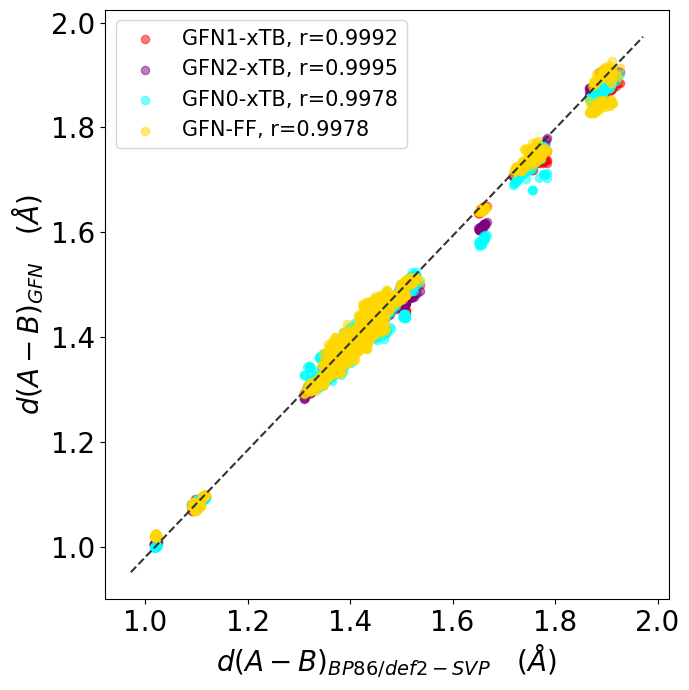}
		\subfloat[CEP samples]{\includegraphics[width=.49\textwidth]{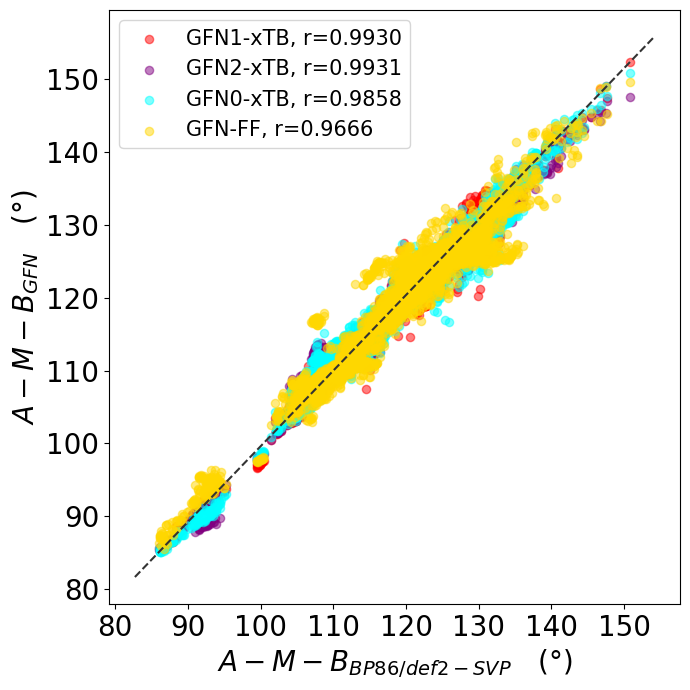}}
		\caption{The correlation plots for bond lengths and angles, measured in \unit{\angstrom} and degree, respectively, calculated between the GFN optimized structures and (a) B3LYP/6-31G(2df, p) level for the optimized structures of small $\pi$-systems from the QM9 sample set and (b) BP86/def2-SVP level for the optimized structures of extended $\pi$-systems from the CEP sample set.}
		\label{fig:Bonds_Angles}
	\end{center}
\end{figure}

\Cref{fig:Bonds_Angles} presents correlation graphs and statistical summaries for the lengths and angles of the bonds. For QM9 bonds (\Cref{fig:Bonds_Angles}(a)), \gfnb demonstrates the highest correlation ($r=0.9976$) and lowest MAD (\qty{0.0089}{\angstrom}), showing excellent agreement with B3LYP. \gfnc and \gfna also perform well ($r > 0.99$), while \gfnf shows a lower correlation ($r=0.9887$) despite analyzing the majority of bonds. For QM9 angles (\Cref{fig:Bonds_Angles}(a)), \gfnb again leads ($r=0.9788$, MAD=\qty{1.6141}{\degree}), followed by \gfnc ($r=0.9656$, MAD=\qty{1.9872}{\degree}). \gfna is less accurate (\qty{2.1665}{\degree}), and \gfnf performs least well ($r=0.9536$, MAD=\qty{2.7803}{\degree}), indicating challenges in capturing angular relationships.

For CEP bonds (\Cref{fig:Bonds_Angles}(b)), all GFN methods show markedly higher correlations ($r > 0.99$) compared to QM9, indicating stronger linear relationships with BP86. \gfnb retains the lowest MAD (\qty{0.0179}{\angstrom}), although the deviations are slightly larger than for QM9 bonds, probably due to increased SIE in extended $\pi$-systems. \gfnf shows consistent bond error behavior in both data sets (MAD =\qty{0.0194}{\angstrom}). For CEP angles (\Cref{fig:Bonds_Angles}(b)), \gfnc (MAD=\qty{0.7000}{\degree}) and \gfnb (MAD=\qty{0.7219}{\degree}) exhibit the highest accuracy and correlations ($r \approx 0.993$). \gfna performs reasonably well ($r=0.9858$), while \gfnf again has larger errors (MAD=\qty{1.5128}{\degree}). In general, the GFN methods exhibit better angular agreement for CEP compared to QM9 (average difference in MAD $>\qty{1}{\degree}$), indicating an improved capability for handling angles in extended systems. These results, particularly the accurate bond lengths for small systems, underscore the robustness of GFN parameterization for specific types of chemical interactions. The refined treatment of electrostatics and dispersion in SCC GFN methods further contributes to improved agreement on bond lengths and angles with DFT compared to non-iterative methods.

In summary of structural fidelity (\Cref{tab:ResultsSummary}), \gfnb and \gfnc consistently provide the highest agreement with the reference geometries of DFT, particularly for the more localized small $\pi$-systems in QM9, where they clearly outperform the non-iterative (\gfna) and atomistic (\gfnf) GFN variants. \gfnf shows competitive accuracy for CEP, especially with respect to Rg and connectivity preservation, but struggles with angular accuracy. \gfna offers reasonable structural precision, which is between the iterative and \gfnf methods. This nuanced performance profile highlights the strengths and weaknesses of each method's theoretical model and parameterization across different structural metrics and system types, offering valuable insights for method developers and users. The superior performance of iterative SCC methods for CEP is likely due to their explicit treatment of charge interactions and enhanced dispersion model, which are essential to describe delocalized electrons in extended $\pi$-systems. For these extended conjugated systems, the robust force-field nature of GFN-FF and its efficient usage of EEQ charge model may offer a more stable and accurate description of the average structural properties, even with its simplified angular treatment.

\subsubsection{Molecular orbitals - HOMO-LUMO gap}

\begin{figure}[htbp]
	\begin{center}
		\subfloat[\centering QM9]{\includegraphics[width=.9\textwidth]{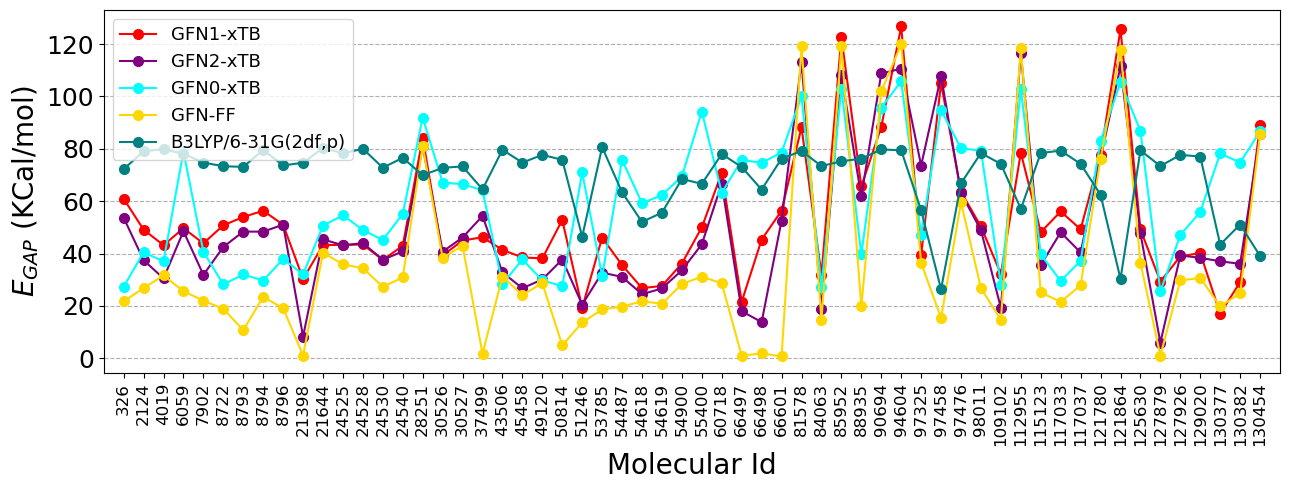}} \\
		\subfloat[\centering CEP]{\includegraphics[width=.9\textwidth]{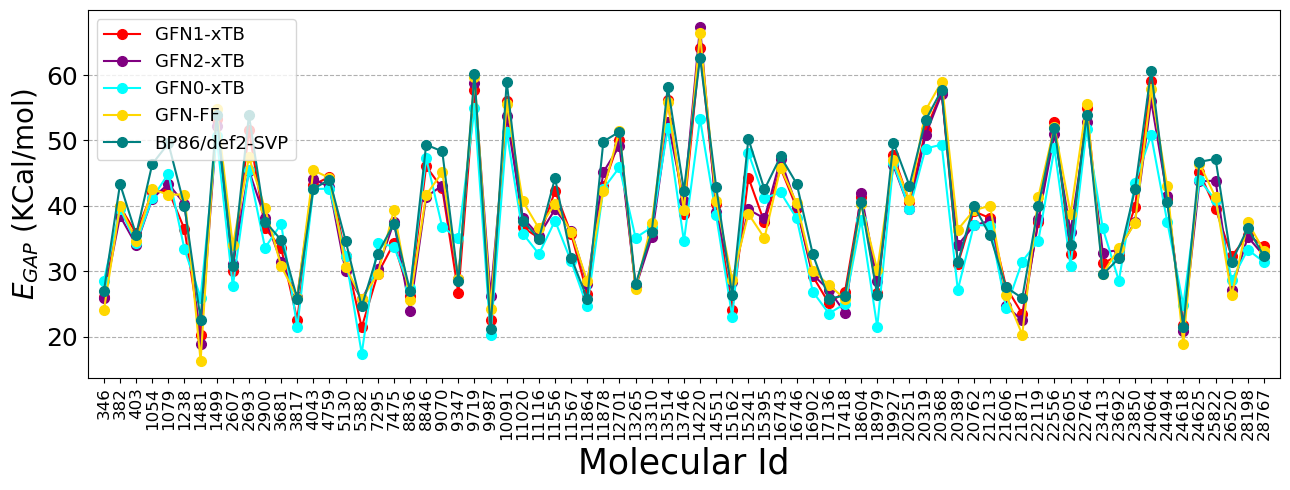}}
		\caption{HOMO-LUMO gap energies in \qty{}{\kilo\calorie\per\mole} of GFN optimized structures and (a) the B3LYP/6-31G(2df, p) level for optimized structures of small $\pi$-systems from the QM9 sample set and (b) the BP86/def2-SVP level for optimized structures of extended $\pi$-systems from the CEP sample set. Note that \qty{1}{\kilo\calorie\per\mole} is approximately \qty{0.04336}{\electronvolt}.}
		\label{fig:HL_Gap}
	\end{center}
\end{figure}

The HOMO-LUMO gap is a pivotal electronic property of semiconductors \cite{fukui_molecular_1952}, and we evaluated the ability of GFN methods to predict it. While density functional theory (DFT), even with hybrid (B3LYP) or generalized gradient approximation (GGA) such as BP86 functionals, has known limitations in reliably predicting absolute LUMO eigenvalues and consequently the gap \cite{zhang_comparison_2007, zhan_ionization_2003, allen_eigenvalues_2002}, these DFT results serve as a valuable computational benchmark in the absence of readily available experimental data for our large datasets. It is important to note that the absolute HOMO-LUMO gaps from DFT (particularly with GGA functionals) are generally approximate; achieving quantitative agreement with experimental optical gaps or ionization potentials typically requires further refinement using methods such as $\Delta$SCF, TD-DFT, or empirical calibration.

\Cref{fig:HL_Gap} presents the per-molecule HOMO-LUMO gaps obtained from our calculations. For the QM9 set (\Cref{fig:HL_Gap}(a)), GFN methods generally underestimate the gaps compared to B3LYP. Surprisingly, despite being non-self-consistent, \gfna yields the lowest MAD (\qty{1.256}{\electronvolt}) for the QM9 set, demonstrating the greatest proximity to the B3LYP reference values (\Cref{tab:ResultsSummary}). \gfnb follows (\qty{1.320}{\electronvolt}), while \gfnc (\qty{1.537}{\electronvolt}) and \gfnf (\qty{2.027}{\electronvolt}) exhibit larger deviations. This systematic underestimation by GFN methods for smaller, rigid systems is consistent with the known challenges of tight-binding approaches in accurately describing localized electronic features, a task where B3LYP's inclusion of exact Hartree-Fock exchange provides a distinct advantage.

For the CEP set (\Cref{fig:HL_Gap}(b)), extended systems generally exhibit significantly narrower gaps, ranging from approximately \qtyrange[range-phrase = --]{0.43}{2.82}{\electronvolt} (equivalent to \qtyrange[range-phrase = --]{10}{65}{\kilo\calorie\per\mole}), in contrast to QM9's broader range of approximately \qtyrange[range-phrase = --]{0.43}{5.20}{\electronvolt} (\qtyrange[range-phrase = --]{10}{120}{\kilo\calorie\per\mole}). Here, \gfnb demonstrates superior performance (MAD=\qty{0.0906}{\electronvolt}), followed by \gfnf (MAD=\qty{0.1294}{\electronvolt}) and \gfnc (MAD=\qty{0.1432}{\electronvolt}). \gfna shows the largest deviation (MAD=\qty{0.2775}{\electronvolt}). The improved performance of self-consistent charge (SCC) GFN methods and \gfnf (with SCC post-processing for molecular orbitals) for these extended systems can be attributed to their more effective handling of electron delocalization and dispersion, which are more prominent in larger $\pi$-systems. These results provide further support for the \gfnf method as a valuable tool for the simulation of extended $\pi$-systems. Our benchmarking thus provides a valuable baseline for the expected accuracy of GFN methods for this essential electronic property across different system sizes and complexities, underscoring that SCC-based methods and \gfnf with post-processing capabilities offer improved performance for delocalized $\pi$-systems compared to non-self-consistent \gfna.

\subsubsection{Computational cost}

\begin{figure}[htbp]
	\begin{center}
		\subfloat[\centering QM9]{\includegraphics[width=.9\textwidth]{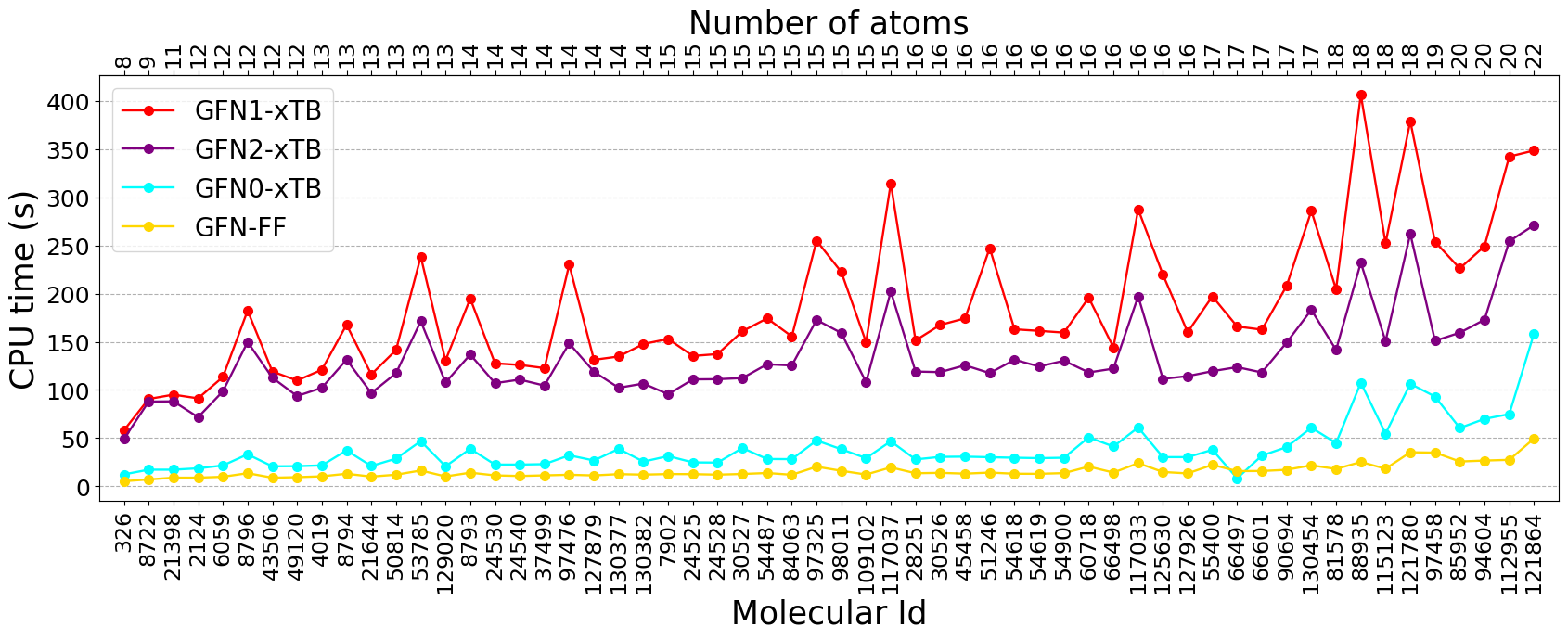}} \\
		\subfloat[\centering CEP]{\includegraphics[width=.9\textwidth]{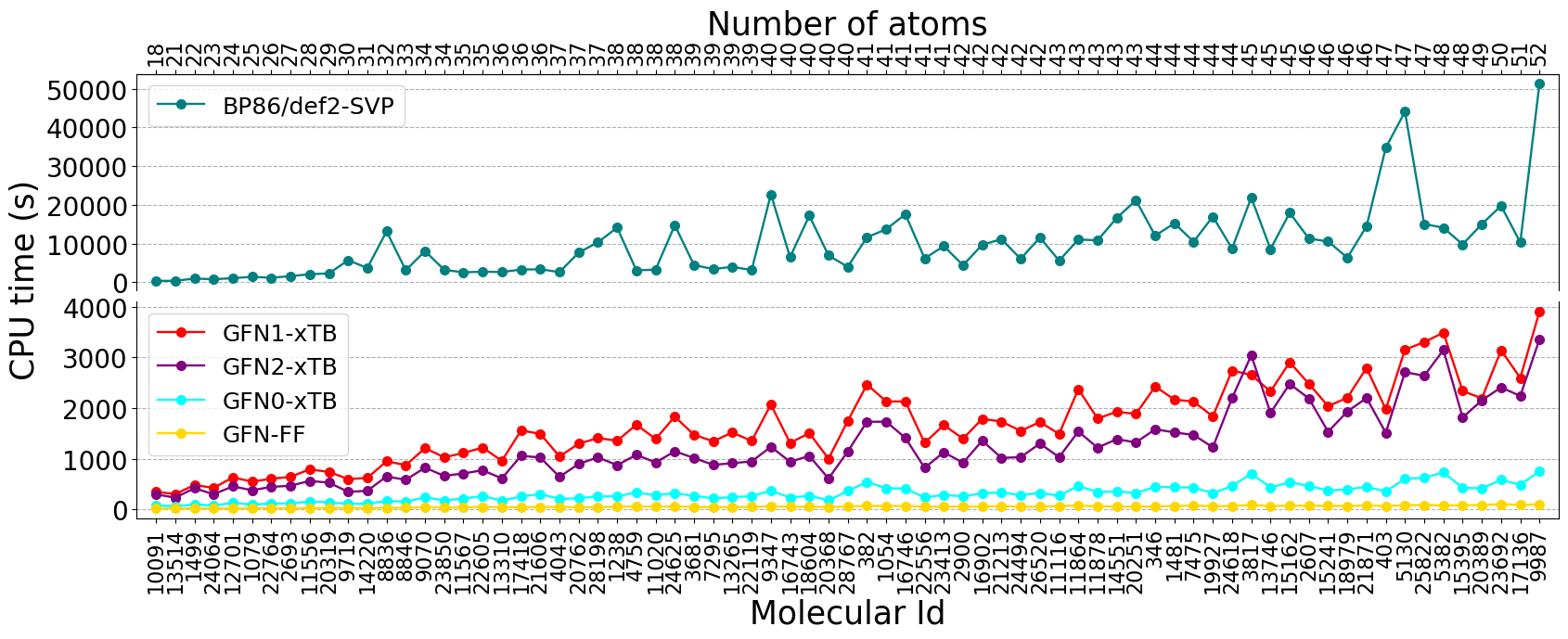}}
		\caption{The CPU times measured in seconds for the geometry optimizations of the QM9 samples (a) and the CEP samples (b). The number of atoms for each sample is indicated at the top of the abscissa. Graph (b) also shows the CPU times required for the geometry optimizations of extended $\pi$-systems of the CEP sample set using the BP86/def2-SVP level. Calculations were performed using a high-performance computer with two 16-core Intel Xeon Gold 6142@2.6 GHz processors and 192 GB of RAM, using 8 cores per task with 20 GB of memory per core.}
		\label{fig:CPU}
	\end{center}
\end{figure}

A pivotal aspect for high-throughput applications is computational efficiency. \Cref{fig:CPU} shows the CPU times required for geometry optimization. As anticipated, the computational cost is substantially higher for larger CEP molecules (average $\sim \qtyrange[range-phrase = \text{ to }]{56}{1600}{\second}$) compared to small QM9 molecules (average $\sim \qtyrange[range-phrase = \text{ to }]{15}{170}{\second}$). This order-of-magnitude difference profoundly underscores the challenges associated with simulating larger systems using higher-level quantum chemical methods.

Within the GFN family, \gfna and \gfnf are significantly faster than the iterative SCC methods (\gfnb and \gfnc) for both datasets (\Cref{tab:ResultsSummary}). For QM9, \gfnf is the fastest (average $\qty{14.85}{\second}$), followed by \gfna ($\qty{40.97}{\second}$). \gfnc ($\qty{130.93}{\second}$) and \gfnb ($\qty{168.13}{\second}$) are substantially slower. For CEP, \gfnf is also the fastest ($\qty{56.28}{\second}$), showing a dramatic speed-up (more than $\sim 170$ times faster) compared to the average CPU time for DFT at the BP86/def2-SVP level ($\sim \qty{9931.45}{\second}$). \gfna ($\qty{302.21}{\second}$) is the next fastest GFN, followed by \gfnc ($\qty{1358.46}{\second}$) and \gfnb ($\qty{1608.87}{\second}$), highlighting the significant computational cost of the SCC iteration for larger systems.

Analysis of algorithmic scaling behavior (illustrated with complexity fits in \ref{fig:SI_S9_QM9_Scaling} for QM9 and \ref{fig:SI_S10_CEP_Scaling} for CEP) provides a deeper insight into the efficiency of these quantum chemical approaches for varying system sizes. For QM9, \gfna and \gfnf exhibit primarily cubic scaling $\mathcal{O}(\mathcal{N}^3)$, while \gfnb and \gfnc show closer to quadratic $\mathcal{O}(\mathcal{N}^2)$ or linearithmic $\mathcal{O}(\mathcal{N}\log\mathcal{N})$ behavior in some fits, although cubic fits also have high $R^2$ values. It is important to acknowledge that the $R^2$ values for some fits, particularly for QM9 (e.g., GFN1-xTB with $R^2=\num{0.4983}$), indicate that the scaling trends are approximate and can be influenced by system-specific factors and conformational complexity within the dataset. For CEP, all electronic GFN methods (\gfnb, \gfnc, \gfna) consistently show cubic scaling $\mathcal{O}(\mathcal{N}^3)$, while the non-electronic \gfnf method achieves a more favorable quadratic scaling $\mathcal{O}(\mathcal{N}^2)$. This confirms that \gfnf's speed advantage stems from its force-field nature and efficient charge determination via the semiclassical electronegativity equilibration (EEQ) model, making it particularly well-suited for larger extended systems, aligning with findings for protein systems \cite{bannwarth_extended_2021}. Understanding these scaling characteristics is therefore essential for judiciously selecting appropriate methods for studies involving large chemical libraries or extended molecular systems, representing a key consideration in computational quantum chemistry explorations of vast chemical spaces.

{\renewcommand{\arraystretch}{1.2}
	\begin{center}
		\begin{longtable}[H]{@{}l l cccc@{}}
			\caption{Summary of comprehensive benchmarking results for the GFN family of semiempirical methods compared to DFT reference calculations. This table presents a detailed overview of the performance of \gfnb, \gfnc, \gfna, and \gfnf when evaluated against DFT on the QM9 and CEP organic molecular datasets. Key metrics for assessing structural fidelity (including Mean Absolute Deviations for Radius of Gyration, heavy-atom RMSD, rotational constants, bond lengths, and angles), electronic property prediction (Mean Absolute Deviation for HOMO-LUMO Gap), and computational efficiency (average CPU time) are summarized, providing a clear picture of each method's capabilities and associated computational cost for these systems.\label{tab:ResultsSummary}}\\
			\toprule
			\textbf{Metrics} & \textbf{Units} & \multicolumn{4}{c}{\textbf{Methods}} \\
			\cmidrule(l){3-6}
			& & \textbf{\gfnb} & \textbf{\gfnc} & \textbf{\gfna} & \textbf{\gfnf} \\
			\midrule
			\endfirsthead
			
			\multicolumn{6}{@{}l}{\textit{Continued from previous page}} \\
			\toprule
			\textbf{Metrics} & \textbf{Units} & \multicolumn{4}{c}{\textbf{Methods}} \\
			\cmidrule(l){3-6}
			& & \textbf{GFN2-xTB} & \textbf{GFN-FF} & \textbf{GFN1-xTB} & \textbf{GFN0-xTB} \\
			\midrule
			\endhead
			
			\bottomrule
			\multicolumn{6}{@{}r@{}}{\textit{Continued on next page}} \\
			\endfoot
			
			\bottomrule
			\endlastfoot
			
			\multicolumn{6}{c}{\textbf{QM9}} \\\midrule
			
			Rg MAD & \unit{\angstrom} & $0.0985\pm0.001124$ & $\bm{0.0921\pm0.001477}$ & $0.1158\pm0.000344$ & $0.1054\pm0.000428$ \\
			
			hRMSD & \unit{\angstrom} & $\bm{0.4915\pm0.056503}$ & $0.4988\pm0.054616$ & $0.5231\pm0.052283$ & $0.5235\pm0.054673$ \\
			
			$A_e$ MAD & $\unit{\per\centi\meter}$ & $4.0949\pm0.000364$ & $\bm{4.0939\pm0.000334}$ & $4.0965\pm0.000170$ & $4.0944\pm0.000044$ \\
			
			$B_e$ MAD & $\unit{\per\centi\meter}$ & $1.2431\pm0.000044$ & $1.2438\pm0.000060$ & $\bm{1.2413\pm0.000037}$ & $1.2435\pm0.000033$ \\
			
			$C_e$ MAD & $\unit{\per\centi\meter}$ & $\bm{0.9890\pm0.000027}$ & $0.9897\pm0.000037$ & $0.9892\pm0.000036$ & $0.9901\pm0.000023$ \\
			
			Bonds MAD & \unit{\angstrom} & $\bm{0.0089\pm0.000024}$ & $0.0117\pm0.000031$ & $0.0181\pm0.000006$ & $0.0194\pm0.000003$ \\
			
			Angles MAD & $\unit{\degree}$ & $\bm{1.6141\pm0.006083}$ & $1.9872\pm0.007022$ & $2.1665\pm0.001190$ & $2.7803\pm0.001845$ \\
			
			Gap MAD & \unit{\electronvolt} & $1.3207\pm0.001759$ & $1.5373\pm0.004676$ & $\bm{1.2560\pm0.000316}$ & $2.0271\pm0.000885$ \\
			
			CPU time & \unit{\second} & $168.1297\pm8.905793$ & $130.9318\pm6.067700$ & $40.9685\pm3.364115$ & $\bm{14.8524\pm0.929212}$ \\
			
			\midrule
			\multicolumn{6}{c}{\textbf{CEP}} \\\midrule
			Rg MAD & \unit{\angstrom} & $0.0691\pm0.000017$ & $0.0737\pm0.000102$ & $0.0714\pm0.000351$ & $\bm{0.0611\pm0.000244}$ \\
			
			hRMSD & \unit{\angstrom} & $\bm{0.7648\pm0.077280}$ & $0.7707\pm0.069165$ & $0.8282\pm0.079761$ & $0.7930\pm0.074828$ \\
			
			$A_e$ MAD & $\unit{\per\centi\meter}$ & $0.0121\pm0.000000$ & $0.0122\pm0.000022$ & $\bm{0.0119\pm0.000029}$ & $0.0124\pm0.000004$ \\
			
			$B_e$ MAD & $\unit{\per\centi\meter}$ & $0.0031\pm0.000000$ & $0.0031\pm0.000000$ & $0.0031\pm0.000001$ & $\bm{0.0030\pm0.000000}$ \\
			
			$C_e$ MAD & $\unit{\per\centi\meter}$ & $0.0012\pm0.000000$ & $0.0012\pm0.000000$ & $0.0012\pm0.000000$ & $\bm{0.0011\pm0.000000}$ \\
			
			Bonds MAD & \unit{\angstrom} & $\bm{0.0179\pm0.000002}$ & $0.0209\pm0.000003$ & $0.0199\pm0.000006$ & $0.0194\pm0.000000$ \\
			
			Angles MAD & $\unit{\degree}$ & $0.7219\pm0.001034$ & $\bm{0.7000\pm0.001366}$ & $1.1132\pm0.001328$ & $1.5128\pm0.000372$ \\
			
			Gap MAD & \unit{\electronvolt} & $\bm{0.0906\pm0.000644}$ & $0.1243\pm0.004093$ & $0.1749\pm0.000727$ & $0.1238\pm0.000089$ \\
			
			CPU time & \unit{\second} & $1608.8737\pm87.680572$ & $1358.4581\pm80.832675$ & $302.2100\pm17.597610$ & $\bm{56.2762\pm2.515636}$ \\
			
		\end{longtable}
	\end{center}
}

\section{Conclusions}\label{sec:conclusions}
%Completely rewriting

This work has presented a detailed benchmarking study of the GFN family of semiempirical methods (\gfnb, \gfnc, \gfna, and \gfnf) against density functional theory (DFT) for geometry optimization and electronic property prediction across two representative datasets: QM9-derived small $\pi$-systems and CEP-derived extended $\pi$-systems. These datasets were selected to reflect chemically diverse systems relevant to organic electronics and materials discovery. The results elucidate the inherent trade-offs between computational cost and accuracy across GFN levels and offer practical guidance for method selection in high-throughput quantum chemical applications.

For small $\pi$-systems in the QM9 dataset, iterative self-consistent charge (SCC) methods \gfnb and \gfnc demonstrated the highest structural fidelity, particularly for sensitive metrics such as heavy-atom RMSD, rotational constants, and bond angles. \gfnb provided marginally improved accuracy over \gfnc, including more consistent HOMO–LUMO gap predictions, albeit with a substantially higher computational cost. This reflects the general trend that improved accuracy in SCC-based approaches comes at the expense of scalability, which becomes critical for larger systems.

In the CEP dataset, comprising extended conjugated structures, the GFN methods showed better agreement with the BP86 / def2-SVP geometries than with B3LYP / 6-31G (2df, p), probably due to the greater importance of the dispersion and delocalization effects captured more effectively by the underlying GFN parameterization. In this regime, \gfnb maintained superior performance in geometric and electronic descriptors, while \gfnc offered a competitive compromise between accuracy and efficiency. The non-SCC method \gfna proved adequate for the QM9 set but was less reliable for extended systems, indicating its more limited applicability.

A particularly relevant outcome concerns the performance of \gfnf. As the fastest method among those examined, it exhibits favorable $\mathcal{O}(\mathcal{N}^2)$ scaling and up to 20-fold speedups relative to SCC methods in large systems. In practical terms, for molecules exceeding 100 atoms, such as those in the CEP set, the cubic scaling of SCC-based GFN methods becomes a limiting factor in high-throughput workflows, making \gfnf an attractive alternative where rapid geometry optimization is essential. However, its simplified angular parameterization results in higher deviations in bond angles (up to \qty{2.78}{\degree} for QM9 and \qty{1.51}{\degree} for CEP), which may limit its utility in applications requiring accurate conformational modeling. Despite these limitations, when paired with post-optimization SCC calculations (e.g., GFN2-xTB), \gfnf provides access to reliable electronic structure predictions at a fraction of the computational cost. This hybrid approach is well-suited to hierarchical screening protocols.

It is also important to emphasize that the benchmarking of HOMO–LUMO gaps conducted herein is relative, as the DFT reference values, derived from Kohn–Sham eigenvalues, systematically underestimate fundamental gaps. Therefore, deviations reported for the GFN methods should be interpreted in terms of consistency with DFT rather than agreement with experimental or quasiparticle-level data.

Additionally, while GFN calculations were performed using the ALPB implicit solvation model (with toluene), the DFT references correspond to gas-phase calculations. This methodological mismatch reflects common practice in semiempirical workflows and primarily impacts absolute quantities. Our preliminary tests did not show significant alterations in the relative performance ranking due to this solvation model, which justified its use for comparative benchmarking.

Taken together, these findings reinforce the value of GFN-based methods as practical tools for accelerating quantum chemical simulations, particularly in the context of materials discovery and organic semiconductor design. The choice among the GFN levels should be guided by the target system size, the accuracy required, and the computational constraints. Future developments may benefit from improved angular parameterization, refined solvation models, and systematic treatment of edge cases to further broaden the applicability of these methods. The insights provided by this study are also expected to inform the design of Quantitative Structure-Property Relationships (QSPR) and machine learning models for property prediction in organic electronics and to support the development of next-generation semiempirical quantum chemical methods.

%\par\null
\section*{Acknowledgements}\label{sec:Aknowledgements}

The authors thank Professor Gian-Marco Rignanese of Université catholique de Louvain for his support, especially in providing access to essential high-performance computing resources. Computation was supported by Université catholique de Louvain’s supercomputing facilities (CISM/UCL) and the Consortium des Équipements de Calcul Intensif en Fédération Wallonie Bruxelles (CÉCI), funded by the Fond de la Recherche Scientifique de Belgique (F.R.S.-FNRS) under convention 2.5020.11 and the Walloon Region. The authors also thank Professor Stefan Grimme and his team for developing the xTB software, a reliable tool for semiempirical quantum chemistry calculations, which significantly aided our research.

%\par \null
\section*{Conflict of Interest}\label{sec:Conflict_of_interest}

The authors have declared no conflicts of interest for this article.

%\par\null
\section*{Author Contributions}\label{sec:Author_contributions}

\textbf{Steve Cabrel Teguia Kouam:} Conceptualization; data curation; statistical analysis; methodology; writing-original draft; writing-review and editing. \textbf{Jean-Pierre Tchapet Njafa:} Statistical analysis; methodology; supervision; writing-original draft; writing-review and editing. \textbf{Raoult Dabou Teukam:} Data curation; methodology; supervision; writing-review and editing. \textbf{Patrick Mvoto Kongo:} Data curation; statistical analysis; methodology; writing-original draft. \textbf{Jean-Pierre Nguenang:} Conceptualization; methodology; supervision; writing-review and editing. \textbf{Serge Guy Nana Engo:} Conceptualization; methodology; supervision; writing-review and editing.

%\par\null
\section*{Data Availability}\label{sec:Data_availability}

Data supporting the findings of this study are available from the corresponding author.

%----------------------------------------------------------------------------------------
%	 REFERENCES
%----------------------------------------------------------------------------------------

\printbibliography % Output the bibliography

@misc{nigam_tartarus_2023,
	title = {Tartarus: {A} {Benchmarking} {Platform} for {Realistic} {And} {Practical} {Inverse} {Molecular} {Design}},
	shorttitle = {Tartarus},
	url = {http://arxiv.org/abs/2209.12487},
	urldate = {2023-09-10},
	publisher = {arXiv},
	author = {Nigam, AkshatKumar and Pollice, Robert and Tom, Gary and Jorner, Kjell and Thiede, Luca A. and Kundaje, Anshul and Aspuru-Guzik, Alan},
	month = jul,
	year = {2023},
	note = {arXiv:2209.12487 [cs]},
}

@article{rogers_computer_1960,
	title = {A {Computer} {Program} for {Classifying} {Plants}},
	volume = {132},
	issn = {0036-8075},
	url = {https://www.jstor.org/stable/1706749},
	number = {3434},
	urldate = {2023-10-22},
	journal = {Science},
	author = {Rogers, David J. and Tanimoto, Taffee T.},
	year = {1960},
	note = {Publisher: American Association for the Advancement of Science},
	pages = {1115--1118},
}

@article{rousseeuw_silhouettes_1987,
	title = {Silhouettes: {A} graphical aid to the interpretation and validation of cluster analysis},
	volume = {20},
	issn = {03770427},
	shorttitle = {Silhouettes},
	url = {https://linkinghub.elsevier.com/retrieve/pii/0377042787901257},
	doi = {10.1016/0377-0427(87)90125-7},
	language = {en},
	urldate = {2024-01-30},
	journal = {Journal of Computational and Applied Mathematics},
	author = {Rousseeuw, Peter J.},
	month = nov,
	year = {1987},
	pages = {53--65},
}

@incollection{macqueen_methods_1967,
	title = {Some methods for classification and analysis of multivariate observations},
	volume = {5.1},
	url = {https://projecteuclid.org/ebooks/berkeley-symposium-on-mathematical-statistics-and-probability/Proceedings-of-the-Fifth-Berkeley-Symposium-on-Mathematical-Statistics-and/chapter/Some-methods-for-classification-and-analysis-of-multivariate-observations/bsmsp/1200512992},
	urldate = {2024-02-12},
	booktitle = {Proceedings of the {Fifth} {Berkeley} {Symposium} on {Mathematical} {Statistics} and {Probability}, {Volume} 1: {Statistics}},
	publisher = {University of California Press},
	author = {MacQueen, J.},
	month = jan,
	year = {1967},
	pages = {281--298},
}

@article{ringner_what_2008,
	title = {What is principal component analysis?},
	volume = {26},
	copyright = {http://www.springer.com/tdm},
	issn = {1087-0156, 1546-1696},
	url = {https://www.nature.com/articles/nbt0308-303},
	doi = {10.1038/nbt0308-303},
	language = {en},
	number = {3},
	urldate = {2024-03-26},
	journal = {Nat Biotechnol},
	author = {Ringnér, Markus},
	month = mar,
	year = {2008},
	pages = {303--304},
}

@article{bannwarth_extended_2021,
	title = {Extended tight‐binding quantum chemistry methods},
	volume = {11},
	issn = {1759-0876, 1759-0884},
	shorttitle = {Extended},
	url = {https://wires.onlinelibrary.wiley.com/doi/10.1002/wcms.1493},
	doi = {10.1002/wcms.1493},
	language = {en},
	number = {2},
	urldate = {2024-11-29},
	journal = {WIREs Comput Mol Sci},
	author = {Bannwarth, Christoph and Caldeweyher, Eike and Ehlert, Sebastian and Hansen, Andreas and Pracht, Philipp and Seibert, Jakob and Spicher, Sebastian and Grimme, Stefan},
	month = mar,
	year = {2021},
	pages = {e1493},
}

@article{porezag_construction_1995,
	title = {Construction of tight-binding-like potentials on the basis of density-functional theory: {Application} to carbon},
	volume = {51},
	shorttitle = {Construction of tight-binding-like potentials on the basis of density-functional theory},
	url = {https://link.aps.org/doi/10.1103/PhysRevB.51.12947},
	doi = {10.1103/PhysRevB.51.12947},
	number = {19},
	urldate = {2024-12-10},
	journal = {Phys. Rev. B},
	author = {Porezag, D. and Frauenheim, Th. and Köhler, Th. and Seifert, G. and Kaschner, R.},
	month = may,
	year = {1995},
	note = {Publisher: American Physical Society},
	pages = {12947--12957},
}

@article{elstner_self-consistent-charge_1998,
	title = {Self-consistent-charge density-functional tight-binding method for simulations of complex materials properties},
	volume = {58},
	url = {https://link.aps.org/doi/10.1103/PhysRevB.58.7260},
	doi = {10.1103/PhysRevB.58.7260},
	number = {11},
	urldate = {2024-12-10},
	journal = {Phys. Rev. B},
	author = {Elstner, M. and Porezag, D. and Jungnickel, G. and Elsner, J. and Haugk, M. and Frauenheim, Th. and Suhai, S. and Seifert, G.},
	month = sep,
	year = {1998},
	note = {Publisher: American Physical Society},
	pages = {7260--7268},
}

@article{gaus_dftb3_2011,
	title = {{DFTB3}: {Extension} of the {Self}-{Consistent}-{Charge} {Density}-{Functional} {Tight}-{Binding} {Method} ({SCC}-{DFTB})},
	volume = {7},
	issn = {1549-9618, 1549-9626},
	shorttitle = {{DFTB3}},
	url = {https://pubs.acs.org/doi/10.1021/ct100684s},
	doi = {10.1021/ct100684s},
	language = {en},
	number = {4},
	urldate = {2024-12-10},
	journal = {J. Chem. Theory Comput.},
	author = {Gaus, Michael and Cui, Qiang and Elstner, Marcus},
	month = apr,
	year = {2011},
	pages = {931--948},
}

@article{christensen_semiempirical_2016,
	title = {Semiempirical {Quantum} {Mechanical} {Methods} for {Noncovalent} {Interactions} for {Chemical} and {Biochemical} {Applications}},
	volume = {116},
	copyright = {http://pubs.acs.org/page/policy/authorchoice\_termsofuse.html},
	issn = {0009-2665, 1520-6890},
	url = {https://pubs.acs.org/doi/10.1021/acs.chemrev.5b00584},
	doi = {10.1021/acs.chemrev.5b00584},
	language = {en},
	number = {9},
	urldate = {2024-12-10},
	journal = {Chem. Rev.},
	author = {Christensen, Anders S. and Kubař, Tomáš and Cui, Qiang and Elstner, Marcus},
	month = may,
	year = {2016},
	pages = {5301--5337},
}

@article{grimme_robust_2017,
	title = {A {Robust} and {Accurate} {Tight}-{Binding} {Quantum} {Chemical} {Method} for {Structures}, {Vibrational} {Frequencies}, and {Noncovalent} {Interactions} of {Large} {Molecular} {Systems} {Parametrized} for {All} spd-{Block} {Elements} ( \textit{{Z}} = 1–86)},
	volume = {13},
	copyright = {http://pubs.acs.org/page/policy/authorchoice\_termsofuse.html},
	issn = {1549-9618, 1549-9626},
	url = {https://pubs.acs.org/doi/10.1021/acs.jctc.7b00118},
	doi = {10.1021/acs.jctc.7b00118},
	language = {en},
	number = {5},
	urldate = {2024-12-10},
	journal = {J. Chem. Theory Comput.},
	author = {Grimme, Stefan and Bannwarth, Christoph and Shushkov, Philip},
	month = may,
	year = {2017},
	pages = {1989--2009},
}

@article{bannwarth_gfn2-xtbaccurate_2019,
	title = {{GFN2}-{xTB}—{An} {Accurate} and {Broadly} {Parametrized} {Self}-{Consistent} {Tight}-{Binding} {Quantum} {Chemical} {Method} with {Multipole} {Electrostatics} and {Density}-{Dependent} {Dispersion} {Contributions}},
	volume = {15},
	copyright = {http://pubs.acs.org/page/policy/authorchoice\_termsofuse.html},
	issn = {1549-9618, 1549-9626},
	url = {https://pubs.acs.org/doi/10.1021/acs.jctc.8b01176},
	doi = {10.1021/acs.jctc.8b01176},
	language = {en},
	number = {3},
	urldate = {2024-12-10},
	journal = {J. Chem. Theory Comput.},
	author = {Bannwarth, Christoph and Ehlert, Sebastian and Grimme, Stefan},
	month = mar,
	year = {2019},
	pages = {1652--1671},
}

@misc{pracht_robust_2019,
	title = {A {Robust} {Non}-{Self}-{Consistent} {Tight}-{Binding} {Quantum} {Chemistry} {Method} for large {Molecules}},
	url = {https://chemrxiv.org/engage/chemrxiv/article-details/60c742abbdbb890c7ba3851a},
	doi = {10.26434/chemrxiv.8326202.v1},
	language = {en},
	urldate = {2024-12-10},
	publisher = {ChemRxiv},
	author = {Pracht, Philipp and Caldeweyher, Eike and Ehlert, Sebastian and Grimme, Stefan},
	month = jun,
	year = {2019},
}

@article{hadipour_deep_2022,
	title = {Deep clustering of small molecules at large-scale via variational autoencoder embedding and {K}-means},
	volume = {23},
	issn = {1471-2105},
	url = {https://bmcbioinformatics.biomedcentral.com/articles/10.1186/s12859-022-04667-1},
	doi = {10.1186/s12859-022-04667-1},
	language = {en},
	number = {S4},
	urldate = {2024-12-10},
	journal = {BMC Bioinformatics},
	author = {Hadipour, Hamid and Liu, Chengyou and Davis, Rebecca and Cardona, Silvia T. and Hu, Pingzhao},
	month = apr,
	year = {2022},
	pages = {132},
}

@article{ramakrishnan_quantum_2014,
	title = {Quantum chemistry structures and properties of 134 kilo molecules},
	volume = {1},
	issn = {2052-4463},
	url = {https://www.nature.com/articles/sdata201422},
	doi = {10.1038/sdata.2014.22},
	language = {en},
	number = {1},
	urldate = {2024-12-11},
	journal = {Sci Data},
	author = {Ramakrishnan, Raghunathan and Dral, Pavlo O. and Rupp, Matthias and Von Lilienfeld, O. Anatole},
	month = aug,
	year = {2014},
	pages = {140022},
}

@article{hachmann_harvard_2011,
	title = {The {Harvard} {Clean} {Energy} {Project}: {Large}-{Scale} {Computational} {Screening} and {Design} of {Organic} {Photovoltaics} on the {World} {Community} {Grid}},
	volume = {2},
	issn = {1948-7185, 1948-7185},
	shorttitle = {The {Harvard} {Clean} {Energy} {Project}},
	url = {https://pubs.acs.org/doi/10.1021/jz200866s},
	doi = {10.1021/jz200866s},
	language = {en},
	number = {17},
	urldate = {2024-12-11},
	journal = {J. Phys. Chem. Lett.},
	author = {Hachmann, Johannes and Olivares-Amaya, Roberto and Atahan-Evrenk, Sule and Amador-Bedolla, Carlos and Sánchez-Carrera, Roel S. and Gold-Parker, Aryeh and Vogt, Leslie and Brockway, Anna M. and Aspuru-Guzik, Alán},
	month = sep,
	year = {2011},
	pages = {2241--2251},
}

@article{costa_optical_2016,
	title = {Optical band gaps of organic semiconductor materials},
	volume = {58},
	issn = {09253467},
	url = {https://linkinghub.elsevier.com/retrieve/pii/S0925346716301483},
	doi = {10.1016/j.optmat.2016.03.041},
	language = {en},
	urldate = {2024-12-11},
	journal = {Optical Materials},
	author = {Costa, José C.S. and Taveira, Ricardo J.S. and Lima, Carlos F.R.A.C. and Mendes, Adélio and Santos, Luís M.N.B.F.},
	month = aug,
	year = {2016},
	pages = {51--60},
}

@article{weininger_smiles_1988,
	title = {{SMILES}, a chemical language and information system. 1. {Introduction} to methodology and encoding rules},
	volume = {28},
	issn = {0095-2338, 1520-5142},
	url = {https://pubs.acs.org/doi/abs/10.1021/ci00057a005},
	doi = {10.1021/ci00057a005},
	language = {en},
	number = {1},
	urldate = {2024-12-11},
	journal = {J. Chem. Inf. Comput. Sci.},
	author = {Weininger, David},
	month = feb,
	year = {1988},
	pages = {31--36},
}

@article{greenacre_principal_2022,
	title = {Principal component analysis},
	volume = {2},
	copyright = {2022 Springer Nature Limited},
	issn = {2662-8449},
	url = {https://www.nature.com/articles/s43586-022-00184-w},
	doi = {10.1038/s43586-022-00184-w},
	language = {en},
	number = {1},
	urldate = {2024-12-12},
	journal = {Nat Rev Methods Primers},
	author = {Greenacre, Michael and Groenen, Patrick J. F. and Hastie, Trevor and D’Enza, Alfonso Iodice and Markos, Angelos and Tuzhilina, Elena},
	month = dec,
	year = {2022},
	note = {Publisher: Nature Publishing Group},
	pages = {1--21},
}

@inproceedings{liu_understanding_2010,
	address = {Sydney, Australia},
	title = {Understanding of {Internal} {Clustering} {Validation} {Measures}},
	isbn = {978-1-4244-9131-5},
	url = {http://ieeexplore.ieee.org/document/5694060/},
	doi = {10.1109/ICDM.2010.35},
	urldate = {2024-12-12},
	booktitle = {2010 {IEEE} {International} {Conference} on {Data} {Mining}},
	publisher = {IEEE},
	author = {Liu, Yanchi and Li, Zhongmou and Xiong, Hui and Gao, Xuedong and Wu, Junjie},
	month = dec,
	year = {2010},
	pages = {911--916},
}

@incollection{singh_stratified_1996,
	address = {Dordrecht},
	title = {Stratified {Sampling}},
	volume = {15},
	isbn = {978-90-481-4703-8 978-94-017-1404-4},
	url = {http://link.springer.com/10.1007/978-94-017-1404-4_5},
	urldate = {2024-12-13},
	booktitle = {Elements of {Survey} {Sampling}},
	publisher = {Springer Netherlands},
	author = {Singh, Ravindra and Mangat, Naurang Singh},
	collaborator = {Singh, Ravindra and Mangat, Naurang Singh},
	year = {1996},
	doi = {10.1007/978-94-017-1404-4_5},
	note = {Series Title: Kluwer Texts in the Mathematical Sciences},
	pages = {102--144},
}

@incollection{nicolotti_molecular_2018,
	address = {New York, NY},
	title = {Molecular {Descriptors} for {Structure}–{Activity} {Applications}: {A} {Hands}-{On} {Approach}},
	volume = {1800},
	isbn = {978-1-4939-7898-4 978-1-4939-7899-1},
	shorttitle = {Molecular {Descriptors} for {Structure}–{Activity} {Applications}},
	url = {http://link.springer.com/10.1007/978-1-4939-7899-1_1},
	urldate = {2024-12-15},
	booktitle = {Computational {Toxicology}},
	publisher = {Springer New York},
	author = {Grisoni, Francesca and Ballabio, Davide and Todeschini, Roberto and Consonni, Viviana},
	editor = {Nicolotti, Orazio},
	year = {2018},
	doi = {10.1007/978-1-4939-7899-1_1},
	note = {Series Title: Methods in Molecular Biology},
	pages = {3--53},
}

@article{spicher_robust_2020,
	title = {Robust {Atomistic} {Modeling} of {Materials}, {Organometallic}, and {Biochemical} {Systems}},
	volume = {59},
	issn = {1433-7851, 1521-3773},
	url = {https://onlinelibrary.wiley.com/doi/10.1002/anie.202004239},
	doi = {10.1002/anie.202004239},
	language = {en},
	number = {36},
	urldate = {2024-12-24},
	journal = {Angew Chem Int Ed},
	author = {Spicher, Sebastian and Grimme, Stefan},
	month = sep,
	year = {2020},
	pages = {15665--15673},
}

@article{kohn_self-consistent_1965,
	title = {Self-{Consistent} {Equations} {Including} {Exchange} and {Correlation} {Effects}},
	volume = {140},
	copyright = {http://link.aps.org/licenses/aps-default-license},
	issn = {0031-899X},
	url = {https://link.aps.org/doi/10.1103/PhysRev.140.A1133},
	doi = {10.1103/PhysRev.140.A1133},
	language = {en},
	number = {4A},
	urldate = {2024-12-30},
	journal = {Phys. Rev.},
	author = {Kohn, W. and Sham, L. J.},
	month = nov,
	year = {1965},
	pages = {A1133--A1138},
}

@article{rogers_extended-connectivity_2010,
	title = {Extended-{Connectivity} {Fingerprints}},
	volume = {50},
	issn = {1549-9596, 1549-960X},
	url = {https://pubs.acs.org/doi/10.1021/ci100050t},
	doi = {10.1021/ci100050t},
	language = {en},
	number = {5},
	urldate = {2024-12-30},
	journal = {J. Chem. Inf. Model.},
	author = {Rogers, David and Hahn, Mathew},
	month = may,
	year = {2010},
	pages = {742--754},
}

@article{puzzarini_connections_2023,
	title = {Connections between the accuracy of rotational constants and equilibrium molecular structures},
	volume = {25},
	issn = {1463-9076, 1463-9084},
	url = {https://xlink.rsc.org/?DOI=D2CP04706C},
	doi = {10.1039/D2CP04706C},
	language = {en},
	number = {3},
	urldate = {2025-03-08},
	journal = {Phys. Chem. Chem. Phys.},
	author = {Puzzarini, Cristina and Stanton, John F.},
	year = {2023},
	pages = {1421--1429},
}

@misc{li_machine_2023,
	title = {Machine learning photodynamics reveals the role of solvent and pressure on the [2+2]- cycloadditions toward cubanes},
	copyright = {https://creativecommons.org/licenses/by-nc-nd/4.0/},
	url = {https://chemrxiv.org/engage/chemrxiv/article-details/6504ee8599918fe53704f5cd},
	doi = {10.26434/chemrxiv-2023-xswwp},
	urldate = {2025-03-10},
	publisher = {Chemistry},
	author = {Li, Jingbai and Lopez, Steven},
	month = sep,
	year = {2023},
}

@article{chen_reorganization_2022,
	title = {Reorganization energies of flexible organic molecules as a challenging target for machine learning enhanced virtual screening},
	volume = {1},
	issn = {2635-098X},
	url = {https://xlink.rsc.org/?DOI=D1DD00038A},
	doi = {10.1039/D1DD00038A},
	language = {en},
	number = {2},
	urldate = {2025-03-10},
	journal = {Digital Discovery},
	author = {Chen, Ke and Kunkel, Christian and Reuter, Karsten and Margraf, Johannes T.},
	year = {2022},
	pages = {147--157},
}

@article{chen_physics-inspired_2023,
	title = {Physics-inspired machine learning of localized intensive properties},
	volume = {14},
	issn = {2041-6520, 2041-6539},
	url = {https://xlink.rsc.org/?DOI=D3SC00841J},
	doi = {10.1039/D3SC00841J},
	language = {en},
	number = {18},
	urldate = {2025-03-10},
	journal = {Chem. Sci.},
	author = {Chen, Ke and Kunkel, Christian and Cheng, Bingqing and Reuter, Karsten and Margraf, Johannes T.},
	year = {2023},
	pages = {4913--4922},
}

@misc{anstine_aimnet2_2024,
	title = {{AIMNet2}: {A} {Neural} {Network} {Potential} to {Meet} your {Neutral}, {Charged}, {Organic}, and {Elemental}-{Organic} {Needs}},
	copyright = {https://creativecommons.org/licenses/by-nc/4.0/},
	shorttitle = {{AIMNet2}},
	url = {https://chemrxiv.org/engage/chemrxiv/article-details/6763b51281d2151a022fb6a5},
	doi = {10.26434/chemrxiv-2023-296ch-v3},
	urldate = {2025-03-10},
	publisher = {Chemistry},
	author = {Anstine, Dylan and Zubatyuk, Roman and Isayev, Olexandr},
	month = dec,
	year = {2024},
}

@article{riniker_similarity_2013,
	title = {Similarity maps - a visualization strategy for molecular fingerprints and machine-learning methods},
	volume = {5},
	copyright = {http://creativecommons.org/licenses/by/2.0},
	issn = {1758-2946},
	url = {https://jcheminf.biomedcentral.com/articles/10.1186/1758-2946-5-43},
	doi = {10.1186/1758-2946-5-43},
	language = {en},
	number = {1},
	urldate = {2025-03-12},
	journal = {J Cheminform},
	author = {Riniker, Sereina and Landrum, Gregory A},
	month = dec,
	year = {2013},
	pages = {43},
}

@Article{sun_pyscf_2018,
  author     = {Sun, Qiming and Berkelbach, Timothy C. and Blunt, Nick S. and Booth, George H. and Guo, Sheng and Li, Zhendong and Liu, Junzi and McClain, James D. and Sayfutyarova, Elvira R. and Sharma, Sandeep and Wouters, Sebastian and Chan, Garnet Kin‐Lic},
  journal    = {WIREs Comput Mol Sci},
  title      = {PySCF: the {Python}‐based simulations of chemistry framework},
  year       = {2018},
  issn       = {1759-0876, 1759-0884},
  month      = jan,
  number     = {1},
  pages      = {e1340},
  volume     = {8},
  copyright  = {http://onlinelibrary.wiley.com/termsAndConditions\#vor},
  doi        = {10.1002/wcms.1340},
  language   = {en},
  shorttitle = {P {\textless}span style="font-variant},
  url        = {https://wires.onlinelibrary.wiley.com/doi/10.1002/wcms.1340},
  urldate    = {2025-03-12},
}

@article{wang_geometry_2016,
	title = {Geometry optimization made simple with translation and rotation coordinates},
	volume = {144},
	issn = {0021-9606, 1089-7690},
	url = {https://pubs.aip.org/jcp/article/144/21/214108/313176/Geometry-optimization-made-simple-with-translation},
	doi = {10.1063/1.4952956},
	language = {en},
	number = {21},
	urldate = {2025-03-12},
	journal = {The Journal of Chemical Physics},
	author = {Wang, Lee-Ping and Song, Chenchen},
	month = jun,
	year = {2016},
	pages = {214108},
}

@article{bero_similarity_2017,
	title = {Similarity {Measure} for {Molecular} {Structure}: {A} {Brief} {Review}},
	volume = {892},
	copyright = {http://iopscience.iop.org/info/page/text-and-data-mining},
	issn = {1742-6588, 1742-6596},
	shorttitle = {Similarity {Measure} for {Molecular} {Structure}},
	url = {https://iopscience.iop.org/article/10.1088/1742-6596/892/1/012015},
	doi = {10.1088/1742-6596/892/1/012015},
	urldate = {2025-03-12},
	journal = {J. Phys.: Conf. Ser.},
	author = {Bero, S A and Muda, A K and Choo, Y H and Muda, N A and Pratama, S F},
	month = sep,
	year = {2017},
	pages = {012015},
}

@article{yoshikawa_fast_2019,
	title = {Fast, efficient fragment-based coordinate generation for {Open} {Babel}},
	volume = {11},
	issn = {1758-2946},
	url = {https://jcheminf.biomedcentral.com/articles/10.1186/s13321-019-0372-5},
	doi = {10.1186/s13321-019-0372-5},
	language = {en},
	number = {1},
	urldate = {2025-03-12},
	journal = {J Cheminform},
	author = {Yoshikawa, Naruki and Hutchison, Geoffrey R.},
	month = dec,
	year = {2019},
	pages = {49},
}

@article{halgren_mmff_1999,
	title = {{MMFF} {VI}. {MMFF94s} option for energy minimization studies},
	volume = {20},
	copyright = {http://doi.wiley.com/10.1002/tdm\_license\_1.1},
	issn = {0192-8651, 1096-987X},
	url = {https://onlinelibrary.wiley.com/doi/10.1002/(SICI)1096-987X(199905)20:7<720::AID-JCC7>3.0.CO;2-X},
	doi = {10.1002/(SICI)1096-987X(199905)20:7<720::AID-JCC7>3.0.CO;2-X},
	language = {en},
	number = {7},
	urldate = {2025-03-12},
	journal = {J. Comput. Chem.},
	author = {Halgren, Thomas A.},
	month = may,
	year = {1999},
	pages = {720--729},
}

@article{rappe_uff_1992,
	title = {{UFF}, a full periodic table force field for molecular mechanics and molecular dynamics simulations},
	volume = {114},
	issn = {0002-7863, 1520-5126},
	url = {https://pubs.acs.org/doi/abs/10.1021/ja00051a040},
	doi = {10.1021/ja00051a040},
	language = {en},
	number = {25},
	urldate = {2025-03-12},
	journal = {J. Am. Chem. Soc.},
	author = {Rappe, A. K. and Casewit, C. J. and Colwell, K. S. and Goddard, W. A. and Skiff, W. M.},
	month = dec,
	year = {1992},
	pages = {10024--10035},
}

@article{ehlert_robust_2021,
	title = {Robust and {Efficient} {Implicit} {Solvation} {Model} for {Fast} {Semiempirical} {Methods}},
	volume = {17},
	copyright = {https://doi.org/10.15223/policy-029},
	issn = {1549-9618, 1549-9626},
	url = {https://pubs.acs.org/doi/10.1021/acs.jctc.1c00471},
	doi = {10.1021/acs.jctc.1c00471},
	language = {en},
	number = {7},
	urldate = {2025-03-12},
	journal = {J. Chem. Theory Comput.},
	author = {Ehlert, Sebastian and Stahn, Marcel and Spicher, Sebastian and Grimme, Stefan},
	month = jul,
	year = {2021},
	pages = {4250--4261},
}

@article{lopez_harvard_2016,
	title = {The {Harvard} organic photovoltaic dataset},
	volume = {3},
	issn = {2052-4463},
	url = {https://www.nature.com/articles/sdata201686},
	doi = {10.1038/sdata.2016.86},
	language = {en},
	number = {1},
	urldate = {2025-03-12},
	journal = {Sci Data},
	author = {Lopez, Steven A. and Pyzer-Knapp, Edward O. and Simm, Gregor N. and Lutzow, Trevor and Li, Kewei and Seress, Laszlo R. and Hachmann, Johannes and Aspuru-Guzik, Alán},
	month = sep,
	year = {2016},
	pages = {160086},
}

@article{bajusz_why_2015,
	title = {Why is {Tanimoto} index an appropriate choice for fingerprint-based similarity calculations?},
	volume = {7},
	issn = {1758-2946},
	url = {https://jcheminf.biomedcentral.com/articles/10.1186/s13321-015-0069-3},
	doi = {10.1186/s13321-015-0069-3},
	language = {en},
	number = {1},
	urldate = {2025-03-14},
	journal = {J Cheminform},
	author = {Bajusz, Dávid and Rácz, Anita and Héberger, Károly},
	month = dec,
	year = {2015},
	pages = {20},
}

@article{cordella_subgraph_2004,
	title = {A (sub)graph isomorphism algorithm for matching large graphs},
	volume = {26},
	copyright = {https://ieeexplore.ieee.org/Xplorehelp/downloads/license-information/IEEE.html},
	issn = {0162-8828},
	url = {http://ieeexplore.ieee.org/document/1323804/},
	doi = {10.1109/TPAMI.2004.75},
	language = {en},
	number = {10},
	urldate = {2025-03-25},
	journal = {IEEE Trans. Pattern Anal. Machine Intell.},
	author = {Cordella, L.P. and Foggia, P. and Sansone, C. and Vento, M.},
	month = oct,
	year = {2004},
	pages = {1367--1372},
}

@article{fukui_molecular_1952,
	title = {A {Molecular} {Orbital} {Theory} of {Reactivity} in {Aromatic} {Hydrocarbons}},
	volume = {20},
	issn = {0021-9606, 1089-7690},
	url = {https://pubs.aip.org/jcp/article/20/4/722/73673/A-Molecular-Orbital-Theory-of-Reactivity-in},
	doi = {10.1063/1.1700523},
	language = {en},
	number = {4},
	urldate = {2025-03-27},
	journal = {The Journal of Chemical Physics},
	author = {Fukui, Kenichi and Yonezawa, Teijiro and Shingu, Haruo},
	month = apr,
	year = {1952},
	pages = {722--725},
}

@article{pople_approximate_1965-1,
	title = {Approximate {Self}-{Consistent} {Molecular} {Orbital} {Theory}. {I}. {Invariant} {Procedures}},
	volume = {43},
	issn = {0021-9606, 1089-7690},
	url = {https://pubs.aip.org/jcp/article/43/10/S129/83581/Approximate-Self-Consistent-Molecular-Orbital},
	doi = {10.1063/1.1701475},
	language = {en},
	number = {10},
	urldate = {2025-04-01},
	journal = {The Journal of Chemical Physics},
	author = {Pople, J. A. and Santry, D. P. and Segal, G. A.},
	month = nov,
	year = {1965},
	pages = {S129--S135},
}

@article{yilmazer_comparison_2013,
	title = {Comparison of {Molecular} {Mechanics}, {Semi}-{Empirical} {Quantum} {Mechanical}, and {Density} {Functional} {Theory} {Methods} for {Scoring} {Protein}–{Ligand} {Interactions}},
	volume = {117},
	issn = {1520-6106, 1520-5207},
	url = {https://pubs.acs.org/doi/10.1021/jp402719k},
	doi = {10.1021/jp402719k},
	language = {en},
	number = {27},
	urldate = {2025-04-02},
	journal = {J. Phys. Chem. B},
	author = {Yilmazer, Nusret Duygu and Korth, Martin},
	month = jul,
	year = {2013},
	pages = {8075--8084},
}

@article{tortorella_benchmarking_2016,
	title = {Benchmarking {DFT} and semi-empirical methods for a reliable and cost-efficient computational screening of benzofulvene derivatives as donor materials for small-molecule organic solar cells},
	volume = {28},
	issn = {0953-8984, 1361-648X},
	url = {https://iopscience.iop.org/article/10.1088/0953-8984/28/7/074005},
	doi = {10.1088/0953-8984/28/7/074005},
	number = {7},
	urldate = {2025-04-02},
	journal = {J. Phys.: Condens. Matter},
	author = {Tortorella, Sara and Talamo, Maurizio Mastropasqua and Cardone, Antonio and Pastore, Mariachiara and De Angelis, Filippo},
	month = feb,
	year = {2016},
	pages = {074005},
}

@article{zheng_performance_2005,
	title = {Performance of the {DFTB} method in comparison to {DFT} and semiempirical methods for geometries and energies of {C20}–{C86} fullerene isomers},
	volume = {412},
	copyright = {https://www.elsevier.com/tdm/userlicense/1.0/},
	issn = {00092614},
	url = {https://linkinghub.elsevier.com/retrieve/pii/S0009261405009334},
	doi = {10.1016/j.cplett.2005.06.105},
	language = {en},
	number = {1-3},
	urldate = {2025-04-02},
	journal = {Chemical Physics Letters},
	author = {Zheng, Guishan and Irle, Stephan and Morokuma, Keiji},
	month = aug,
	year = {2005},
	pages = {210--216},
}

@article{schenker_assessment_2011,
	title = {Assessment of {Popular} {DFT} and {Semiempirical} {Molecular} {Orbital} {Techniques} for {Calculating} {Relative} {Transition} {State} {Energies} and {Kinetic} {Product} {Distributions} in {Enantioselective} {Organocatalytic} {Reactions}},
	volume = {7},
	issn = {1549-9618, 1549-9626},
	url = {https://pubs.acs.org/doi/10.1021/ct2002013},
	doi = {10.1021/ct2002013},
	language = {en},
	number = {11},
	urldate = {2025-04-02},
	journal = {J. Chem. Theory Comput.},
	author = {Schenker, Sebastian and Schneider, Christopher and Tsogoeva, Svetlana B. and Clark, Timothy},
	month = nov,
	year = {2011},
	pages = {3586--3595},
}

@article{schmitz_quantum_2020,
	title = {Quantum {Chemical} {Calculation} of {Molecular} and {Periodic} {Peptide} and {Protein} {Structures}},
	volume = {124},
	copyright = {https://doi.org/10.15223/policy-029},
	issn = {1520-6106, 1520-5207},
	url = {https://pubs.acs.org/doi/10.1021/acs.jpcb.0c00549},
	doi = {10.1021/acs.jpcb.0c00549},
	language = {en},
	number = {18},
	urldate = {2025-04-02},
	journal = {J. Phys. Chem. B},
	author = {Schmitz, Sarah and Seibert, Jakob and Ostermeir, Katja and Hansen, Andreas and Göller, Andreas H. and Grimme, Stefan},
	month = may,
	year = {2020},
	pages = {3636--3646},
}

@article{bursch_structure_2019,
	title = {Structure {Optimisation} of {Large} {Transition}‐{Metal} {Complexes} with {Extended} {Tight}‐{Binding} {Methods}},
	volume = {58},
	issn = {1433-7851, 1521-3773},
	url = {https://onlinelibrary.wiley.com/doi/10.1002/anie.201904021},
	doi = {10.1002/anie.201904021},
	language = {en},
	number = {32},
	urldate = {2025-04-02},
	journal = {Angew Chem Int Ed},
	author = {Bursch, Markus and Neugebauer, Hagen and Grimme, Stefan},
	month = aug,
	year = {2019},
	pages = {11078--11087},
}

@article{menzel_silico_2022,
	title = {In {Silico} {Optimization} of {Charge} {Separating} {Dyes} for {Solar} {Energy} {Conversion}},
	volume = {15},
	issn = {1864-5631, 1864-564X},
	url = {https://chemistry-europe.onlinelibrary.wiley.com/doi/10.1002/cssc.202200594},
	doi = {10.1002/cssc.202200594},
	language = {en},
	number = {15},
	urldate = {2025-04-02},
	journal = {ChemSusChem},
	author = {Menzel, Jan Paul and Boeije, Yorrick and Bakker, Tijmen M. A. and Belić, Jelena and Reek, Joost N. H. and De Groot, Huub J. M. and Visscher, Lucas and Buda, Francesco},
	month = aug,
	year = {2022},
	pages = {e202200594},
}

@article{kohn_efficient_2023,
	title = {Efficient calculation of electronic coupling integrals with the dimer projection method via a density matrix tight-binding potential},
	volume = {159},
	issn = {0021-9606, 1089-7690},
	url = {https://pubs.aip.org/jcp/article/159/14/144106/2916100/Efficient-calculation-of-electronic-coupling},
	doi = {10.1063/5.0167484},
	language = {en},
	number = {14},
	urldate = {2025-04-02},
	journal = {The Journal of Chemical Physics},
	author = {Kohn, J. T. and Gildemeister, N. and Grimme, S. and Fazzi, D. and Hansen, A.},
	month = oct,
	year = {2023},
	pages = {144106},
}

@article{kohn_quantum_2023,
	title = {Quantum {Chemistry} {Insight} into the {Multifaceted} {Structural} {Properties} of {Two}-{Dimensional} {Covalent} {Organic} {Frameworks}},
	volume = {35},
	copyright = {https://doi.org/10.15223/policy-029},
	issn = {0897-4756, 1520-5002},
	url = {https://pubs.acs.org/doi/10.1021/acs.chemmater.2c03555},
	doi = {10.1021/acs.chemmater.2c03555},
	language = {en},
	number = {7},
	urldate = {2025-04-02},
	journal = {Chem. Mater.},
	author = {Kohn, Julia T. and Li, Hong and Evans, Austin M. and Brédas, Jean-Luc and Grimme, Stefan},
	month = apr,
	year = {2023},
	pages = {2820--2826},
}

@article{kohn_semi-automated_2024,
	title = {A semi-automated quantum-mechanical workflow for the generation of molecular monolayers and aggregates},
	volume = {161},
	issn = {0021-9606, 1089-7690},
	url = {https://pubs.aip.org/jcp/article/161/12/124707/3313995/A-semi-automated-quantum-mechanical-workflow-for},
	doi = {10.1063/5.0230341},
	language = {en},
	number = {12},
	urldate = {2025-04-02},
	journal = {The Journal of Chemical Physics},
	author = {Kohn, J. T. and Grimme, S. and Hansen, A.},
	month = sep,
	year = {2024},
	pages = {124707},
}

@article{kohn_quickstart_2022,
	title = {Quickstart guide to model structures and interactions of artificial molecular muscles with efficient computational methods},
	volume = {58},
	issn = {1359-7345, 1364-548X},
	url = {https://xlink.rsc.org/?DOI=D1CC05759F},
	doi = {10.1039/D1CC05759F},
	language = {en},
	number = {2},
	urldate = {2025-04-02},
	journal = {Chem. Commun.},
	author = {Kohn, Julia and Spicher, Sebastian and Bursch, Markus and Grimme, Stefan},
	year = {2022},
	pages = {258--261},
}

@article{parr_method_1952,
	title = {A {Method} for {Estimating} {Electronic} {Repulsion} {Integrals} {Over} {LCAO} {MO}'{S} in {Complex} {Unsaturated} {Molecules}},
	volume = {20},
	issn = {0021-9606, 1089-7690},
	url = {https://pubs.aip.org/jcp/article/20/9/1499/202458/A-Method-for-Estimating-Electronic-Repulsion},
	doi = {10.1063/1.1700802},
	language = {en},
	number = {9},
	urldate = {2025-04-03},
	journal = {The Journal of Chemical Physics},
	author = {Parr, Robert G.},
	month = sep,
	year = {1952},
	pages = {1499--1499},
}

@article{zhang_comparison_2007,
	title = {Comparison of {DFT} {Methods} for {Molecular} {Orbital} {Eigenvalue} {Calculations}},
	volume = {111},
	issn = {1089-5639, 1520-5215},
	url = {https://pubs.acs.org/doi/10.1021/jp061633o},
	doi = {10.1021/jp061633o},
	language = {en},
	number = {8},
	urldate = {2025-04-20},
	journal = {J. Phys. Chem. A},
	author = {Zhang, Gang and Musgrave, Charles B.},
	month = mar,
	year = {2007},
	pages = {1554--1561},
}

@Article{zhan_ionization_2003,
  author     = {Zhan, Chang-Guo and Nichols, Jeffrey A. and Dixon, David A.},
  title      = {Ionization {Potential}, {Electron} {Affinity}, {Electronegativity}, {Hardness}, and {Electron} {Excitation} {Energy}: {Molecular} {Properties} from {Density} {Functional} {Theory} {Orbital} {Energies}},
  journal    = {J. Phys. Chem. A},
  year       = {2003},
  volume     = {107},
  number     = {20},
  pages      = {4184--4195},
  month      = may,
  issn       = {1089-5639, 1520-5215},
  doi        = {10.1021/jp0225774},
  language   = {en},
  publisher  = {American Chemical Society (ACS)},
  shorttitle = {Ionization {Potential}, {Electron} {Affinity}, {Electronegativity}, {Hardness}, and {Electron} {Excitation} {Energy}},
  url        = {https://pubs.acs.org/doi/10.1021/jp0225774},
  urldate    = {2025-04-21},
}

@article{allen_eigenvalues_2002,
	title = {Eigenvalues, integer discontinuities and {NMR} shielding constants in {Kohn}—{Sham} theory},
	volume = {100},
	issn = {0026-8976, 1362-3028},
	url = {http://www.tandfonline.com/doi/abs/10.1080/00268970110078335},
	doi = {10.1080/00268970110078335},
	language = {en},
	number = {4},
	urldate = {2025-04-21},
	journal = {Molecular Physics},
	author = {Allen, Mark J. and Tozer, David J.},
	month = feb,
	year = {2002},
	pages = {433--439},
}

@Article{oboyle_open_2011,
  author     = {O'Boyle, Noel M and Banck, Michael and James, Craig A and Morley, Chris and Vandermeersch, Tim and Hutchison, Geoffrey R},
  title      = {Open {Babel}: {An} open chemical toolbox},
  journal    = {J Cheminform},
  year       = {2011},
  volume     = {3},
  number     = {1},
  pages      = {33},
  month      = dec,
  issn       = {1758-2946},
  doi        = {10.1186/1758-2946-3-33},
  language   = {en},
  shorttitle = {Open {Babel}},
  url        = {https://jcheminf.biomedcentral.com/articles/10.1186/1758-2946-3-33},
  urldate    = {2024-12-10},
}

@Article{pedregosa2011scikit,
  author  = {Pedregosa, Fabian and Varoquaux, Ga{\"e}l and Gramfort, Alexandre and Michel, Vincent and Thirion, Bertrand and Grisel, Olivier and Blondel, Mathieu and Prettenhofer, Peter and Weiss, Ron and Dubourg, Vincent and others},
  journal = {Journal of machine learning research},
  title   = {Scikit-learn: Machine learning in Python},
  year    = {2011},
  number  = {Oct},
  pages   = {2825--2830},
  volume  = {12},
}

@Article{Perdew1996,
  author    = {Perdew, John P. and Burke, Kieron and Ernzerhof, Matthias},
  journal   = {Physical Review Letters},
  title     = {Generalized Gradient Approximation Made Simple},
  year      = {1996},
  issn      = {1079-7114},
  month     = oct,
  number    = {18},
  pages     = {3865--3868},
  volume    = {77},
  doi       = {10.1103/physrevlett.77.3865},
  publisher = {American Physical Society (APS)},
}

@Article{weigend_balanced_2005,
  author    = {Florian Weigend and Reinhart Ahlrichs},
  journal   = {Physical Chemistry Chemical Physics},
  title     = {Balanced basis sets of split valence, triple zeta valence and quadruple zeta valence quality for H to Rn: Design and assessment of accuracy},
  year      = {2005},
  issn      = {1463-9076},
  number    = {18},
  pages     = {3297-3305},
  volume    = {7},
  doi       = {10.1039/b508541a},
  issue     = {16},
  publisher = {Royal Society of Chemistry (RSC)},
}

@Article{schuchardt_basis_2007,
  author    = {Karen L. Schuchardt and Brett T. Didier and Todd Elsethagen and Lisong Sun and Vidhya Gurumoorthi and Jared Chase and Jun Li and Theresa L. Windus},
  journal   = {Journal of Chemical Information and Modeling},
  title     = {Basis Set Exchange: A Community Database for Computational Sciences},
  year      = {2007},
  issn      = {1549-9596},
  number    = {3},
  pages     = {1045-1052},
  volume    = {47},
  doi       = {10.1021/ci600510j},
  issue     = {3},
  publisher = {American Chemical Society (ACS)},
}

@Article{pracht_crest_2024,
  author    = {Philipp Pracht and Stefan Grimme and Christoph Bannwarth and Fabian Bohle and Sebastian Ehlert and Gereon Feldmann and Johannes Gorges and Marcel Müller and Tim Neudecker and Christoph Plett and Sebastian Spicher and Pit Steinbach and Patryk A. Wesołowski and Felix Zeller},
  journal   = {The Journal of Chemical Physics},
  title     = {CREST—A program for the exploration of low-energy molecular chemical space},
  year      = {2024},
  issn      = {0021-9606},
  number    = {11},
  pages     = {114110},
  volume    = {160},
  doi       = {10.1063/5.0197592},
  issue     = {11},
  publisher = {AIP Publishing},
  url       = {https://pubs.aip.org/jcp/article/160/11/114110/3278084/CREST-A-program-for-the-exploration-of-low-energy},
}
%---------------------------------------------------------------------------------------
\newpage
\setcounter{table}{0}
\setcounter{figure}{0}
\newsecnumstyle

\raggedbottom

\section*{\centering Supplementary Information}

This Supplementary Information provides additional details on the computational methods and presents supplementary results visualizations supporting the findings reported in the main text.

\subsection{Methods}\label{sec:Suppl_Methods}

\subsubsection{Molecular featurization}\label{sec:SI_Featurization}

The molecular featurization framework integrates the calculation of both global molecular descriptors and local atom and bond features. A set of \num{22} global descriptors, encompassing various levels of complexity (including constitutional, connectivity, and topological descriptors), was computed for each molecule in the databases using \texttt{OpenBabel} \cite{oboyle_open_2011}. These descriptors, detailed in \Cref{tab:SI_S1_GlobFeat}, were standardized using a $Z$-score transformation, which scales each feature to have a mean of zero and a standard deviation of one.

Next, molecular graphs were constructed as sets of atomic and bond matrices using local atomic and bond features, as summarized in \Cref{tab:SI_S2_LocFeat}. To handle categorical data representation and avoid potential model bias due to differences in feature scales, one-hot encoding was applied to both atomic and bond features. The exception was atomic mass features, which were scaled by dividing by \num{100}. All other atomic features were one-hot encoded.

After extracting global and local molecular features, these features were aggregated and embedded into a unified feature vector for subsequent tasks. To achieve this, a principal component analysis (PCA)-based approach was employed to reduce the dimensionality of the data and combine local molecular features. PCA is a mathematical algorithm that reduces the number of dimensions by identifying uncorrelated principal component variables, which are linear combinations of the original variables, while retaining the majority of the variance in the dataset \cite{ringner_what_2008, greenacre_principal_2022}. The atomic and bond transposed matrices were aggregated into a one-dimensional subspace using PCA, resulting in linear atomic and bond feature vectors that capture the greatest variance of the original feature matrices. Finally, the global and local feature vectors were concatenated, yielding a single feature vector of length \num{226} for small $\pi$-systems of the QM9 set, and \num{210} for extended $\pi$-systems of the CEP set.

\begin{center}
	\begin{longtable}[H]{>{\small}m{3.25cm}>{\small}m{5cm}>{\small}p{6cm}}
		\caption{Descriptions of global molecular features\label{tab:SI_S1_GlobFeat}}\\
		\toprule
		\textbf{Level of complexity} & \textbf{Description} & \textbf{Attributes} \\
		\midrule
		\endfirsthead
		\multicolumn{3}{@{}l}{\textit{Continued from previous page}}\\
		\toprule
		\textbf{Level of complexity} & \textbf{Description} & \textbf{Attributes} \\
		\midrule
		\endhead
		%\bottomrule
		\endfoot \\ \\ \\
		\bottomrule
		\endlastfoot
		\multirow{7}{3cm}{0-Dimensional} & \multirow{7}{5cm}{These descriptors are based on the chemical formula, specifying chemical species and their occurrence, and capturing bulk properties of molecules} & Number of atoms \\
		& & Number of heavy atoms \\
		& & Number of fluorine atoms \\
		& & Molecular weight \\
		& & Exact mass \\
		& & Total charge \\
		& & Total spin multiplicity \\
		& & Energy \\
		& & Octanol/water partition coefficient \\
		& & Melting point \\
		& & Molar refractivity \\
		\cmidrule(r){2-3}
		
		\multirow{9}{3cm}{2-Dimensional} & \multirow{9}{5cm}{This representation addresses the connectivity of atoms in terms of both the presence and nature of chemical bonds, as well as topological properties through topostructural and topochemical indices} & Number of bonds \\
		& & Number of aromatic bonds \\
		& & Number of single bonds \\
		& & Number of double bonds \\
		& & Number of triple bonds \\
		& & Number of hydrogen bond acceptors 1 \\
		& & Number of hydrogen bond acceptors 2 \\
		& & Number of hydrogen bond donors \\
		& & Topological polar surface area \\
		\cmidrule(r){2-3}
		
		\multirow{6}{3cm}{3-Dimensional} & \multirow{6}{5cm}{This level of complexity enables the perception of molecules as geometric objects in space, characterized by the spatial arrangement of atoms ($xyz$ Cartesian coordinates)} & Number of rotors/rotatable bonds \\
		& & Periodicity \\
		
	\end{longtable}
\end{center}

%\newpage % Use judiciously
\begin{center}
	\begin{longtable}[H]{m{2cm} m{2.6cm} m{1.1cm} m{1.1cm} m{6.5cm}}
		\caption{Descriptions of local atomic and bond features\label{tab:SI_S2_LocFeat}}\\
		\toprule
		\textbf{Feature type} & \textbf{Attributes} & \textbf{QM9 size} & \textbf{CEP size} & \textbf{Description} \\
		\midrule\endfirsthead
		\multicolumn{5}{@{}l}{\textit{Continued from previous page}}\\
		\toprule
		\textbf{Feature type} & \textbf{Attributes} & \textbf{QM9 size} & \textbf{CEP size} & \textbf{Description} \\
		\midrule\endhead
		\bottomrule
		\endfoot
		\bottomrule
		\endlastfoot
		\multirow{31}{=}{Atomic features} & Atom type & $118$ & $118$ & Type of atom (ex. C,N,O), by
		atomic number \\
		
		& Formal charge & $3$ & $1$ & Electronic charge assigned to atom \\
		
		& Total degree & $4$ & $4$ & Number of bonds the atom is involved in \\
		
		& Total valence & $5$ & $4$ &Number of bonds the atom can be involved in \\
		
		& Hybridization & $3$ & $3$ & Atom hybridization (sp, sp2, sp3) \\
		
		& Explicit Hydrogen & $5$ & $4$ & Number of bonded hydrogen atoms \\
		
		& Heavy degree & $5$ & $3$ & Number of non-hydrogen atoms attached \\
		
		& Hetero degree & $4$ & $3$ & Number of heteroatoms attached \\
		
		& Number of rings & $5$ & $3$ & Number of rings the atom is a member of \\
		
		& Ring size & $8$ & $3$ & size (number of atoms) of the ring(s) the atom is a part of \\
		
		& Number of ring bonds & $5$ & $3$ & Number of bonds the atom has that are part of a ring
		structure \\
		
		& Atomic mass & $1$ & $1$ & Mass of the atom \\
		
		& Exact mass & $1$ & $1$ & Exact mass of the atom \\
		
		& Partial charge & $1$ & $1$ & Partial charge of the atom \\
		
		& Aromaticity & $1$ & $1$ & If the atom is aromatic \\
		
		& Is in ring & $1$ & $1$ & If the atom is in ring \\
		
		& Heteroatom & $1$ & $1$ & If the atom is an heteroatom \\
		
		& Chirality & $1$ & $1$ & If the atom is chiral \\
		
		& Periodicity & $1$ & $1$ & If the atom is periodic \\
		
		& Has non-single bond & $1$ & $1$ & If the atom has non-single bond \\
		
		& Has single bond & $1$ & $1$ & If the atom has single bond \\
		
		& Has double bond & $1$ & $1$ & If the atom has double bond \\
		
		& Has aromatic bond & $1$ & $1$ & If the atom has aromatic bond \\
		
		& Axial & $1$ & $1$ & If the atom is axial \\
		
		& Carboxyl oxygen & $1$ & $1$ & If the atom is carboxyl oxygen \\
		
		& Sulfate oxygen & $1$ & $1$ & If the atom is sulfate oxygen \\
		
		& Nitro oxygen & $1$ & $1$ & If the atom is nitro oxygen \\
		
		& Amide nitrogen & $1$ & $1$ & If the atom is amide nitrogen \\
		
		& Polar hydrogen & $1$ & $1$ & If the atom is polar hydrogen \\
		
		& Non-polar hydrogen & $1$ & $1$ & If the atom is non-polar hydrogen \\
		
		& Aromatic Noxide & $1$ & $1$ & If the atom is aromatic Noxide \\
		
		\midrule
		\multirow{17}{=}{Bond features} & Bond order & $3$ & $3$ & If the bond is simple, double or
		triple \\
		
		& Aromaticity & $1$ & $1$ & If the bond is aromatic \\
		
		& Ring & $1$ & $1$ & If the bond is part of a ring \\
		
		& Rotor & $1$ & $1$ & If the bond is a rotor \\
		
		& Periodicity & $1$ & $1$ & If the bond is periodic \\
		
		& Amide & $1$ & $1$ & If the bond is amid \\
		
		& Primary amide & $1$ & $1$ & If the bond is primary amid \\
		
		& Secondary amide & $1$ & $1$ & If the bond is secondary amid \\
		
		& Tertiary amide & $1$ & $1$ & If the bond is tertiary amid \\
		
		& Ester & $1$ & $1$ & If the bond is ester \\
		
		& Carbonyl & $1$ & $1$ & If the bond is carbonyl \\
		
		& Closure & $1$ & $1$ & If the bond is a closure \\
		
		& Wedge & $1$ & $1$ & If the bond is wedge \\
		
		& Hash & $1$ & $1$ & If the bond is hash \\
		
		& Wedge or hash & $1$ & $1$ & If the bond is wedge or hash \\
		
		& Cis or trans & $1$ & $1$ & If the bond is cis or trans \\
		
		& Double bond geometry & $1$ & $1$ & If the bond is a double bond geometry \\
		
	\end{longtable}
\end{center}

\subsubsection{Molecular clustering}\label{sec:SI_Clustering}

Clustering analysis of small organic molecules was performed using the \emph{k} -means algorithm \cite{macqueen_methods_1967}. \emph{K}-means is a vector quantization method that groups $N$ observations $X = \{x_1, x_2, \dots, x_N\} \subseteq \mathfrak{R}^F$, represented by $F$-dimensional real vectors, into $K$ (with $K \leq N$) subsets $S = \{s_1, s_2, \dots, s_K\}$, where $\bigcup_{s_k \in S} s_k = X$ and $s_k \cap s_l = \emptyset$ for all $k \neq l$. This grouping is based on modeling the probability mass function $p$ using the distribution of prototype vectors, or "centroids." The centroid of a cluster $s_k$ is defined as $\bar{s_k} = \frac{1}{|s_k|} \sum_{x_i \in s_k} x_i$, and the centroid of the entire dataset is $\bar{X} = \frac{1}{N} \sum_{x_i \in X} x_i$. The algorithm iteratively minimizes the loss, proportional to the squared error $|z - u_i|^2$, by alternating between assigning molecules (denoted here as $z, z \in X$) to clusters based on the current centroids and updating the centroids (denoted here as $u_i, i = \{1, 2, ..., k\}$) based on the current clusters \cite{hadipour_deep_2022}. Formally, the goal is to find a partition $S$ that minimizes the intracluster variance through the function:
\begin{equation}
	w^2(S) = \sum_{i=1}^{k} \int_{S_i} |z - u_i|^2 dp(z) = \sum_{i=1}^{k} |S_i| Var(S_i),
\end{equation}
where $Var(S_i) = \frac{1}{|S_i|} \int_{S_i} |z - u_i|^2 dp(z)$ is the variance of the $i^{th}$ cluster, and $dp(z)$ is the probability mass at the point $z$.

For an efficient implementation of the \emph{k}-means algorithm, the initial cluster centroids were selected using a sampling method based on an empirical probability distribution of the contribution of the data points to the overall cluster inertia. This iterative approach streamlines the convergence process as it minimizes the intra-cluster sum of squared errors (SSE) by performing several trials at each sampling step and choosing the best centroid among them. Moreover, the optimal number of partitions was determined using the Silhouette validation index \cite{rousseeuw_silhouettes_1987}. The Silhouette score is a clustering validation metric based on the pairwise differences of inter- and intra-cluster distances \cite{liu_understanding_2010}, defined as:
\begin{equation}
	Sil(S) = \frac{1}{N} \sum_{s_k \in S} \sum_{x_i \in s_k} \frac{b(x_i, s_k) - a(x_i, s_k)}{\max\{a(x_i, s_k), b(x_i, s_k)\}},
\end{equation}
where $a(x_i, s_k) = \frac{1}{|s_k|} \sum_{x_j \in s_k} d_e (x_i, x_j)$ and $b(x_i, s_k) = \min_{s_l \in S, s_l \neq s_k} \{\frac{1}{|s_l|} \sum_{x_j \in s_l} d_e (x_i, x_j)\}$ represent the intra-cluster distance (measuring cohesion) and the inter-cluster distance (measuring separation), respectively. Here, $d_e(x_i, x_j)$ is the Euclidean distance between objects $x_i$ and $x_j$. A higher average Silhouette index indicates better clustering, with well-defined and well-separated clusters.

Thus, the algorithm was fitted to the data for different predefined numbers of partitions and the Silhouette score was calculated for each model. The optimal number of clusters was chosen as the minimum value where the Silhouette scores stabilized \cite{hadipour_deep_2022}.

\subsubsection{Quantum chemistry theories}\label{sec:SI_QC_Theories}

This section provides a brief overview of the semiempirical tight-binding methods and force-field approach implemented in the \xtb program and used for geometry optimizations and electronic structure calculations, along with a note on the DFT method employed. The general workflow for these calculations is illustrated in Figures 3 and 4 of the main text.

\paragraph{\textbf{Extended tight-binding theories} \\}\label{sec:SI_xTB}

Semiempirical geometry optimization processes were conducted using the methodologies \gfn (\gfnb, \gfnc, \gfna) and \gfnf. The \xtb methods are based on DFTB theory, which simplifies the Kohn-Sham equations \cite{kohn_self-consistent_1965} of DFT by approximating the total energy $E[\rho]$ using a Taylor expansion around a reference electron density $\rho_0$:
\begin{equation}
	E[\rho] = E^{(0)}[\rho_0] + E^{(1)}[\rho_0, \delta\rho] + E^{(2)}[\rho_0, (\delta\rho)^2] + E^{(3)}[\rho_0, (\delta\rho)^3] + \cdots.
\end{equation}
A frozen-core approximation is applied to consider only the fluctuations of the valence orbitals. The most accurate \xtb variants truncate this Taylor expansion after the third-order term. These approximations significantly reduce computational costs while maintaining reliable accuracy. Below, we briefly discuss the different \gfn and \gfnf methods in chronological order, highlighting the energy terms associated with their respective truncation orders.

\begin{itemize}
	\item \textbf{The \gfnb Hamiltonian}
	
	The initial \xtb methodology, known as \gfnb, is rooted in the Self-Consistent Charge Density Functional Tight-Binding (SCC-DFTB3) scheme, which includes second- and third-order electrostatic interaction terms in the Taylor series expansion of the Hamiltonian. The total energy in \gfnb is calculated as:
	\begin{equation}
		E_{\gfnb} = E_{rep} + E_{disp}^{D3} + E_{XB}^{GFN1} + E_{EHT} + E_{\gamma} + E_{\Gamma}^{GFN1},
	\end{equation}
	where $E_{rep}$ is the classical pairwise repulsion energy, $E_{EHT}$ is the extended Hückel-type energy for covalent bond formation, and $E_{\gamma}$ is the second-order isotropic electrostatic (IES) and exchange-correlation (XC) energy. \gfnb iteratively refines atomic charges based on a monopole description of electrostatic interactions, followed by a third-order on-site correction $E_{\Gamma}^{GFN1}$ to stabilize high partial charges. A Becke–Johnson (BJ)-damped D3 dispersion model ($E_{disp}^{D3}$) and a geometry-dependent halogen-bond (XB) correction ($E_{XB}^{GFN1}$) are included to improve the description of weak halogen bonds. This method avoids element-pair-specific parameters, instead using element-specific parameters typical of HF-based zero-differential overlap \cite{parr_method_1952} (ZDO) approximation. The molecular orbitals are expressed as linear combinations of atom-centered orbitals (LCAO), with partially polarized minimal basis sets consisting of approximate Slater-type orbitals (STO) augmented with a double-$\zeta$ basis for hydrogen.
	
	\item \textbf{The \gfnc Hamiltonian}
	
	The second \xtb version, \gfnc, introduced in 2019, enhances the SCC formulation of \gfnb. The total energy in \gfnc is expressed as:
	\begin{equation}
		E_{\gfnc} = E_{rep} + E_{EHT} + E_{\gamma} + E_{AES} + E_{AXC} + E_{disp}^{D4} + E_{\Gamma}^{GFN2},
	\end{equation}
	where $E_{AES}$ and $E_{AXC}$ represent anisotropic electrostatic and exchange-correlation energies, respectively. \gfnc incorporates a D4-ATM (Axilrod–Teller–Muto) dispersion model ($E_{disp}^{D4}$) and a multipole treatment of electrostatic interactions, improving the accuracy of electron density and energetic properties. The method also employs a shell-resolved approach for stabilizing high partial charges ($E_{\Gamma}^{GFN2}$). The multipole electrostatic treatment with anisotropic exchange-correlation (AXC) energy contribution, when added to the recent charge-dependent (D4) dispersion model as well as the shell-specific third order term involved in the SCC procedure, eliminates the necessity for additional corrections for hydrogen or halogen bonds.
	
	\item \textbf{The \gfna Hamiltonian}
	
	The \gfna method, introduced as a successor to \gfnb and \gfnc, significantly reduces computational costs by a factor of 2–20, depending on system size and complexity. The total energy in \gfna is given by:
	\begin{equation}
		E_{\gfna} = E_{rep} + E_{disp}^{D4} + E_{srb} + E_{EEQ} + E_{EHT},
	\end{equation}
	where $E_{srb}$ is a short-range bond correction, and $E_{EEQ}$ is the electronegativity equilibration (EEQ) atomic charge model. A simplified D4 model is implemented for the description of the dispersion interactions. \gfna is a non-iterative method that includes only the first-order tight-binding energy expansion terms, and avoids convergence issues associated with self-consistent calculations. However, its accuracy is slightly reduced due to the omission of higher-order energy terms.
	
	\item \textbf{The \gfnf Hamiltonian}
	
	The \gfnf method which is an atomistic force field approach, extends \gfna by replacing the extended Hückel-type theory with classical bond, angle, and torsion terms. The total energy in \gfnf is expressed as:
	\begin{equation}
		E_{\gfnf} = E_{cov} + E_{NCI},
	\end{equation}
	where $E_{cov}$ represents covalent interactions (bond stretch, angle, and torsion terms), and $E_{NCI}$ represents non-covalent interactions (electrostatics, dispersion, hydrogen bonds, and halogen bonds). Electrostatic and dispersion interactions are described by the semiclassical EEQ model and a simplified D4 dispersion scheme, respectively, as in \gfna. \gfnf includes charge-dependent correction terms for hydrogen and halogen bonds. Moreover, both covalent and non-covalent interactions incorporate their respective repulsive energies.
\end{itemize}

For a complete derivation of the energy terms, the readers are referred to the original publications by S. Grimme \textit{et al.} \cite{grimme_robust_2017, bannwarth_gfn2-xtbaccurate_2019, pracht_robust_2019, spicher_robust_2020}.

\paragraph{\textbf{Details on DFT calculations} \\}\label{sec:SI_DFT_Details}

Detailed information on the DFT calculations, including the specific functional, basis set, and convergence criteria employed for the CEP reference set, can be found in Section 2.3.2 of the main text. The DFT reference data for the QM9 subset was obtained from Ramakrishnan \textit{et al.} \cite{ramakrishnan_quantum_2014}.

\subsubsection{Theory implemented in the CREST program}\label{sec:SI_CREST}

The exploration of low-energy molecular chemical space was performed using the \crest program. Conformational sampling in \crest follows the iterative metadynamics-genetic crossing (iMTD-GC) workflow, which comprises five main steps. The process begins with system initialization, where the elemental composition, Cartesian coordinates, molecular and nuclear charges, and spin multiplicity are specified. The program then automatically initializes simulations by measuring system size and flexibility, which are used to predict global metadynamics (MTD) execution times and bias potential strengths. An initial geometry optimization is performed to verify topology consistency, and the program halts if any changes in topology are detected.

The input structure is then subjected to MTD simulations, which accelerate conformational sampling by applying bias potentials to escape local energy minima. The bias potential is constructed as a sum of Gaussian hills based on the Cartesian RMSD as the collective variable. Multiple MTD calculations are executed in parallel, each with different bias potential strengths and widths, to explore a wide range of conformations. The lowest-energy conformations are further refined using molecular dynamics (MD) simulations to capture rotamers connected by low-energy barriers.

The generated conformers and rotamers are combined using a genetic crossing (GC) algorithm, which exchanges or blends structural fragments from pairs of conformations to generate new, potentially more stable, or diverse structures. The conformational geometries are optimized and sorted at each step to identify unique conformers and avoid duplicates. If a lower-energy conformer is found, the process is restarted using this new lead conformation. The \crest program employs default optimization (ANCOPT) and ensemble sorting (CREGEN) algorithms to ensure efficient and accurate conformational sampling. For further details, the reader is referred to the original publication by Pracht \emph{et al.} \cite{pracht_crest_2024}.

\subsection{Results}\label{sec:Suppl_Results}

\subsubsection{Molecular sampling visualizations}\label{sec:SI_Sampling_Viz}

This section presents supplementary figures visualizing the molecular sampling process discussed in Section 3.1 of the main text.

\begin{figure}[htbp]
	\centering
	\includegraphics[width=.9\textwidth]{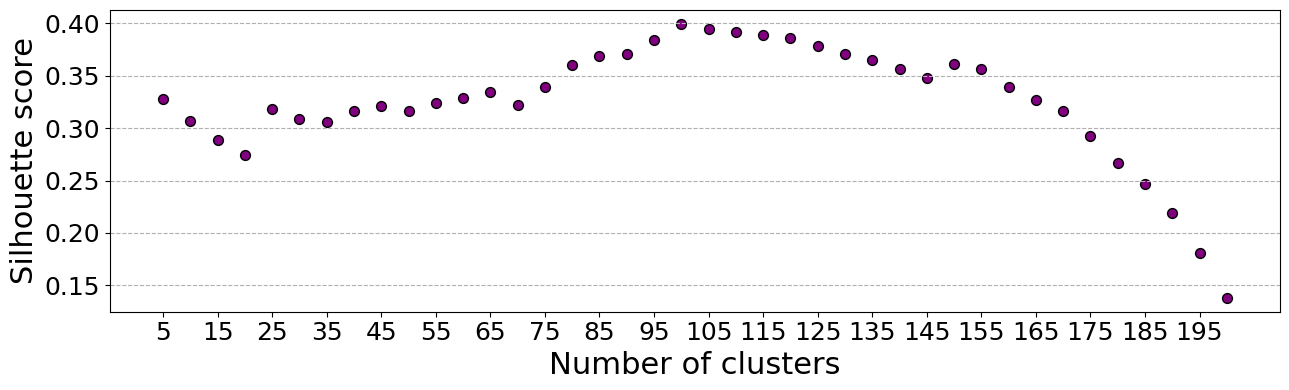}
	\caption{The Silhouette scores for different numbers of molecular clusters, ranging from 5 to 200 with a step size of 5, and using the integration of global molecular descriptors and local bond and angle features. The results are based on embedded feature vectors of length 226 used for training the k-means algorithm. A maximum Silhouette score of \num{0.278} was obtained for $k = 25$, which marks the onset of score stabilization and was thus selected as the optimal number of clusters.}
	\label{fig:SI_S1_Silhouette}
\end{figure}

\begin{figure}[htbp]
	\begin{center}
		\subfloat[\centering QM9 clusters]{\includegraphics[width=.9\textwidth]{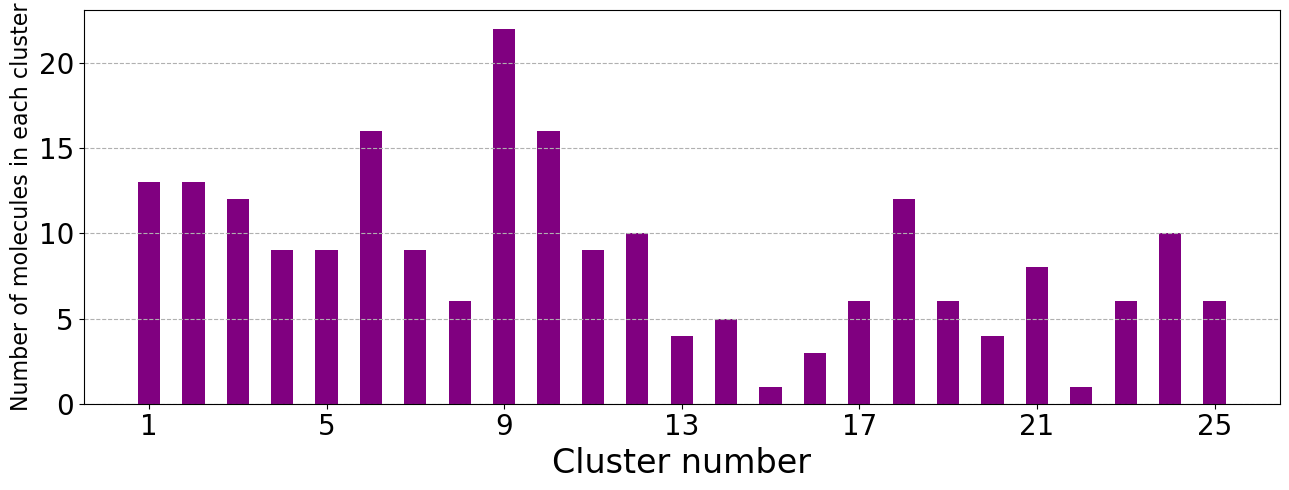}} \\
		\subfloat[\centering CEP strata]{\includegraphics[width=.9\textwidth]{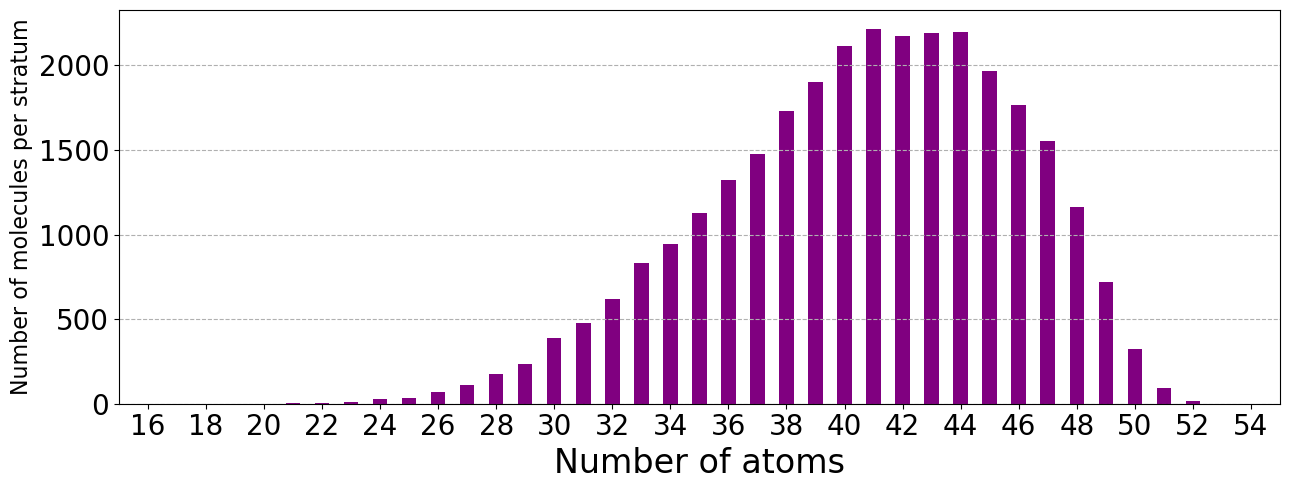}}
		\caption{The distribution of molecules in each (a) cluster for the QM9 subset, and (b) stratum for the CEP subset.}
		\label{fig:SI_S2_Density}
	\end{center}
\end{figure}

\begin{figure}[htbp]
	\begin{center}
		\includegraphics[scale=0.5]{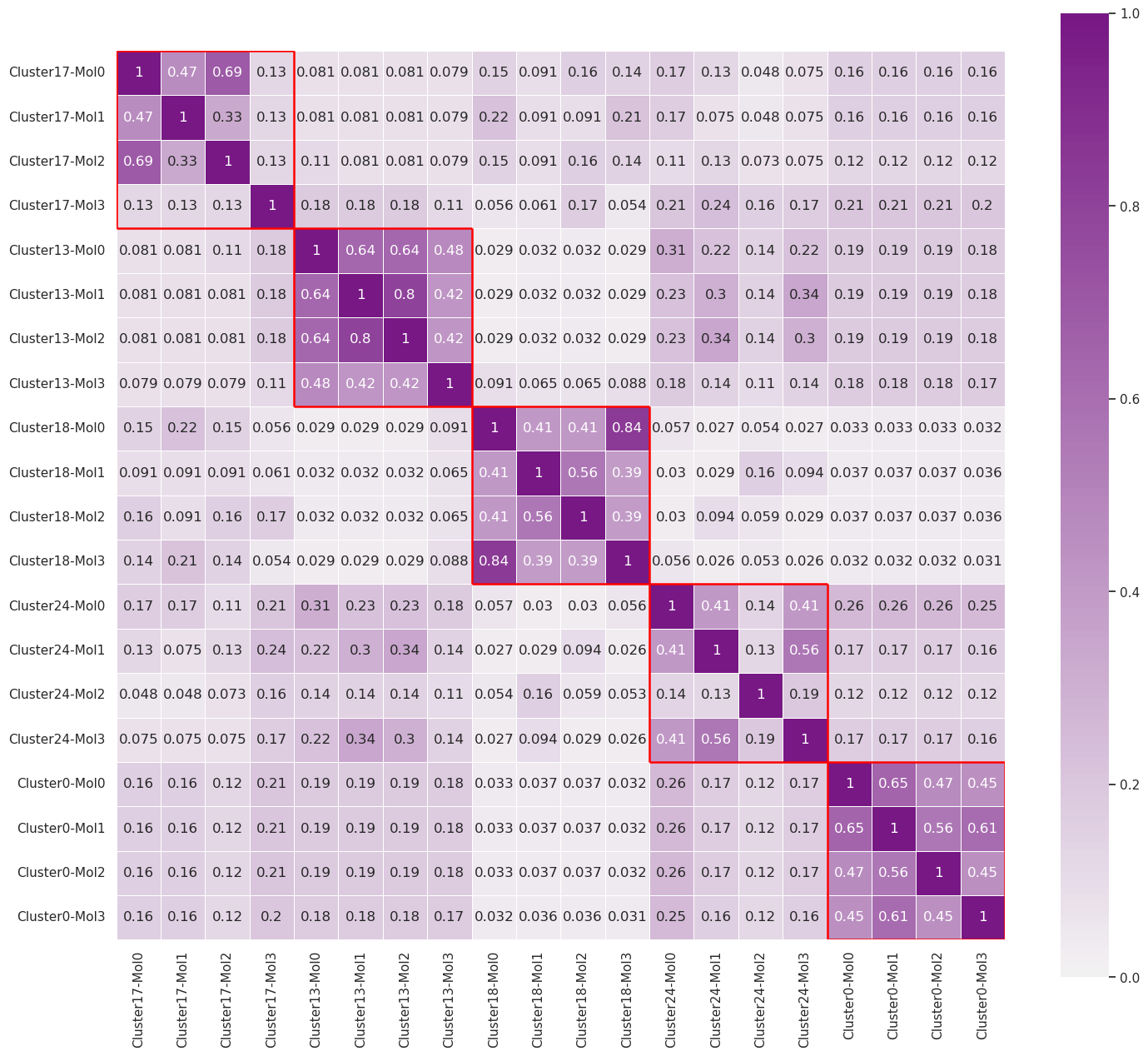}
		\caption{The Tanimoto similarity matrix between four randomly selected molecules from five random clusters comprising the reference compounds or molecular centroids (Mol0) and three test compounds (Mol1, Mol2, Mol3). ECFPs with a radius of 1 and 2048 bits were used as molecular representations for calculating similarity scores. Red rectangles along the diagonal of the similarity matrix indicate molecules that belong to the same cluster.
			\label{fig:SI_S3_Heatmap}
		}
	\end{center}
\end{figure}

\begin{center}
	{\setlength{\arrayrulewidth}{1pt}
		\begin{longtable}[H]{p{0.5cm} p{2.9cm} p{2.9cm} p{2.9cm} p{2.9cm}}
			\caption{Two-dimensional similarity maps for selected QM9 clusters, illustrating structural commonalities within groups (see Section 3.1 in main text). For each cluster, the similarity map between the reference compound and three test compounds was generated using the count-based ECFP, with a radius of 2 and 2048 bits. Molecular fragments with higher weights, increasing the similarity score, are highlighted in green, while the red color denotes the dissimilarities in the similarity maps.\label{tab:SI_S3_QM9_SimMap}}\\
			\toprule
			\centered{\textbf{Cluster}} & \centered{\textbf{Ref. Molecule}} &
			\multicolumn{3}{c}{\textbf{Test Molecules}} \\
			\midrule \endfirsthead\\\midrule
			\centered{\textbf{Cluster}} & \centered{\textbf{Ref. Molecule}} &
			\multicolumn{3}{c}{\textbf{Test Molecules}} \\
			\midrule \endhead
			\bottomrule
			\endfoot
			\bottomrule
			\endlastfoot
			\multicolumn{5}{c}{\textbf{QM9}} \\\midrule
			\centered{\textbf{17}} &
			\centered{\includegraphics[scale=0.2]{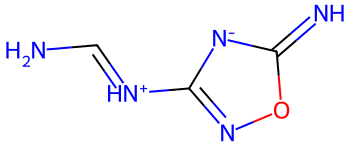}} &
			\centered{\includegraphics[scale=0.12]{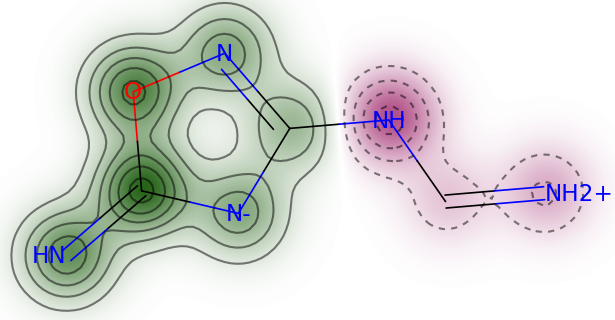}} &
			\centered{\includegraphics[scale=0.12]{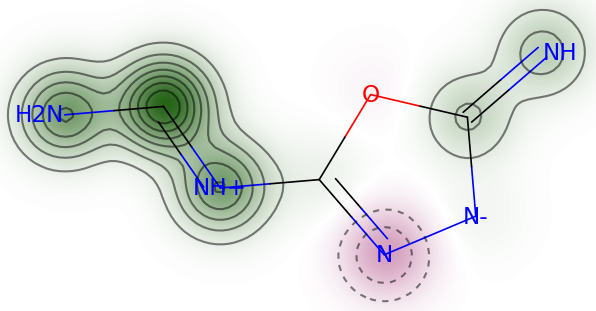}} &
			\centered{\includegraphics[scale=0.12]{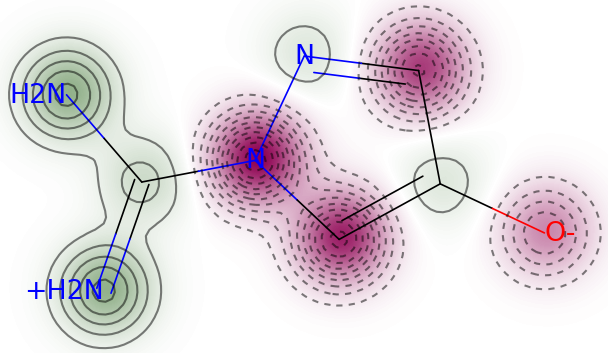}}\\
			\hdashline \\
			
			\centered{\textbf{13}} &
			\centered{\includegraphics[scale=0.2]{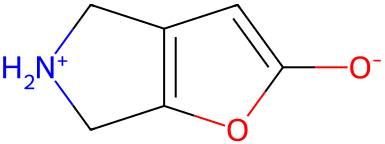}} &
			\centered{\includegraphics[scale=0.12]{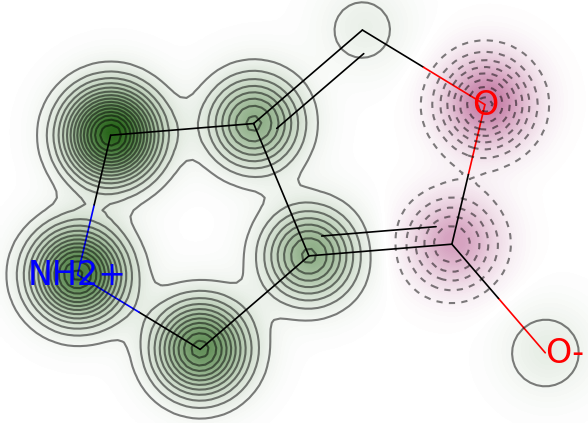}} &
			\centered{\includegraphics[scale=0.12]{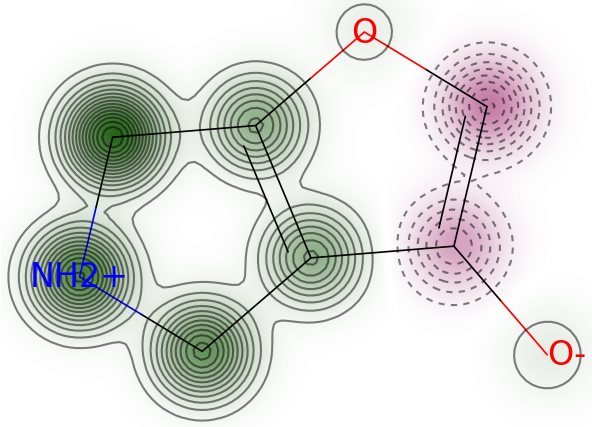}} &
			\centered{\includegraphics[scale=0.12]{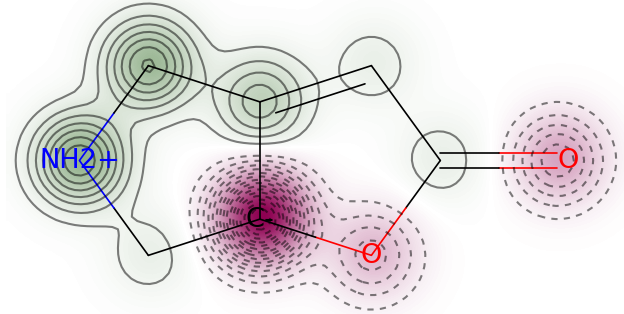}}\\
			\hdashline \\
			
			\centered{\textbf{18}} & \centered{\includegraphics[scale=0.2]{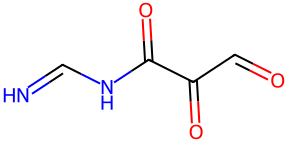}} &
			\centered{\includegraphics[scale=0.12]{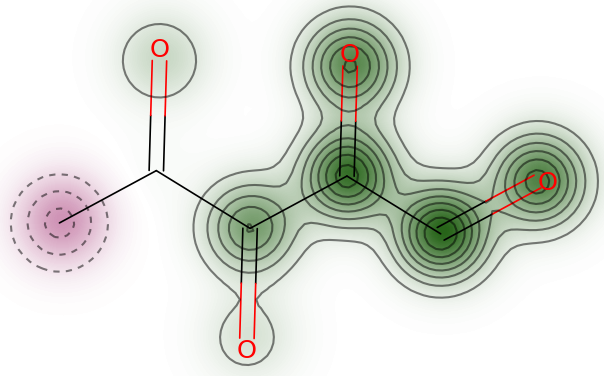}} &
			\centered{\includegraphics[scale=0.12]{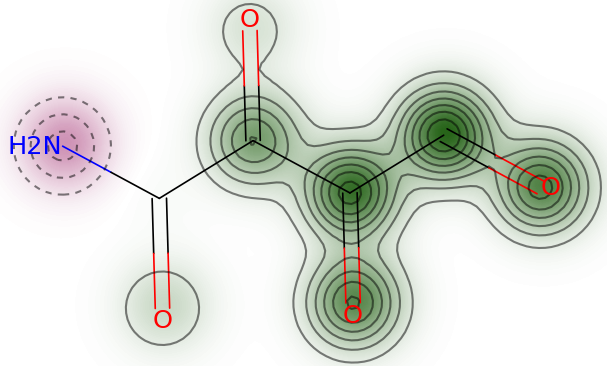}} &
			\centered{\includegraphics[scale=0.12]{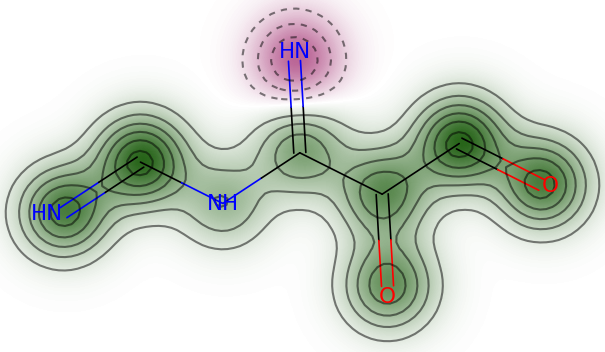}}\\
			\hdashline \\
			
			\centered{\textbf{24}} & \centered{\includegraphics[scale=0.2]{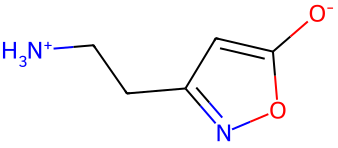}} &
			\centered{\includegraphics[scale=0.12]{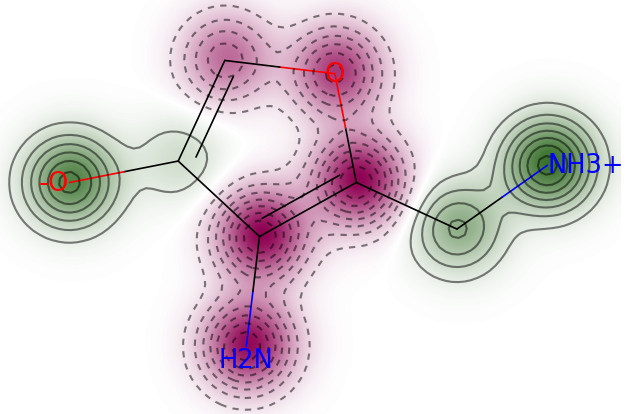}} &
			\centered{\includegraphics[scale=0.12]{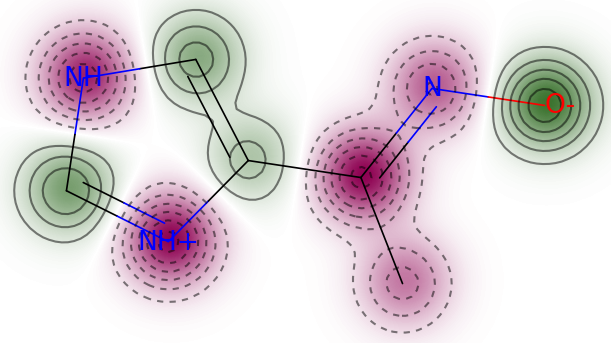}} &
			\centered{\includegraphics[scale=0.12]{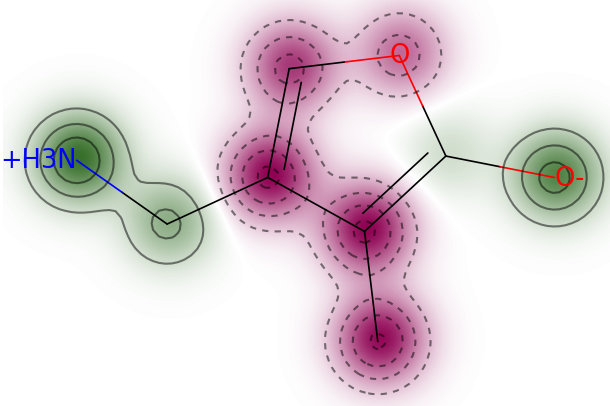}}\\
			\hdashline \\
			
			\centered{\textbf{0}} & \centered{\includegraphics[scale=0.2]{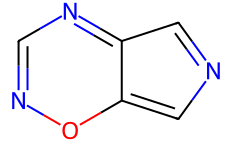}} &
			\centered{\includegraphics[scale=0.12]{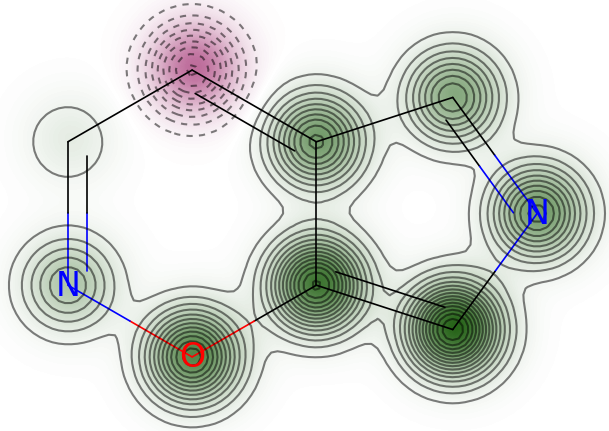}} &
			\centered{\includegraphics[scale=0.12]{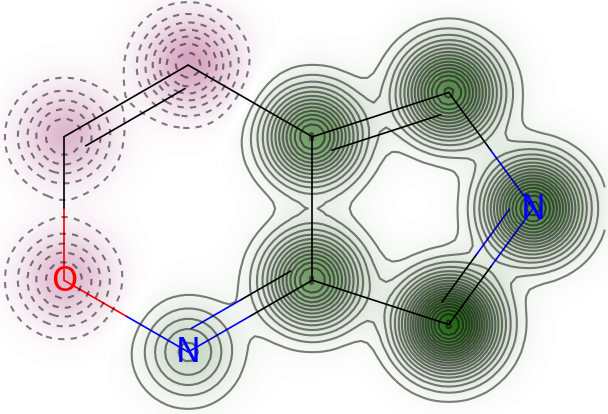}} &
			\centered{\includegraphics[scale=0.12]{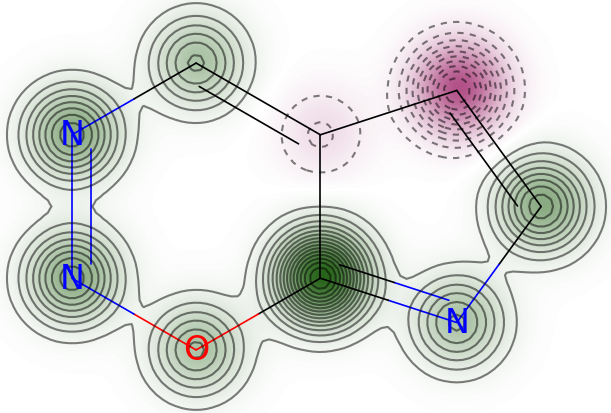}} \\
		\end{longtable}%}
}
\end{center}

\begin{center}
{\setlength{\arrayrulewidth}{1pt}
	\begin{longtable}[H]{p{0.5cm} p{2.9cm} p{2.9cm} p{2.9cm} p{2.9cm}}
		\caption{Two-dimensional similarity maps for selected CEP strata, illustrating structural commonalities within groups (see Section 3.1 in main text). Formatting is as described for Table S3.\label{tab:SI_S4_CEP_SimMap}}\\
		\toprule
		\centered{\textbf{Stratum}} & \centered{\textbf{Ref. Molecule}} &
		\multicolumn{3}{c}{\textbf{Test Molecules}} \\
		\midrule \endfirsthead\\\midrule
		\centered{\textbf{Stratum}} & \centered{\textbf{Ref. Molecule}} &
		\multicolumn{3}{c}{\textbf{Test Molecules}} \\
		\midrule \endhead
		\bottomrule
		\endfoot
		\bottomrule
		\endlastfoot
		\multicolumn{5}{c}{\textbf{CEP}} \\\midrule
		\centered{\textbf{32}} &
		\centered{\includegraphics[scale=0.19]{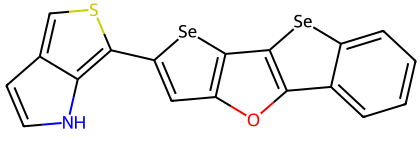}} &
		\centered{\includegraphics[scale=0.16]{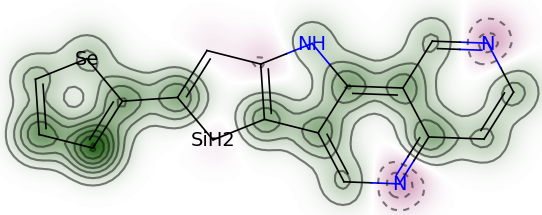}} &
		\centered{\includegraphics[scale=0.16]{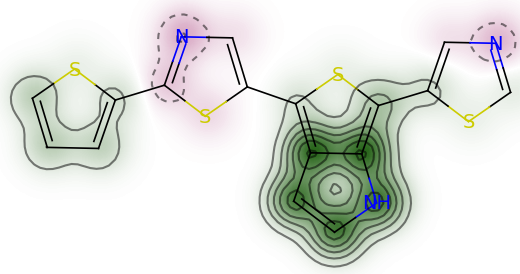}} &
		\centered{\includegraphics[scale=0.16]{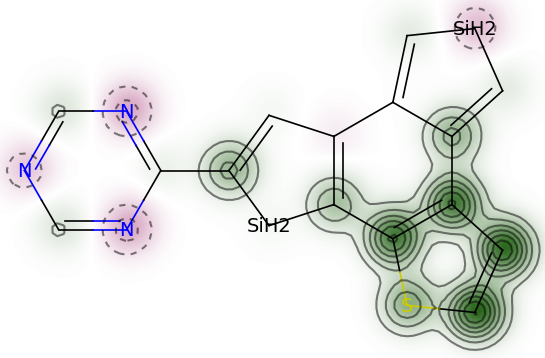}} \\
		\hdashline \\
		
		\centered{\textbf{44}} & \centered{\includegraphics[scale=0.3]{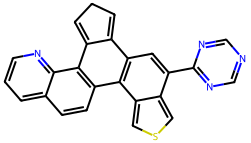}} &
		\centered{\includegraphics[scale=0.15]{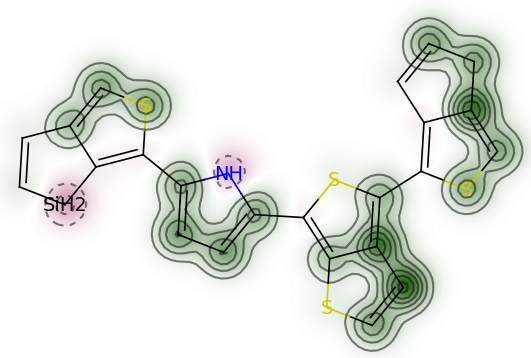}} &
		\centered{\includegraphics[scale=0.15]{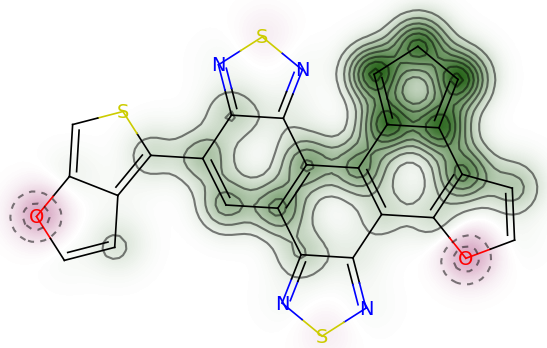}} &
		\centered{\includegraphics[scale=0.15]{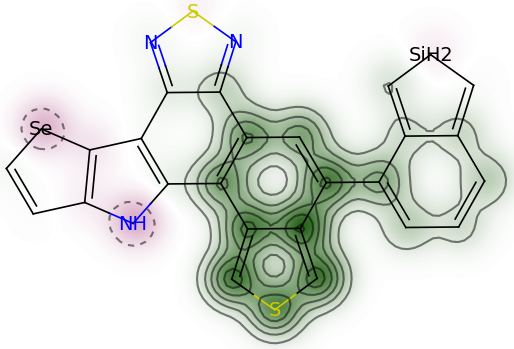}} \\
		\hdashline \\
		
		\centered{\textbf{39}} & \centered{\includegraphics[scale=0.28]{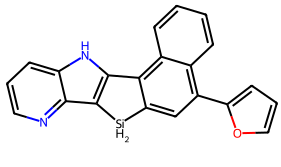}} &
		\centered{\includegraphics[scale=0.16]{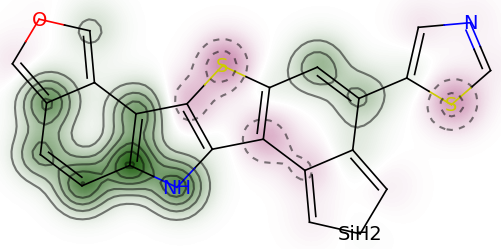}} &
		\centered{\includegraphics[scale=0.16]{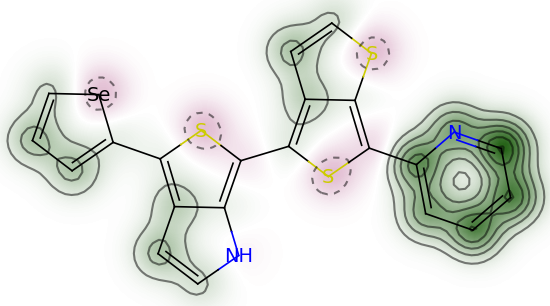}} &
		\centered{\includegraphics[scale=0.16]{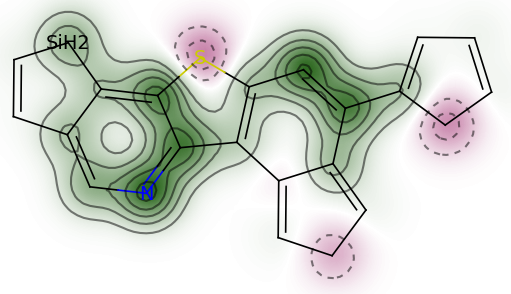}} \\
		\hdashline \\
		
		\centered{\textbf{25}} & \centered{\includegraphics[scale=0.2]{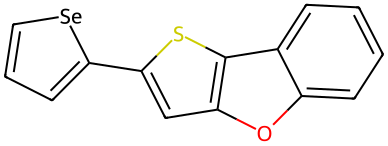}} &
		\centered{\includegraphics[scale=0.17]{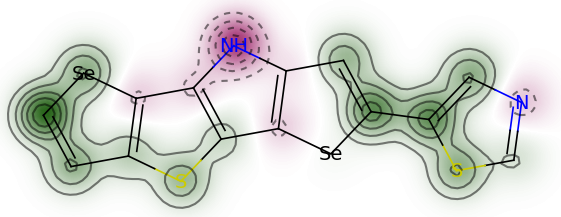}} &
		\centered{\includegraphics[scale=0.17]{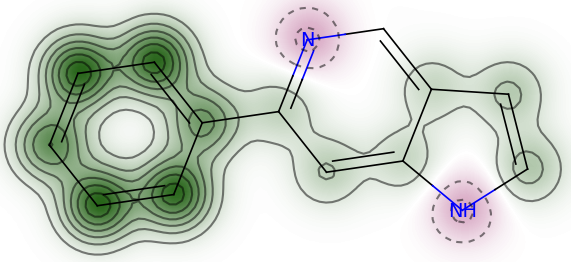}} &
		\centered{\includegraphics[scale=0.17]{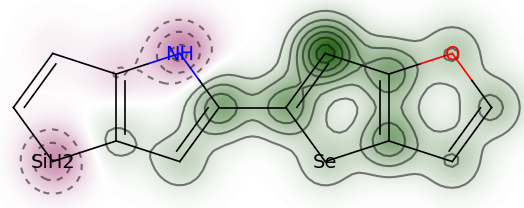}} \\
		\hdashline \\
		
		\centered{\textbf{28}} & \centered{\includegraphics[scale=0.25]{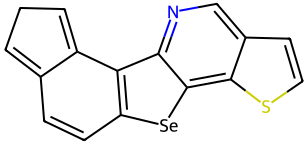}} &
		\centered{\includegraphics[scale=0.17]{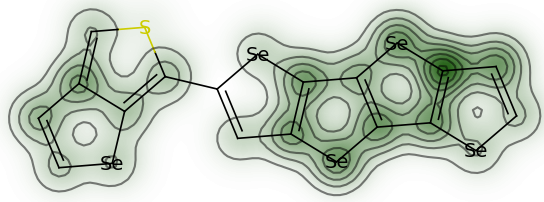}} &
		\centered{\includegraphics[scale=0.17]{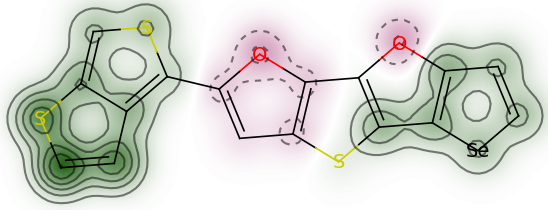}} &
		\centered{\includegraphics[scale=0.17]{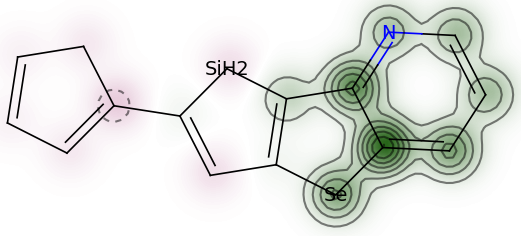}} \\
	\end{longtable}%}
	}
\end{center}

\subsubsection{Geometry and electronic structure visualizations}\label{sec:SI_Geometry_Viz}

This section presents supplementary figures visualizing the results of geometry optimization and electronic structure calculations discussed in Section 3.2 of the main text.

\begin{figure}[H]
\begin{center}
\subfloat[\centering QM9]{\includegraphics[width=.9\textwidth]{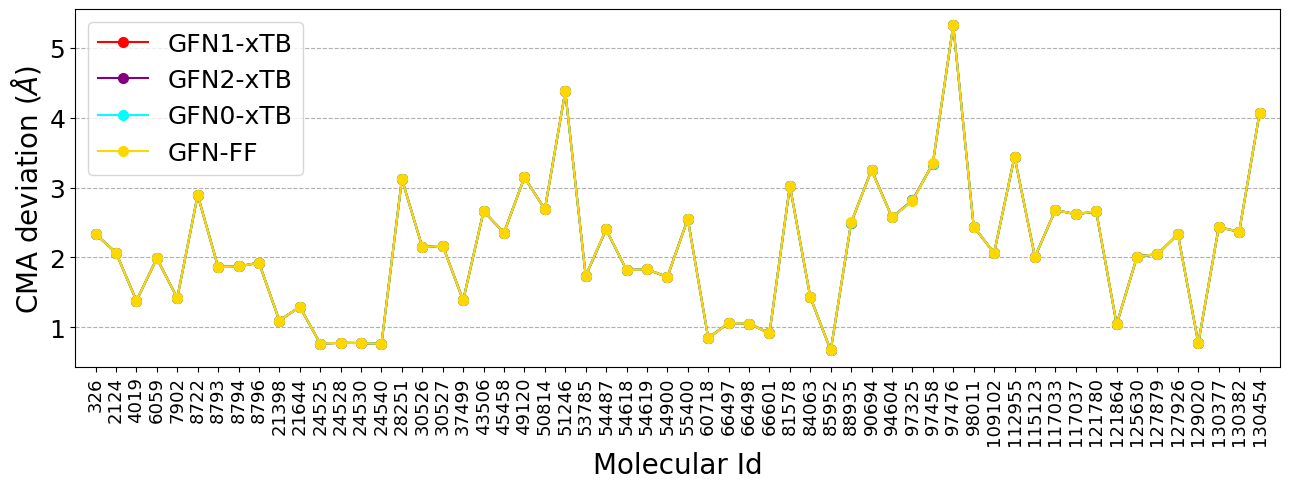}} \\
\subfloat[\centering CEP]{\includegraphics[width=.9\textwidth]{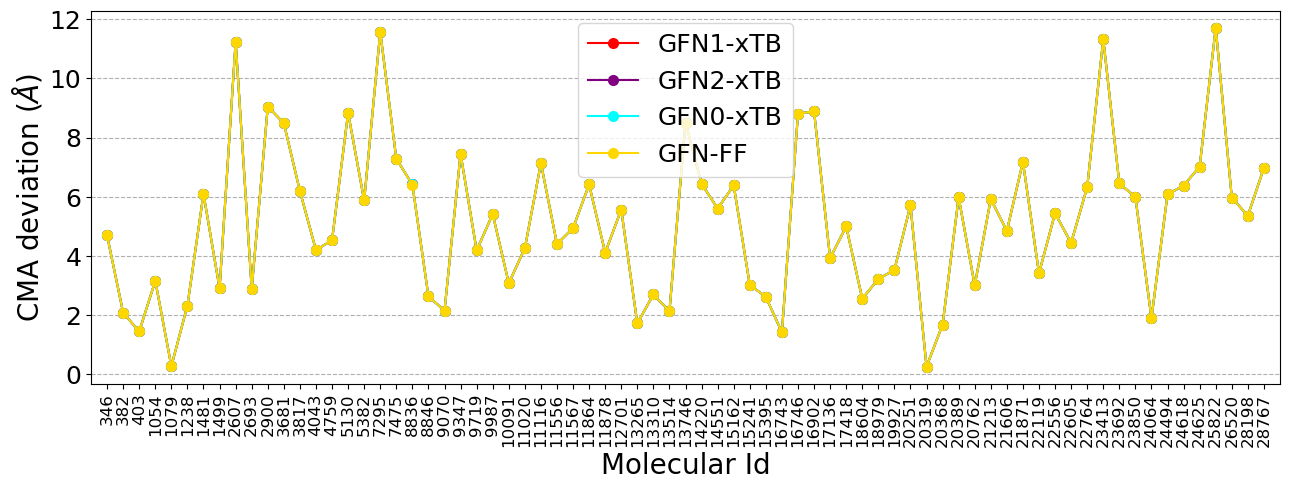}}
\caption{{The center of mass (CMA) deviations measured in \AA of GFN optimized structures with respect to (a) the B3LYP/6-31G(2df, p) level for optimized
		structures of small $\pi$-systems from the QM9 sample set and (b) the BP86/def2-SVP level for optimized structures of extended $\pi$-systems from the CEP
		sample set.
		{\label{fig:SI_S4_CMA}}
}}
\end{center}
\end{figure}

\begin{figure}[H]
\begin{center}
\subfloat[\centering QM9 ]{\includegraphics[width=.9\textwidth]{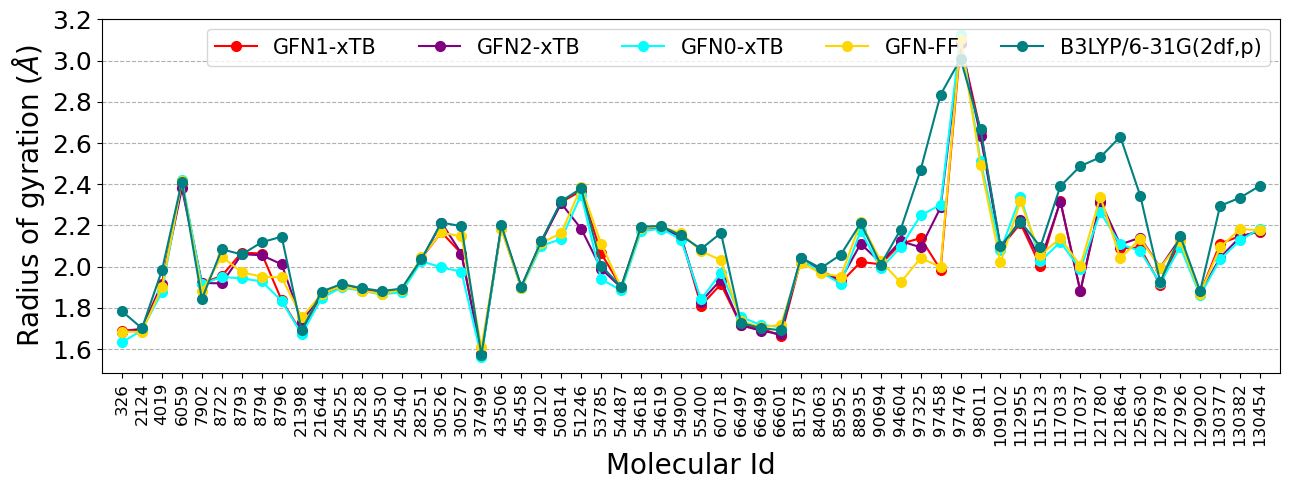}} \\
\subfloat[\centering CEP]{\includegraphics[width=.9\textwidth]{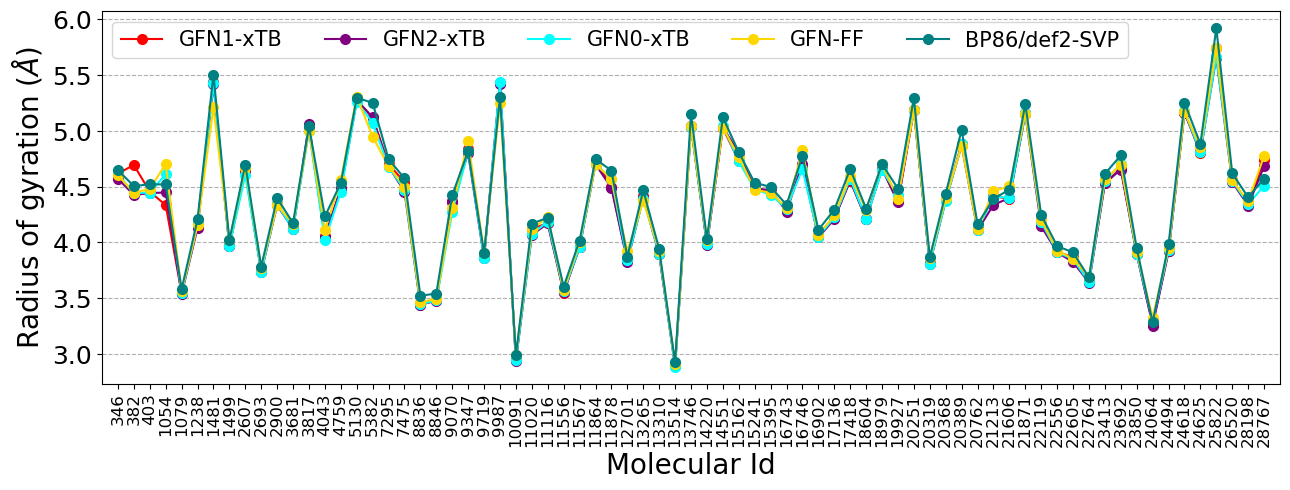}}
\caption{{The molecular radii of gyration (Rg) in \AA of GFN optimized structures and (a) the B3LYP/6-31G(2df, p) level for optimized structures of small
		$\pi$-systems from the QM9 sample set and (b) the BP86/def2-SVP level for optimized structures of extended $\pi$-systems from the CEP sample set.
		{\label{fig:SI_S5_Rg}}
}}
\end{center}
\end{figure}

\begin{figure}[H]
\begin{center}
\includegraphics[width=.9\textwidth]{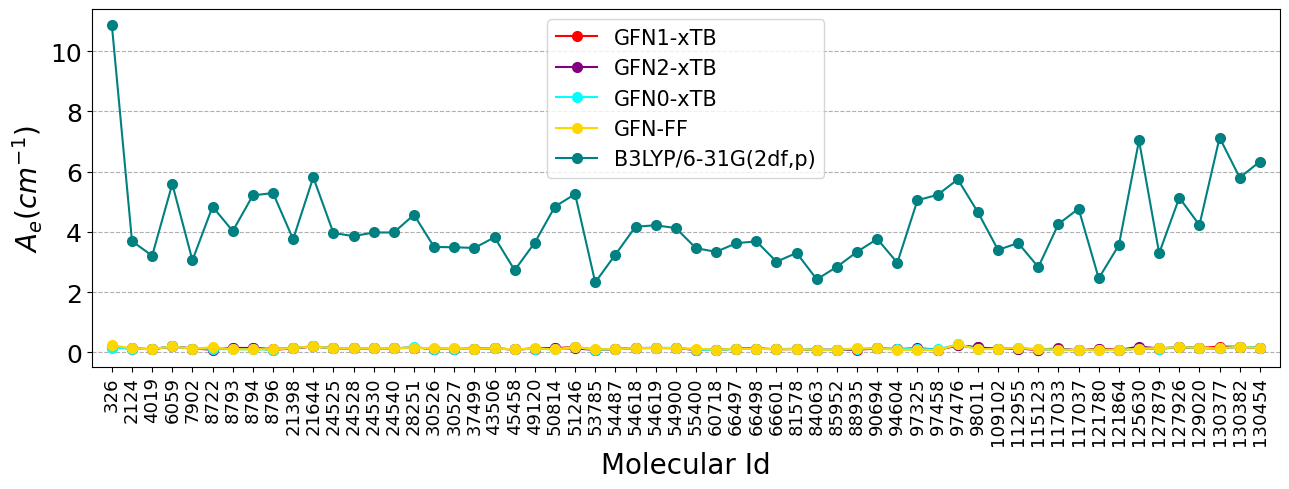} \\
\includegraphics[width=.9\textwidth]{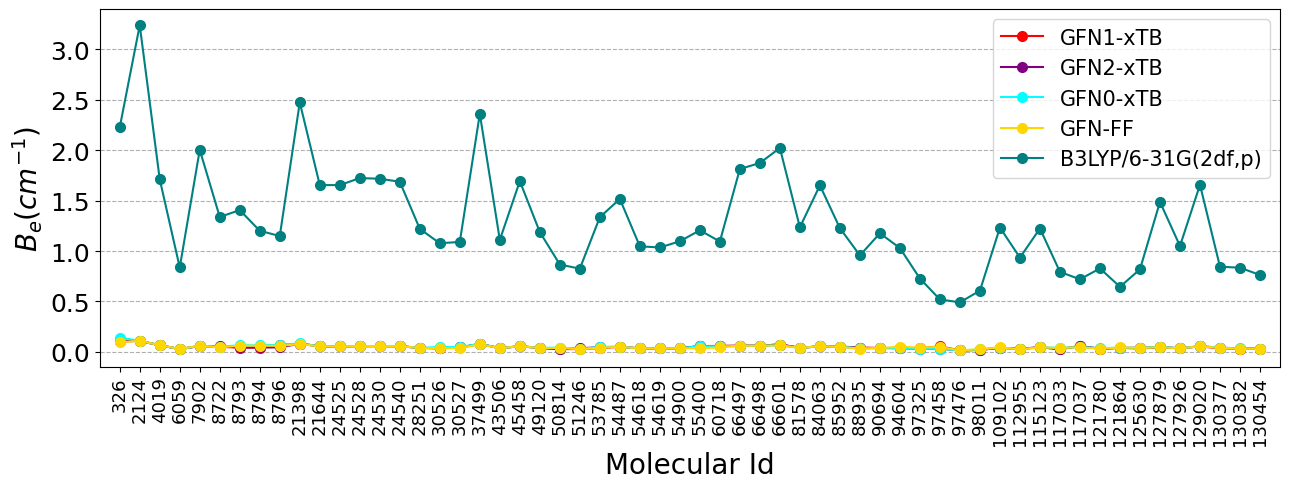} \\
\includegraphics[width=.9\textwidth]{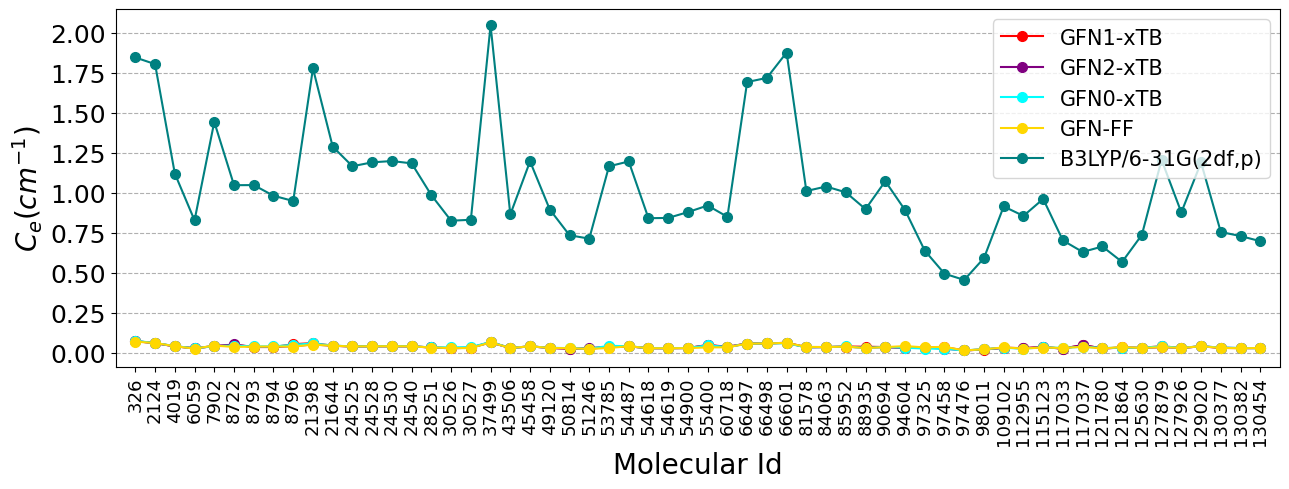}
\caption{{The equilibrium rotational constants $A_e$, $B_e$ and $C_e$, in $\unit{\centi\meter}^{-1}$, computed for the optimized geometries of the
		small $\pi$-systems of the QM9 sample set, using the GFN methods and the B3LYP/6-31G(2df, p) level.
		{\label{fig:SI_S6_QM9_Rot}}
}}
\end{center}
\end{figure}

\begin{figure}[H]
\begin{center}
\includegraphics[width=.9\textwidth]{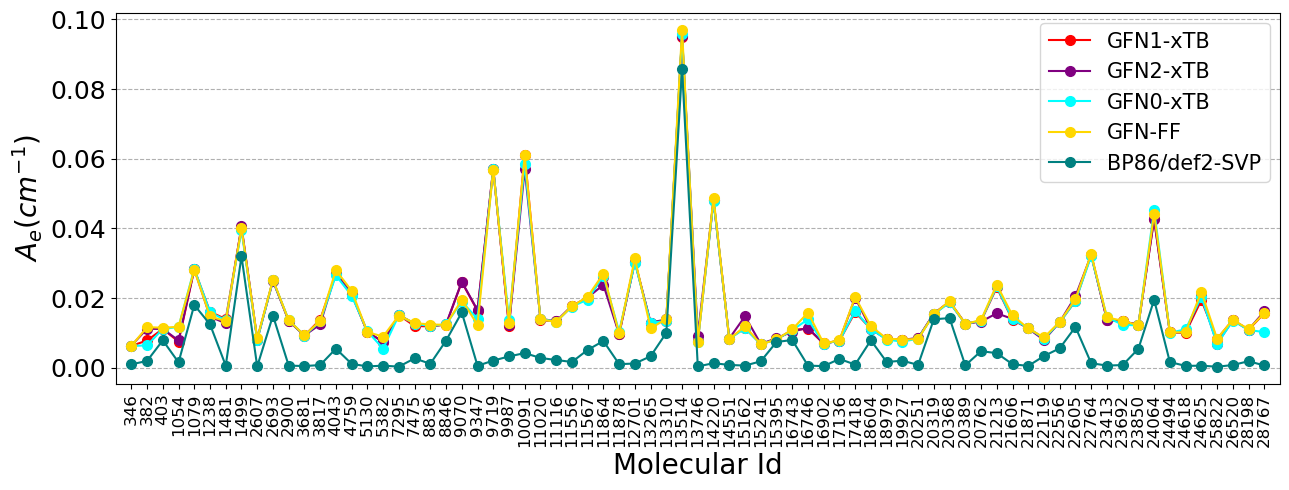} \\
\includegraphics[width=.9\textwidth]{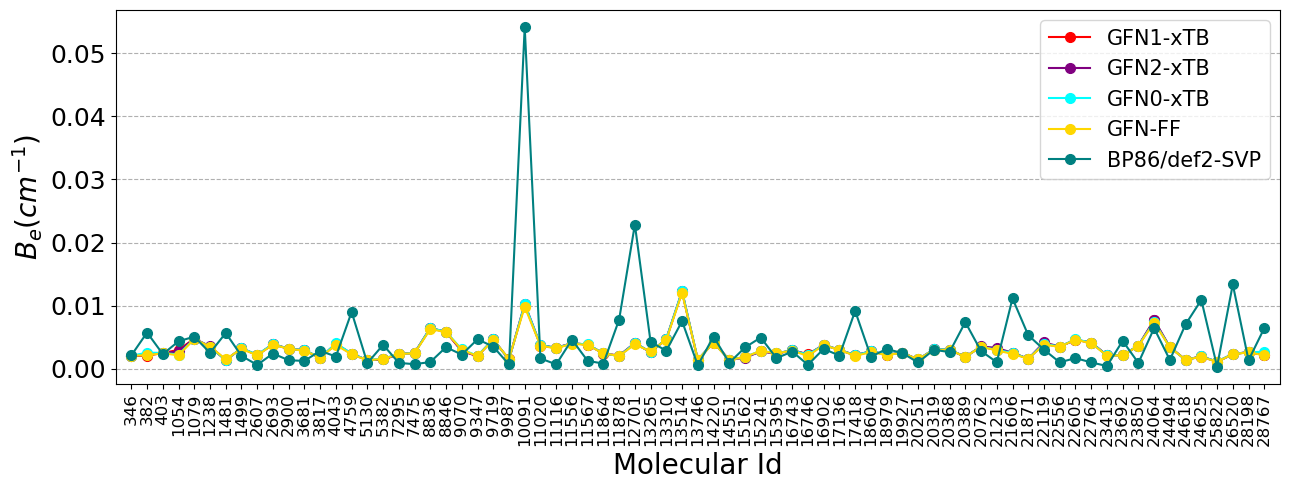} \\
\includegraphics[width=.9\textwidth]{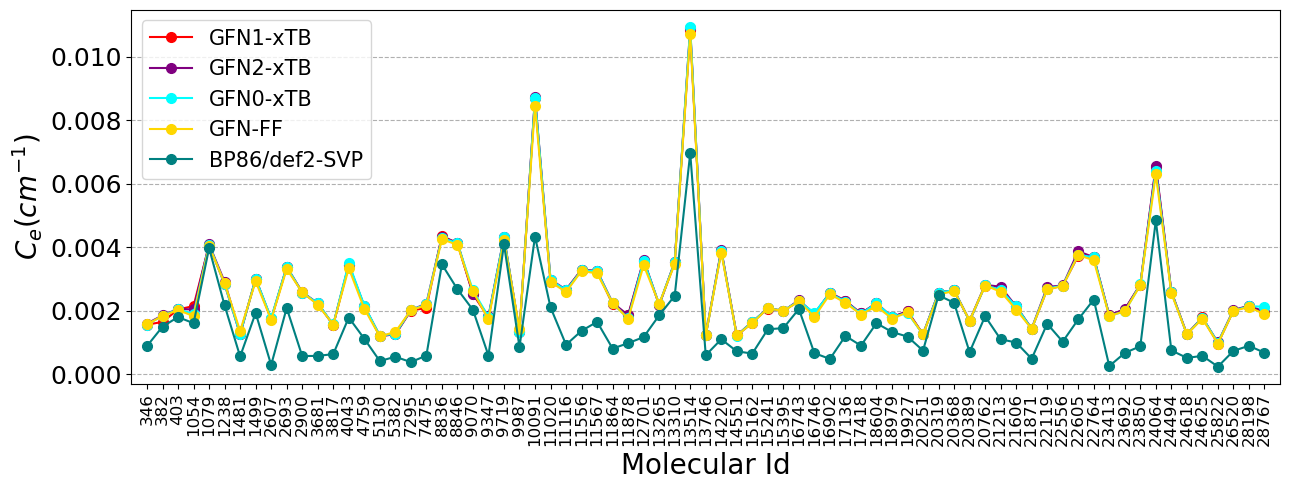}
\caption{{The equilibrium rotational constants $A_e$, $B_e$ and $C_e$, in $\unit{\centi\meter}^{-1}$, computed for the optimized geometries of the
		extended $\pi$-systems of the CEP sample set, using the GFN methods and the BP86/def2-SVP level.
		{\label{fig:SI_S7_CEP_Rot}}
}}
\end{center}
\end{figure}

\begin{figure}[H]
\begin{center}
\subfloat[\centering QM9]{\includegraphics[width=.9\textwidth]{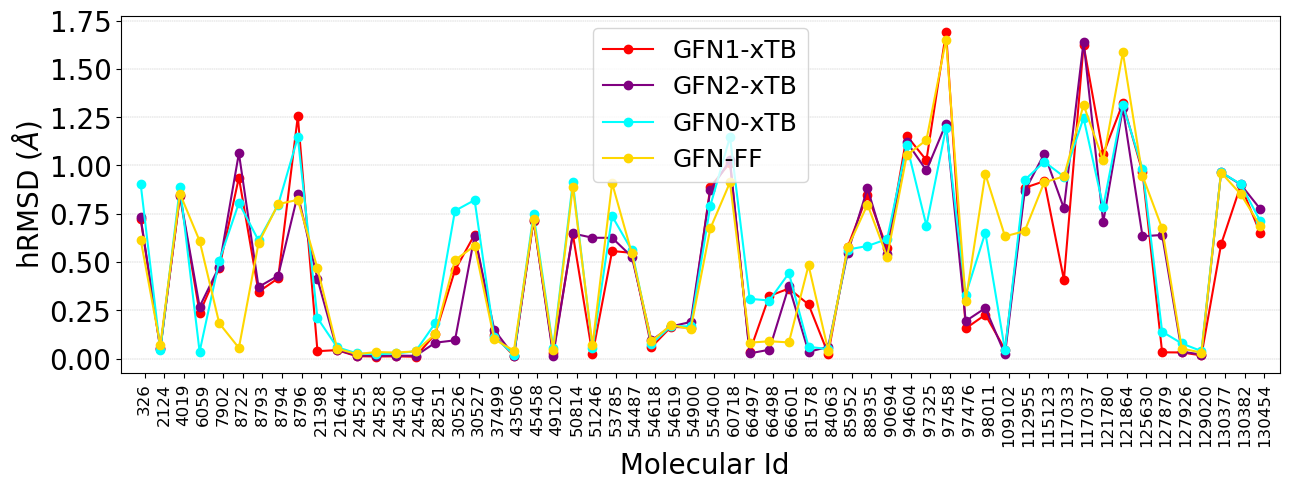}} \\
\subfloat[\centering CEP]{\includegraphics[width=.9\textwidth]{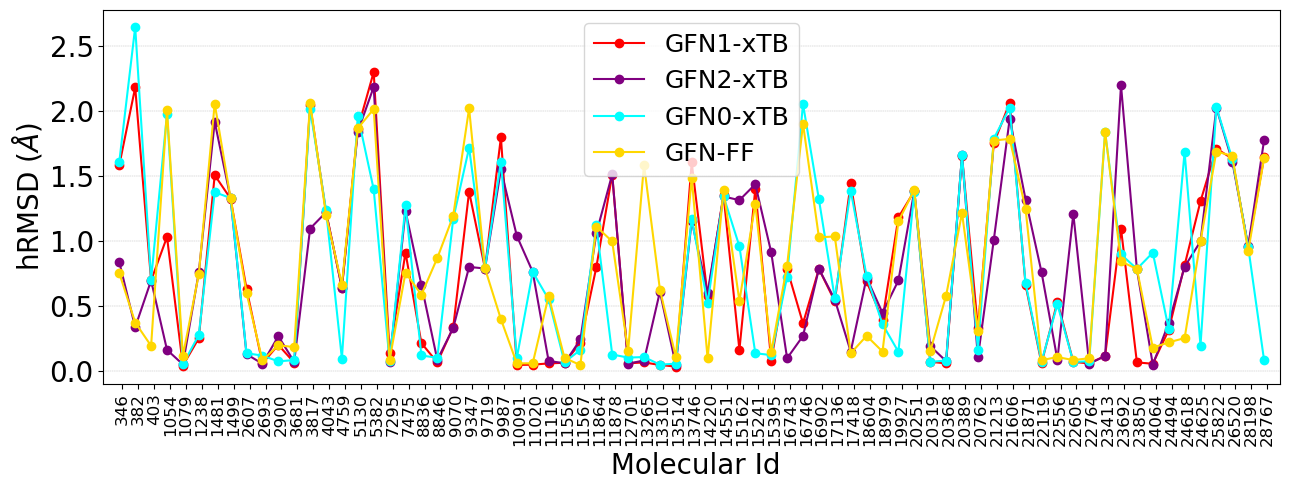}}
\caption{{The heavy-atoms root-mean-square deviations (hRMSD) in \AA of GFN optimized structures with respect to (a) the B3LYP/6-31G(2df, p) level for optimized structures of small $\pi$-systems from the QM9 sample set and (b) the BP86/def2-SVP level for optimized structures of extended $\pi$-systems from the CEP sample set.
		\label{fig:SI_S8_hRMSD}
}}
\end{center}
\end{figure}

\subsubsection{Computational cost and scaling visualizations}\label{sec:SI_Cost_Viz}

This section presents supplementary figures visualizing the computational cost and algorithmic scaling behavior of the GFN methods discussed in Section 3.2.3 of the main text.

\begin{figure}[H]
\begin{center}
\includegraphics[width=.9\textwidth]{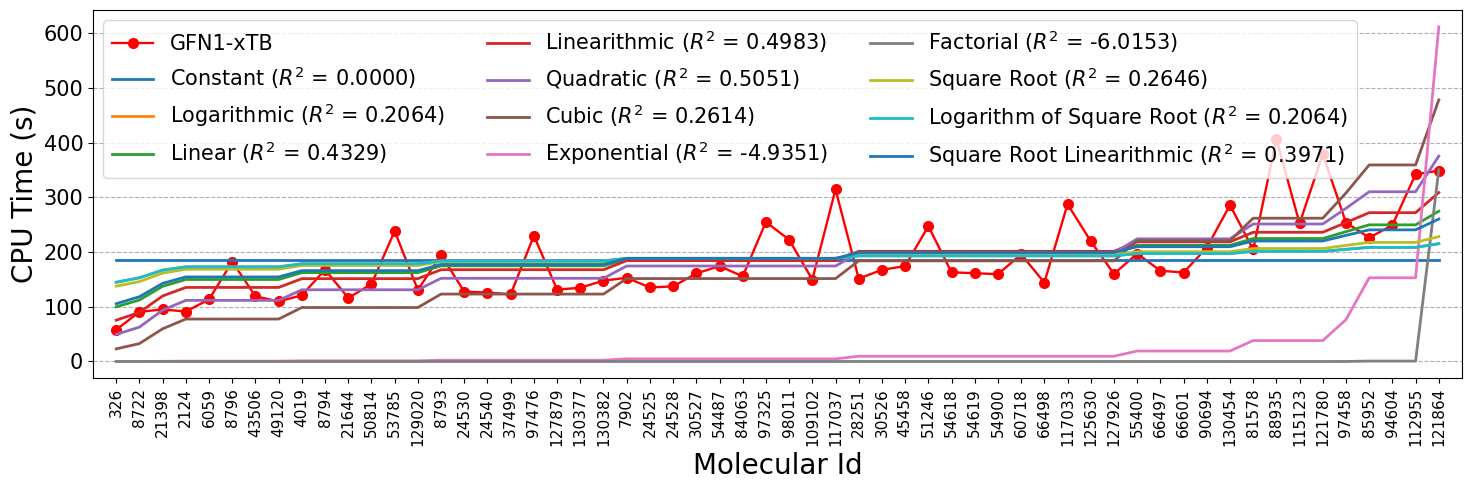} \\
\includegraphics[width=.9\textwidth]{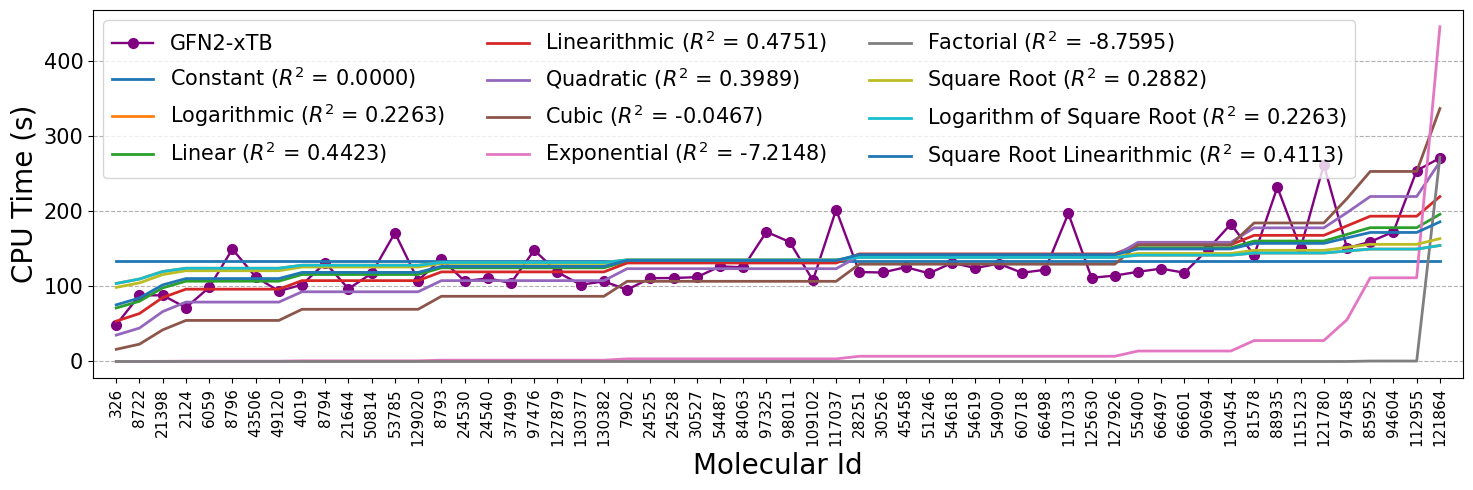} \\
\includegraphics[width=.9\textwidth]{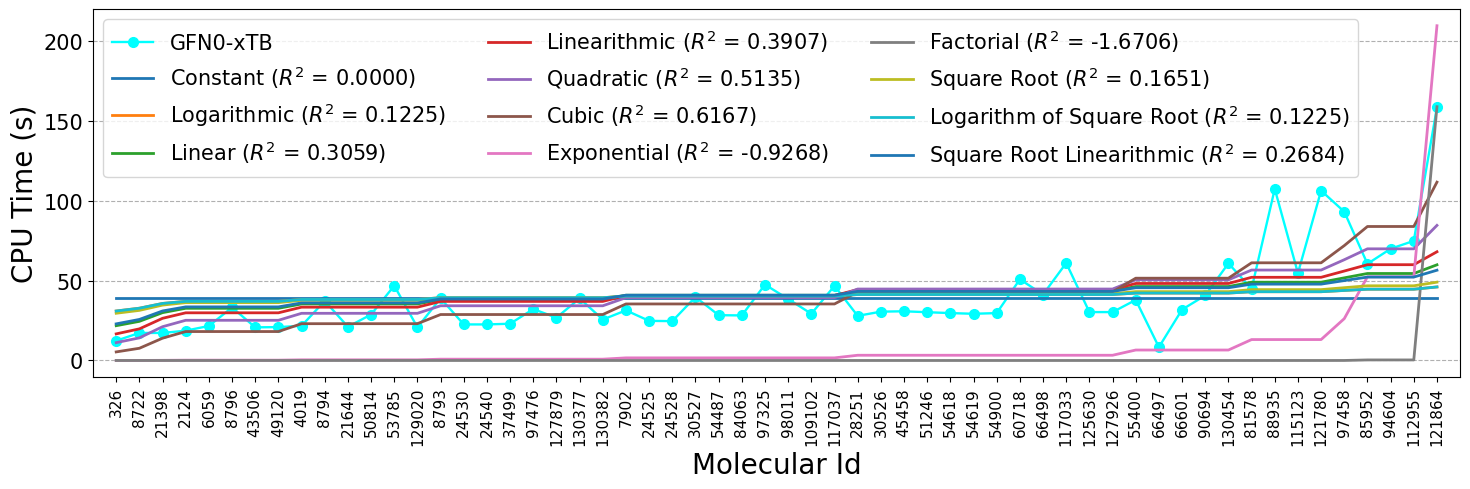} \\
\includegraphics[width=.9\textwidth]{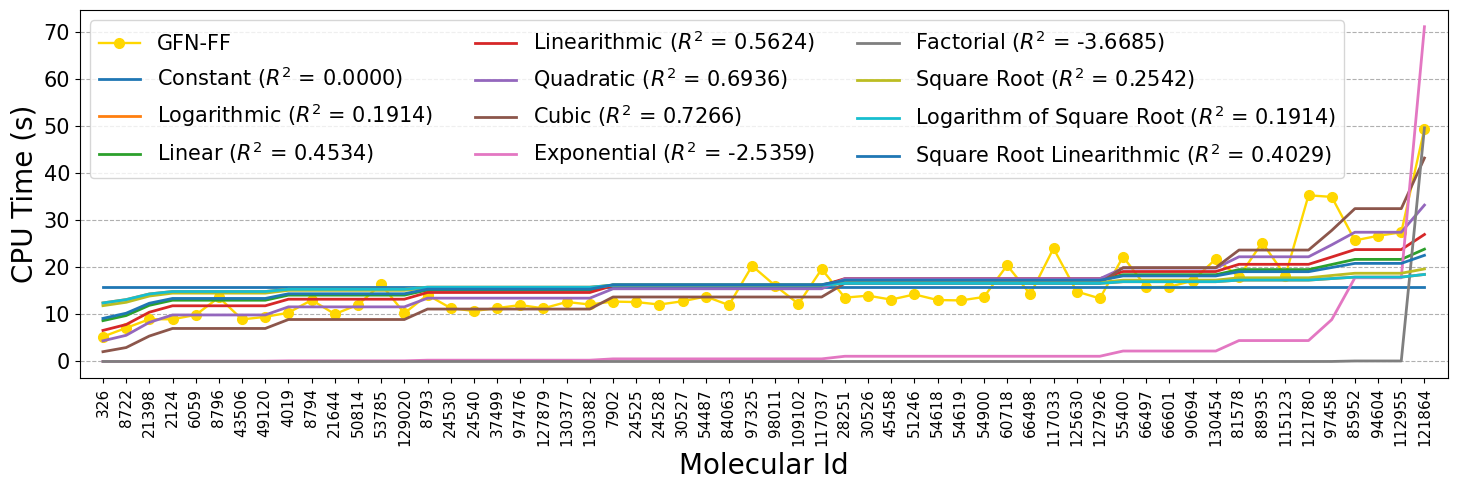}
\caption{{Analysis of the computational complexity of the GFN methods for the QM9 molecular sample set. The four graphs illustrate the CPU time (in seconds) required by each GFN method (from top to bottom: \gfnb, \gfnc, \gfna and \gfnf), sorted by increasing number of molecular atoms. The CPU times of the semiempirical methods are overlaid with several theoretical complexity fits, including constant, logarithmic, linear, linearithmic, quadratic, cubic, exponential, factorial, square root, log square root, and linearithmic square root scaling laws. Each model fit is accompanied by its coefficient of determination ($R^2$), which quantifies the quality of the fit in relation to the observed data.
		\label{fig:SI_S9_QM9_Scaling}
}}
\end{center}
\end{figure}

\begin{figure}[H]
\begin{center}
\includegraphics[width=.9\textwidth]{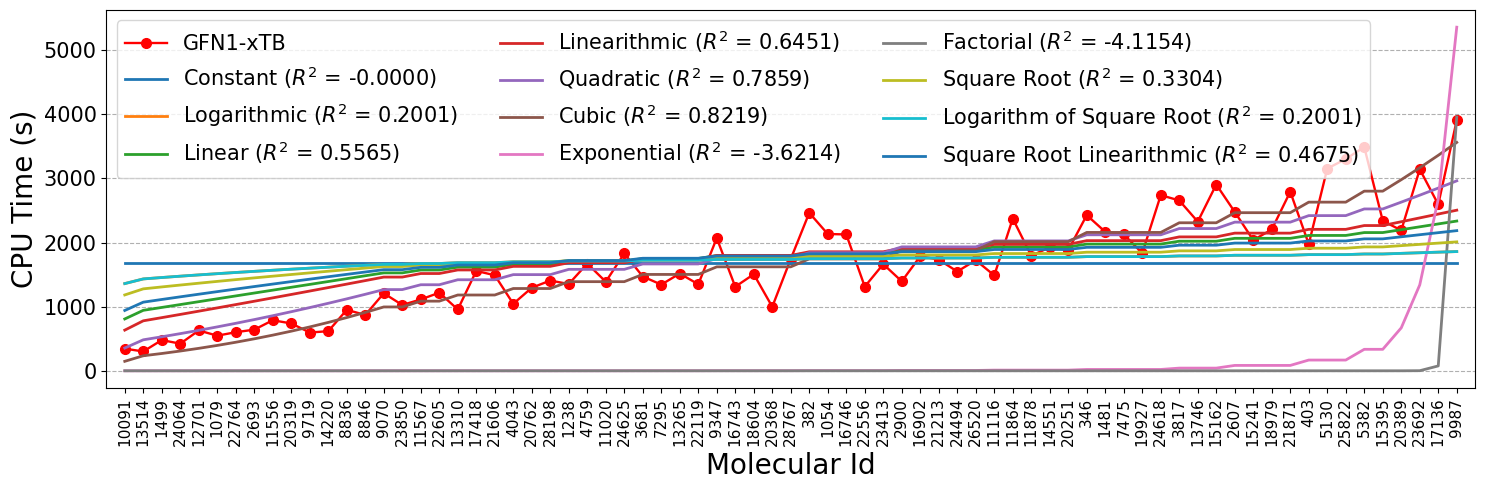} \\
\includegraphics[width=.9\textwidth]{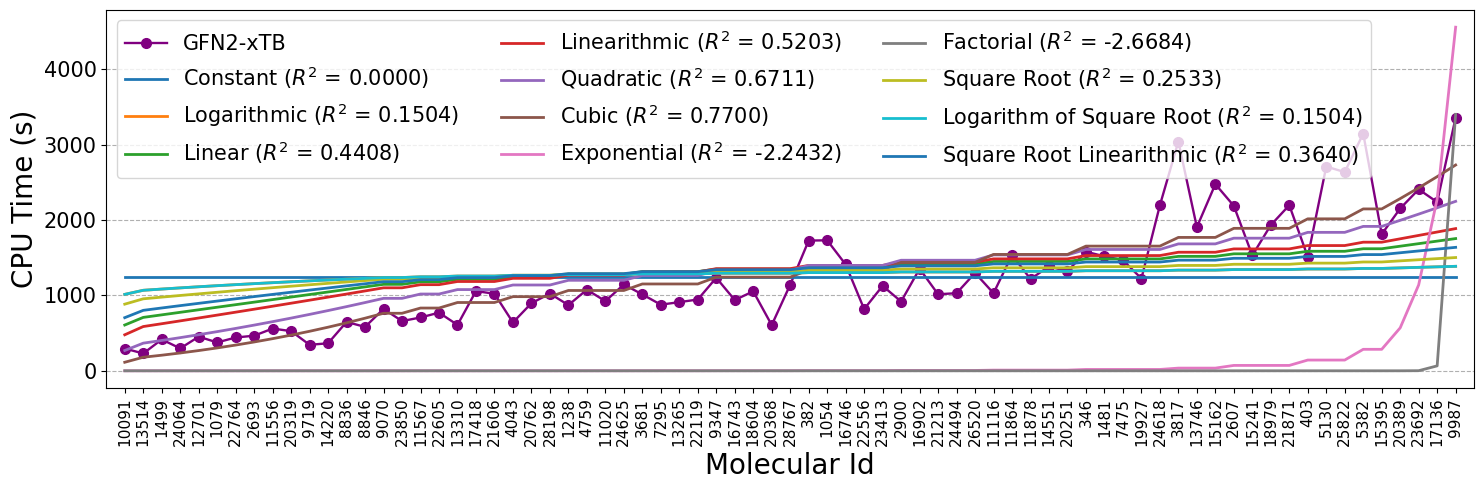} \\
\includegraphics[width=.9\textwidth]{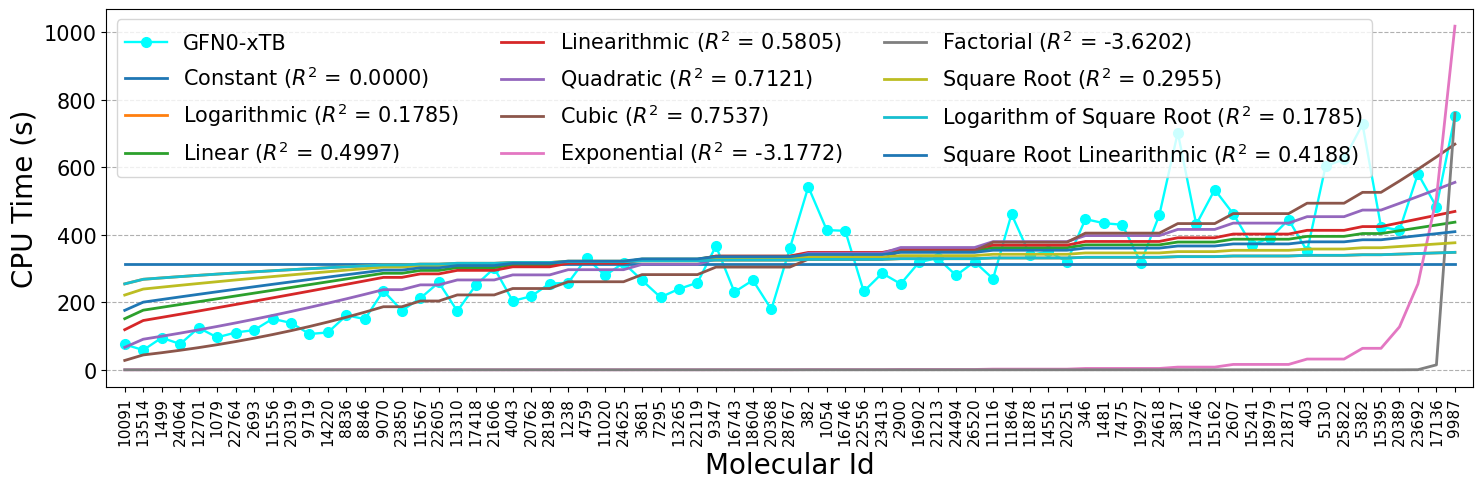} \\
\includegraphics[width=.9\textwidth]{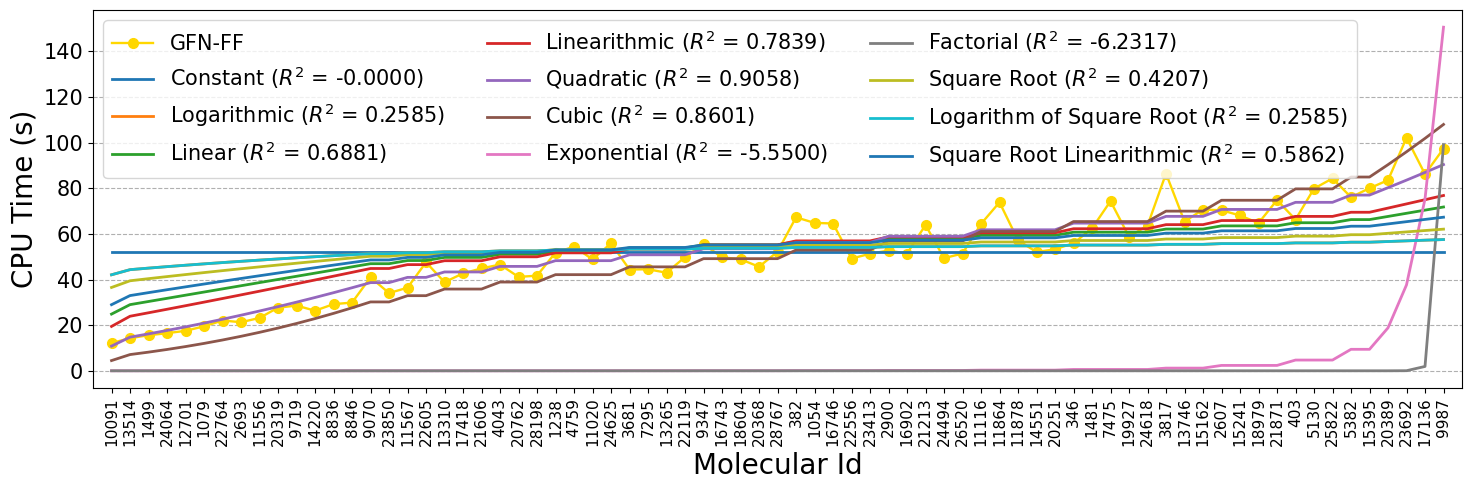}
\caption{{Analysis of the computational complexity of the GFN methods for the CEP molecular sample set. The four graphs illustrate the CPU time (in seconds) required by each GFN method (from top to bottom: \gfnb, \gfnc, \gfna and \gfnf), sorted by increasing number of molecular atoms. The CPU times of the semiempirical methods are overlaid with several theoretical complexity fits, including constant, logarithmic, linear, linearithmic, quadratic, cubic, exponential, factorial, square root, log square root, and linearithmic square root scaling laws. Each model fit is accompanied by its coefficient of determination ($R^2$), which quantifies the quality of the fit in relation to the observed data.
		\label{fig:SI_S10_CEP_Scaling}
}}
\end{center}
\end{figure}

\subsubsection{Excluded molecules}\label{sec:SI_Excluded_Molecules}

This section provides tables listing the IDs of molecules that were excluded from the analysis due to optimization failures or significant structural discrepancies (VF2 failures), as discussed in Section 3.2 of the main text.

\begin{center}
\setlength{\arrayrulewidth}{1pt}
\begin{longtable}[H]{p{1.5cm} p{10cm}}
\caption{List of molecules from the QM9 and CEP sample sets excluded from the entire statistical analysis due to geometry optimization failures.}\label{tab:SI_S5_Optimization_Exclusions}\\
\toprule
\centered{\textbf{Dataset ID}} & \centered{\textbf{Notes on Failure Type}}\\\midrule \endfirsthead\\\midrule
\centered{\textbf{Dataset ID}} & \centered{\textbf{Notes on Failure Type}}\\\midrule \endhead
\bottomrule
\endfoot
\bottomrule
\endlastfoot

\multicolumn{2}{c}{\textbf{QM9 Sample Exclusions (Optimization Failures)}} \\\midrule
\centering 6683 & CREST conformational sampling convergence failure \\

\midrule
\multicolumn{2}{c}{\textbf{CEP Sample Exclusions (Optimization Failures)}} \\\midrule
\centering 161 & \xtb optimization convergence error (gradient threshold exceeded) \\
\centering 18406 & \xtb optimization convergence error (gradient threshold exceeded) \\
\centering 23308 & \xtb optimization convergence error (gradient threshold exceeded) \\
\end{longtable}
\end{center}

\begin{center}
\setlength{\arrayrulewidth}{1pt}
\begin{longtable}[H]{p{1.5cm} p{10cm}}
\caption{List of molecules from the QM9 and CEP sample sets excluded from the statistical analysis due to warnings or persistent issues during structure processing. Molecules exhibiting VF2 algorithm failures were specifically excluded from bond length and angle analyses due to significant topological discrepancies compared to DFT reference structures.\label{tab:SI_S6_VF2_Exclusions}}\\
\toprule
\centered{\textbf{Dataset ID}} & \centered{\textbf{Notes on Discrepancy}}\\\midrule \endfirsthead\\\midrule
\centered{\textbf{Dataset ID}} & \centered{\textbf{Notes on Discrepancy}}\\\midrule \endhead
\bottomrule
\endfoot
\bottomrule
\endlastfoot

\multicolumn{2}{c}{\textbf{QM9 Sample Exclusions}} \\\midrule
\centering 26912 & OpenBabel failed to Kekulize aromatic bonds and to perceive bond orders \\
\centering 30526 & VF2 mapping failure due to discrepancy in bond connectivity \\
\centering 30527 & VF2 mapping failure due to discrepancy in bond connectivity \\
\centering 37499 & VF2 mapping failure due to discrepancy in bond connectivity \\
\centering 54618 & VF2 mapping failure due to discrepancy in bond connectivity \\
\centering 54900 & VF2 mapping failure due to discrepancy in bond connectivity \\
\centering 60718 & VF2 mapping failure due to discrepancy in bond connectivity \\
\centering 66497 & VF2 mapping failure due to discrepancy in bond connectivity \\
\centering 66498 & VF2 mapping failure due to discrepancy in bond connectivity \\
\centering 66599 & OpenBabel failed to set stereochemistry as unable to find an available bond \\
\centering 66601 & VF2 mapping failure due to discrepancy in bond connectivity \\
\centering 74202 & OpenBabel failed to Kekulize aromatic bonds and to perceive bond orders \\
\centering 85952 & VF2 mapping failure due to discrepancy in bond connectivity \\
\centering 94604 & VF2 mapping failure due to discrepancy in bond connectivity \\
\centering 97458 & VF2 mapping failure due to discrepancy in bond connectivity \\
\centering 121864 & VF2 mapping failure due to discrepancy in bond connectivity \\
\centering 130454 & VF2 mapping failure due to discrepancy in bond connectivity \\
\centering 130511 & OpenBabel failed to Kekulize aromatic bonds and to perceive bond orders \\
\centering 130518 & OpenBabel failed to Kekulize aromatic bonds and to perceive bond orders \\
% Add any other QM9 IDs from your data if excluded from bond/angle
\midrule
\multicolumn{2}{c}{\textbf{CEP Sample Exclusions}} \\\midrule
\centering 3681 & VF2 mapping failure due to discrepancy in bond connectivity \\
\centering 18604 & VF2 mapping failure due to discrepancy in bond connectivity \\
\centering 23413 & VF2 mapping failure due to discrepancy in bond connectivity \\
% Add any other CEP IDs from your data if excluded from bond/angle
\end{longtable}
\end{center}

\end{document}